\documentclass[prd,aps,amsfonts,showpacs,nofootinbib]{revtex4}
\usepackage{axodraw}
\usepackage{epsfig}
\usepackage{amsmath}
\usepackage{comment}

\newcommand{\ov}{\overline}
\newcommand{\cd}{\! \cdot \!}
\newcommand{\be}{\begin{equation}}
\newcommand{\ee}{\end{equation}}
\newcommand{\ba}{\begin{eqnarray}}
\newcommand{\ea}{\end{eqnarray}}
\newcommand{\msing}{m_{\hbox{\scriptsize sing}}}

\newcommand{\beqa}{\begin{eqnarray}}
\newcommand{\eeqa}{\end{eqnarray}}
\newcommand{\beq}{\begin{equation}}
\newcommand{\eeq}{\end{equation}}
\newcommand{\pslash}{p\hspace{-.5em}/\hspace{.15em}}
\newcommand{\qslash}{q\hspace{-.5em}/\hspace{.15em}}
\usepackage{graphicx}% Include fig.~files
\usepackage{dcolumn}% Align table columns on decimal point
\usepackage{graphics}
\usepackage{comment}
\usepackage[usenames]{color}

\begin{document}

\title{The quark-gluon vertex in Landau gauge QCD:\\
Its role in dynamical chiral symmetry breaking and quark confinement}

\author{Reinhard Alkofer$^1$, Christian S. Fischer$ ^2$, Felipe J. Llanes-Estrada$^3$
 and Kai Schwenzer$ ^1$}
\affiliation{
$ ^1$ Institut f\"ur Physik der Karl-Franzens Universit\"at Graz, 
A-8010 Graz, Austria. \\
$ ^2$ Institut f\"ur Kernphysik der Technische Universit\"at Darmstadt,
 D-64289 Darmstadt, Germany. \\
$ ^3$ Departamento de F\'{\i}sica Te\'orica I de la  Universidad
Complutense, 28040 Madrid, Spain \\
}

\date{\today}

\begin{abstract}
The infrared behavior of the quark-gluon vertex of quenched Landau gauge  QCD
is studied by analyzing its Dyson-Schwinger equation. Building on previously
obtained results for Green functions in the Yang-Mills sector we analytically
derive the existence of power-law infrared singularities for this vertex. We
establish that dynamical chiral symmetry breaking leads to the self-consistent
generation of components of the quark-gluon vertex forbidden when chiral
symmetry is forced to stay in the Wigner-Weyl mode. In the latter case the
running strong coupling assumes an infrared fixed point. If chiral symmetry is
broken, either dynamically or explicitely, the running coupling is infrared
divergent. Based on a truncation for the
quark-gluon vertex Dyson-Schwinger equation which respects the analytically
determined infrared behavior numerical results for the coupled system of the
quark propagator and vertex Dyson-Schwinger equation  are presented. 
The resulting quark mass function as well as the vertex function show only a
very weak dependence on the current quark mass in the deep infrared. From this
we infer by an analysis of the quark-quark scattering kernel a linearly rising
quark potential with an almost mass independent string tension in the case of
broken chiral symmetry. Enforcing chiral symmetry does lead to a Coulomb type
potential. Therefore we conclude that chiral symmetry breaking and confinement
are closely related. Furthermore we discuss aspects of confinement as the
absence of long-range van-der-Waals forces and Casimir scaling. An examination
of experimental data for quarkonia provides further evidence for the viability
of the presented mechanism for quark confinement in the Landau gauge.
\end{abstract}
\maketitle

%%%%%%%%%%%%%%%%%%%%%%%%%%%%%%%%%%%%%%%%%%%%%%%%%%%%%%%%%%%%%%%%%
\section{Introduction}
%%%%%%%%%%%%%%%%%%%%%%%%%%%%%%%%%%%%%%%%%%%%%%%%%%%%%%%%%%%%%%%%%

Despite the progress seen in the last years the mechanism for confinement of
quarks and gluons into colorless hadrons still lacks a detailed understanding.
Based on the old idea that infrared divergences prevent the emission of colored
states from color--singlet states, see {\it e.g.\/} refs.\ 
\cite{Weinberg:1973un,Marciano:1978su}, different functional approaches have
been employed to study the infrared behavior of QCD Green functions. It has
turned out that Landau gauge is hereby the most suited one. Restricting oneself
to the pure gauge sector it is even possible to make definite statements about
the infrared behavior of all $n$-point functions
\cite{Alkofer:2004it,Fischer:2006vf}. 

An important result of these investigations is the fact that the running
coupling as determined from the Yang-Mills vertex functions shows an infrared
fixed point \cite{vonSmekal:1997is,vonSmekal:1997vx,Lerche:2002ep,Pawlowski:2003hq}. This is
obtained by balancing infrared suppressed and infrared divergent proper Green
functions: Depending on the number of ghost and gluon legs, many
of the one-particle irreducible Green's functions are indeed infrared singular,
others are infrared vanishing.

Hereby especially the infrared suppression of the Landau gauge gluon propagator
immediately raises the question about the cause of the linearly rising quark
potential when expressed in terms of Green functions. Additionally, dynamical
breaking of chiral symmetry requires an effective strength of the quark
interaction which, with an infrared suppressed gluon propagator, necessitates an
enhanced quark-gluon vertex \cite{Fischer:2003rp}. 

Therefore the quark-gluon vertex as the central link between the Yang-Mills and
the quark sector of QCD is at the focus of many contemporary studies.
It is also an important ingredient into the quark-antiquark interaction
that is responsible for the formation and properties of bound states. On a
perturbative level, the quark-gluon vertex has been studied in detail in
arbitrary gauge and dimensions in \cite{davydychev:2000rt}. In this work,
however, we are interested in the nonperturbative properties of this vertex.
Lattice studies have granted some first insights into the relative importance
of several components of the vertex at intermediate momenta
\cite{skullerud:2002ge,Lin:2005zd}. In an attempt to understand these lattice
studies some of us \cite{LlanesEstrada:2004jz,Fischer:2004ym} and others
\cite{Bhagwat:2004kj,Bhagwat:2004hn,Maris:2005tt}  have presented
model studies based on resummations of one-loop results.
As we will explain below such a treatment is not sufficient, consequently, a
fully self-consistent treatment is required.

An, at least at first sight, surprising result of such a self-consistent
treatment is that dynamical chiral symmetry breaking takes place in quite a
non-trivial way also in the quark-gluon vertex: This vertex non-perturbatively
assumes components of a Dirac scalar type in a similar self-consistent fashion as
the quark propagator does. Especially, the assignment of a definite
Lorentz structure is then not plausible. In view of the fact that the Lorentz
nature of confinement, apparently indicated by phenomenological spin splittings,
has worried two generations of QCD practitioners (and is still of interest
\cite{Szczepaniak:1996tk,Bicudo:2003ji,Bicudo:2003km,Nefediev:2005dw,Nefediev:2007pc})
this result appears, a posteriori, less surprising.
 
One of our important findings to be reported in this
paper is that these parts of the quark-gluon vertex have a considerable strength in
the infrared. In particular infrared singularities arise not only in the uniform limit 
when all external momenta tend to zero, but also when only the gluon momentum vanishes. 
Interestingly, this soft-gluon singularity is, in contrast to those in the Yang-Mills 
sector \cite{Alkofer:2008jy}, as strong as the uniform divergence.
This implies, of course, a complicated structure of the effective
low-energy quark-quark interaction with the potential to permanently confine quarks. 
The remarks made above provide more than ample motivation for a concise study of
the Landau gauge quark-gluon vertex. As will become evident below such an
investigation reveals many unexpected qualitative features.

This paper is organized as follows: In section II we introduce our notations and
discuss the structure of the functional equations for the quark-gluon vertex. 
In section III we then develop an analytical infrared analysis for the
vertex-DSE built upon our previous experience with the Yang-Mills sector of QCD.
We discuss selfconsistent power law solutions of these equations in the cases
of broken and restored chiral symmetry, investigate the corresponding running
coupling and determine possible dependencies of the infrared solutions on the 
quark masses. In section IV we then present our numerical results thus confirming
our analytical study. The infrared divergent behavior of the 
quark-gluon 
vertex found in our analysis implies a linear rising quark-antiquark potential 
extracted from the quark four-point function. This is demonstrated in section V.
In the following section VI we deal with theoretical aspects of this confining
potential like Casimir scaling and N-ality of the string tension. Phenomenological
aspects are discussed in section VII, before we end with with a summary and 
conclusions. Many of the details of the corresponding calculations are relegated
to various appendices.

%%%%%%%%%%%%%%%%%%%%%%%%%%%%%%%%%%%%%%%%%%%%%%%%%%%%%%%%%%%%%%%%%%%%%%%%%
\section{The quark-gluon vertex Dyson Schwinger equation \label{introqqg}}
%%%%%%%%%%%%%%%%%%%%%%%%%%%%%%%%%%%%%%%%%%%%%%%%%%%%%%%%%%%%%%%%%%%%%%%%%

The fundamental dynamical element in our improved analysis of the quark 
dynamics in quenched QCD is the quark-gluon vertex. In this section we 
discuss its basic definition, the functional equations it obeys as 
well as suitable approximation schemes.

%%%%%%%%%%%%%%%%%%%%%%%%%%%%%%%%%%%%%%%%%%%%%%%%%%%%%%%%%%%%%%%%%
\subsection{Basics}
%%%%%%%%%%%%%%%%%%%%%%%%%%%%%%%%%%%%%%%%%%%%%%%%%%%%%%%%%%%%%%%%%%

The dressed quark-gluon vertex
\be \label{qqgdressed}
\begin{minipage}[c][1\totalheight]{0.25\columnwidth}
$i g T^a_{ij} \Gamma^\mu(p_1^2,p_2^2,p_1\cd p_2) = \ \ \ $
\end{minipage}
\begin{minipage}[c][1\totalheight]{0.15\columnwidth}
\begin{picture}(80,80)(0,0)
\ArrowLine(0,0)(40,40) \put(0,15){$p_1$}
\ArrowLine(40,40)(80,0) \put(70,15){$p_2$}
\Gluon(40,80)(40,40){3}{6}
\GCirc(40,40){5}{0} 
\end{picture} 
\end{minipage}
\ee
is given by a combination of the twelve independent tensors that 
can be built from the two independent momenta $p_1$ and $p_2$ and the Dirac
gamma matrices. Since the pioneering paper by Ball and Chiu 
\cite{Ball:1980ay}
it is somewhat standard to divide this basis into a part transverse to the 
gluon momentum, $T_{i=1..8}$, and a so called \lq \lq longitudinal'' part $L_{i=1..4}$. 
The latter is not really longitudinal as it contains both longitudinal 
and transverse parts. The resulting basis is complete but not orthonormal 
and is explicitly given in eq. (\ref{app:tensor}) in the appendix. It contains 
the tree level tensor $\gamma_\mu$ that is a vector in Dirac space as well 
as scalar parts like $i(p_1+p_2)^\mu$ that are not chirally symmetric analog 
to a scalar mass term in the quark propagator. Introducing a set of scalar 
functions $\lambda_i$, $\tau_i$ of the three scalar variables $p_1^2$, 
$p_2^2$, $(p_2-p_1)^2$, the vertex reads
\be
\Gamma^\mu=\sum_{i=1}^4 \lambda_i L_i^\mu + \sum_{i=1}^8 \tau_i T_i^\mu \ . \label{eq:tensor-dec}
\ee
In the course of this paper we will give numerical results for the dressing
functions $\lambda_i$ and $\tau_i$. However, to demonstrate certain properties of the
vertex at some places we resort to ans\"atze for these functions. To avoid
confusion in the diagrammatics we represent the quark-gluon vertex by an open circle
\begin{center}
\begin{picture}(80,80)(0,0)
\ArrowLine(0,0)(40,40) \put(0,15){$p_1$}
\ArrowLine(40,40)(80,0) \put(70,15){$p_2$}
\Gluon(40,80)(40,40){3}{6}
\BCirc(40,40){5} 
\end{picture} 
\end{center}
whenever we use an ansatz instead of the fully calculated vertex given in 
eq. (\ref{qqgdressed}).

In addition we use the following diagrammatic expressions for the full 
propagators of QCD. The dressed quark propagator is given by
\be \label{qq}
\begin{picture}(60,20)(0,8)
\ArrowLine(0,10)(60,10) \put(28,0){$p$} \put(0,13){$i$}\put(58,13){$j$}
\GCirc(30,10){3}{0}
\end{picture} \  \  =
(i \!\not p \sigma_v(p^2)+\sigma_s(p^2)) \,\, \delta_{ij}^{\rm color}
\ee
with the vector part $\sigma_v(p^2)$ and the scalar part $\sigma_s(p^2)$.
These functions can be expressed in terms of the quark mass function $M(p^2)$
and the quark wave function $Z_f(p^2)$ via
\beq
\sigma_v(p^2) = \frac{Z_f(p^2)}{p^2 + M^2(p^2)}; \ \
\sigma_s(p^2) = \frac{Z_f(p^2) M(p^2)}{p^2 + M^2(p^2)}
\eeq  
that are also expressed in this work in terms of the functions 
$A(p^2)=1/Z(p^2)$ and $B(p^2)=Z(p^2)M(p^2)$.
The dressed gluon propagator is given by
\be \label{gg}
\begin{picture}(60,20)(0,8)
\Gluon(0,10)(60,10){3}{9}
\GCirc(30,10){3}{0} 
\put(28,0){$p$} \put(0,15){$a$}\put(58,15){$b$}
\put(0,0){$\mu$} \put(58,0){$\nu$}
\end{picture} \  \  = P_{\mu \nu}(p) \frac{Z(p^2)}{p^2}\delta_{ab}^{\rm color} 
\ee
with the gluon dressing function $Z(p^2)$ and the transverse projector 
$P_{\mu \nu}(p) = \left(\delta_{\mu \nu} -\frac{p_\mu p_\nu}{p^2}\right)$.
Finally, the ghost propagator is given by 
\be \label{ghgh}
\begin{picture}(60,20)(0,8)
\DashArrowLine(0,10)(60,10){3}
\GCirc(30,10){3}{0} 
\put(28,0){$p$} \put(0,15){$a$}\put(58,15){$b$}
\put(0,0){$\mu$} \put(58,0){$\nu$}
\end{picture} \  \  =
- \frac{G(p^2)}{p^2}\delta_{ab}^{\rm color} \; .
\ee
It should be stressed that throughout this manuscript all propagators 
appearing in loop diagrams are to be taken as dressed, although we do 
not write the full circles in these cases so as not to clutter the diagrams.
We also frequently refer to higher Green's function with more than three 
external legs. Here we generally denote one-particle irreducible (1PI) and connected functions by
filled circles and ellipses, respectively, whereas connected or two-particle irreducible 
functions (2PI) are denoted by filled rectangles.

%%%%%%%%%%%%%%%%%%%%%%%%%%%%%%%%%%%%%%%%%%%%%%%%%%%%%%%%%%%%%%%%%
\subsection{Functional equations}
%%%%%%%%%%%%%%%%%%%%%%%%%%%%%%%%%%%%%%%%%%%%%%%%%%%%%%%%%%%%%%%%%%

The Dyson-Schwinger equations (DSEs) are given as an infinite tower of coupled 
nonlinear integral equations, see 
\cite{Roberts:dr,Alkofer:2000wg,Maris:2003vk,Fischer:2006ub} for reviews. 
Therefore most explicit computations require
a specification of an approximation scheme. For such a scheme it is desirable
to respect all symmetries of the underlying theory\footnote{Since the 
Lagrangian density with minimum derivative number and all the 
symmetries of QCD is QCD itself, any truncation of the equations of 
motion either 1) breaks at least one symmetry, or 2) is not derivable from a 
Lagrangian density, or 3) includes higher derivative terms.}.

\begin{figure}[h]
\centerline{\epsfig{file=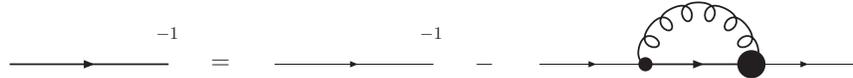,width=12cm}}
\caption{DSE for the quark propagator including one bare and one proper quark-gluon vertex.
\label{quarkdse0}}
\end{figure}

The simplest DSE in the quark sector is the one for the quark propagator 
given in fig.~\ref{quarkdse0} which involves one bare and one dressed 
quark-gluon vertex. This equation has been studied in numerous works in
the past. Of particular relevance for our work is ref.~\cite{Fischer:2003rp},
where it has been shown that the vertex needs to be enhanced in the 
nonperturbative momentum region to provide for dynamical chiral symmetry 
breaking.

\begin{figure}[h]
\centerline{\epsfig{file=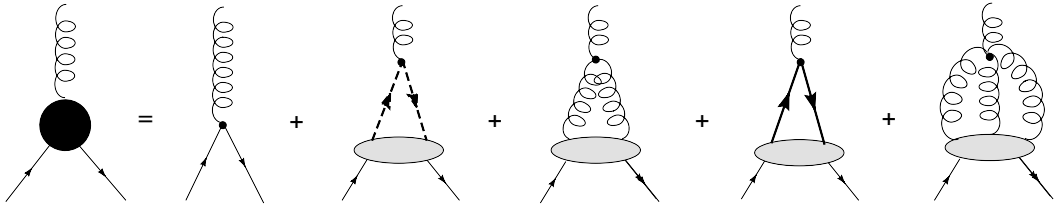,width=140mm}}
\caption{The DSE for the quark-gluon vertex where the r.h.s is written in terms of 
connected Green's functions.
All internal propagators are fully dressed, also in all following 
diagrams. Note symmetry factors (1/2, 1/6) and signs have been omitted.
\label{vertex-DSE0}}
\end{figure}

\begin{figure}[h]
\centerline{\epsfig{file=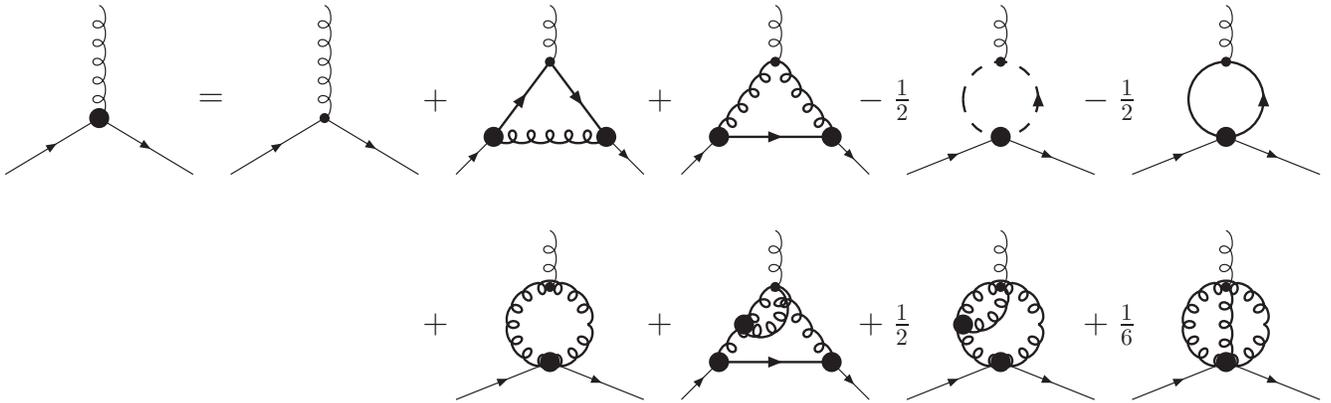,width=18cm}}
\caption{First version of the DSE for the quark-gluon vertex written in terms of 
proper Green's functions. \label{DS1}}
\end{figure}

\begin{figure}[h]
\centerline{\epsfig{file=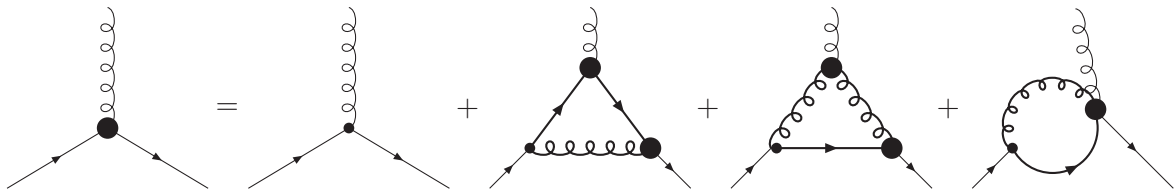,width=16cm}}
\caption{Second version of the DSE for the quark-gluon vertex written entirely in terms of 
proper Green's functions. \label{DS2}}
\end{figure}

The quark-gluon vertex satisfies a full Dyson-Schwinger equation given
in fig.~\ref{vertex-DSE0} (see 
\cite{Marciano:1978su} 
and references therein).
This equation couples the proper quark-gluon vertex to higher 
connected four and five-point functions (the quark-gluon and quark-ghost scattering 
kernel, the two-quark-three gluon scattering kernel, and the quark-quark scattering 
kernel). In terms of one-particle irreducible (1PI) Green's functions the DSE, 
fig.~\ref{vertex-DSE0}, can be rewritten as given in fig.~\ref{DS1}. 
These equations
can also be directly derived in graphical form as outlined in 
\cite{Pawlowski:2005xe}. 
In both equations it is the external gluon leg that is connected to a bare vertex in the
loop diagrams. Corresponding equations with external quark legs attached to the bare 
internal vertex exist. One possibility is given in fig.~\ref{DS2}.
Certainly, in the full theory both equations should give the same quark-gluon vertex.
However, in a truncated theory, one of these two equations may be preferred over the other
and more straightforward to solve. We deal with this point in appendix \ref{app:secondDSE}.

One way to ensure that an approximation scheme respects at least those symmetries that are 
represented linearly, is to derive it from an action formalism. This 
holds for schemes based on  $n$-particle irreducible ($n$PI) actions 
\cite{Berges:2004pu}. 
Such actions depend not only on the averaged field value but also on 
higher averaged $n$-point correlators thereby generating DSE-like equations via 
stationarity conditions. They are generalizations of the 
Cornwall-Jackiw-Tomboulis (CJT) action 
\cite{Luttinger:1960ua,Baym:1962sx,Cornwall:1974vz,Norton:1974bm} 
which 
represents the order $n=2$. Approximations based on the CJT action, 
which are also 
refered to as ``$\Phi$-derivable'', always involve fully dressed 
propagators whereas 
approximations based on an nPI-action involve dressed correlation functions with up 
to $n$ legs. 
 
There are basically two ways to truncate $n$PI actions based either on loop or 
large-$N$ expansions. The loop expansion of the 3PI action of QCD has been studied 
in 
\cite{Berges:2004pu} 
up to 3-loop order. The result has the form of the standard 
CJT action which reads formally in Euclidean space
\begin{equation}
  \Gamma_{3PI}= S + \frac{1}{2} \mathrm{STr} \log \left( {\cal D} \right)
  +\frac{1}{2} \mathrm{STr} \left( {\cal D} \,{\cal D}_0^{-1} \right) +\Gamma^{(2)}
\end{equation}
where STr is a super trace over the generalized two point function $\cal D$ for the 
supermultiplet $\phi$ that includes the fields in the QCD action 
$\phi=(A, \bar c, c, \bar q,q)$. The logarithmic term generates fully dressed 1-loop 
contributions whereas the 2PI-part $\Gamma^{(2)}$ of the action is represented 
graphically in Figs.\ 1-4 of 
ref.\ \cite{Berges:2004pu}. 
Self-consistent equations for the 
Green's functions of the theory are then obtained by variation with respect to 
the full propagators and vertices. 
The interesting result is that the truncated self-consistency equations are nearly 
identical to the DSE system truncated at the 3-point level when neglecting graphs 
with higher order vertices, but with the decisive difference that all vertices that 
enter the truncated self-consistency equations are dressed. 
\begin{figure}[h]
\centerline{\epsfig{file=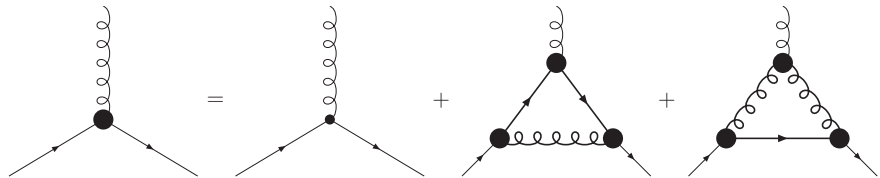 ,width=12cm}}
\caption{Equation of motion of the quark-gluon vertex as derived from a 3PI effective action. 
\label{DS3}}
\end{figure}
We will later on confirm that this difference is not crucial when the
3PI equation is compared to the first version, fig.~\ref{DS1}, of the 
quark-gluon 
vertex DSE. In section \ref{sec:self} we will devise a truncation scheme for this 
DSE which reproduces the infrared behavior of both, the full DSE 
fig.~\ref{vertex-DSE0} 
{\bf and} the 3PI equation. However, things are slightly more 
complicated for the second version, fig.~\ref{DS2}, of the quark-gluon 
vertex DSE. We discuss this point in more detail in appendix 
\ref{app:secondDSE}, where we argue that nevertheless all three 
equations lead to the same result in the 
infrared, in the presence of the ``soft divergences'' studied in 
subsection \ref{softgluondiv} below.

Finally we would like to mention that another important approach to 1PI Green's 
functions is given by the functional renormalization group equations (RGEs; for 
a detailed overview on this technique see 
\cite{Pawlowski:2005xe,Litim:1998nf,Gies:2006wv}). 
These have 
been successfully used in the past as an alternative technique to study the 
infrared behavior of Yang-Mills theory and implications for confinement 
\cite{Fischer:2006vf,Pawlowski:2003hq,Fischer:2004uk,Braun:2007bx}. The functional 
RGE for the quark-gluon vertex comprises of one-loop diagrams with all vertices 
dressed. Since this property is also captured by the $nPI$ equations, we will 
not explicitly discuss these RGEs within this work.

%%%%%%%%%%%%%%%%%%%%%%%%%%%%%%%%%%%%%%%%%%%%%%%%%%%%%%%%%%%%%%%%%%%%%%%%%%
\subsection{Comparison of Abelian and Non-Abelian diagram \label{section:ab-nab}}
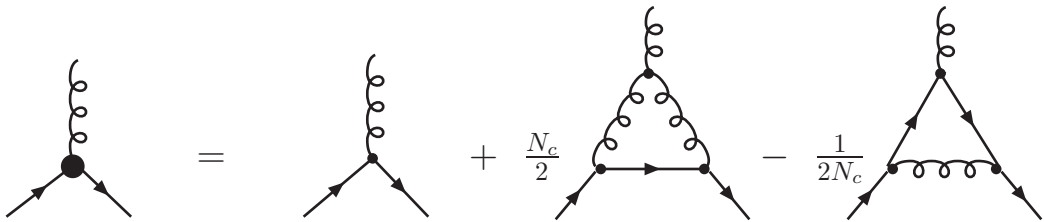
\begin{figure}[t]
\begin{center}
  \begin{picture}(284,70)(60,10)
    \SetWidth{1}
    \ArrowLine(207,0)(222,18)
    \ArrowLine(265,18)(280,0)
    \ArrowLine(224,19)(263,19)
    \Gluon(222,20)(242,55){3}{3}
    \Gluon(242,55)(264,20){3}{3}
    \Gluon(244,55)(244,80){3}{2}
    \ArrowLine(112,1)(138,23)
    \Gluon(139,23)(139,62){3}{3}
    \ArrowLine(138,23)(159,1)
    \put(175,20){\Large{$\mathbf + \ \ \frac{N_c}{2}$}}
    \Vertex(25,20){4.47}
    \Vertex(138,23){2}
    \ArrowLine(0,1)(26,21)
    \Gluon(27,21)(27,60){3}{3}
    \put(72,20){\Large{$\mathbf =$}}
    \ArrowLine(26,21)(47,1)
    \Vertex(263,19){2}
    \Vertex(224,19){2}
    \Vertex(242,55){2}
    \ArrowLine(317,0)(332,18)
    \ArrowLine(375,18)(390,0)
    \Gluon(334,19)(373,19){3}{3}
    \ArrowLine(332,20)(352,55)
    \ArrowLine(352,55)(374,20)
    \Gluon(354,55)(354,80){3}{2}
    \Vertex(373,19){2}
    \Vertex(334,19){2}
    \Vertex(352,55){2}
    \put(285,20){\Large{$\mathbf - \ \ \frac{1}{2N_c}$}}
  \end{picture}
\end{center}
\caption{Semiperturbative approximation to the quark-gluon DSE. The first loop diagram is 
labeled the \lq \lq non-Abelian diagram'' and the second one \lq \lq Abelian diagram''. All
propagators are dressed.\label{ab}}
\end{figure}

%%%%%%%%%%%%%%%%%%%%%%%%%%%%%%%%%%%%%%%%%%%%%%%%%%%%%%%%%%%%%%%%%%%%%%%%%%
The quark-gluon vertex has been investigated in great detail in perturbation 
theory. Here Davydychev, Osland and Saks gave results for arbitrary gauge and
dimension to one loop order and confirmed the validity of the corresponding 
Slavnov-Taylor identity \cite{davydychev:2000rt}. However, as the momentum 
scale decreases and QCD becomes more strongly coupled, perturbation theory 
becomes inappropriate. One then has to resort to resummations and finally 
solve the vertex DSE self-consistently. As a first step it is useful to assess 
the importance of the different diagrams in the DSEs. This can be done within 
a semi-perturbative construction based on loop counting in the spirit of a 
particular type of skeleton expansion \cite{Lu:1991yu} in which all propagators 
are dressed but the internal vertices are still bare. This approximation can 
also be obtained from a two-loop approximation of the 2PI effective action. It 
allows us to assess the importance of the diagrams in fig.~\ref{DS2} on the basis
of both, large $N_c$ counting and dynamical momentum dependences. In this 
approximation the DSE in fig.~\ref{DS2} is given by the diagrams of fig.\ref{ab},
which are conveniently labeled the \lq \lq non-Abelian 
diagram'' and the \lq \lq Abelian diagram''.Whereas the latter diagram has a 
counterpart in the corresponding DSE for the fermion-photon vertex, the 
former has not. 

These two diagrams and the corresponding semi-perturbative analysis is discussed 
in detail in Appendix \ref{section:semipert} and allows to compare the relative 
strength of the Abelian and non-Abelian diagram numerically. 
From the color structure of the two diagrams one immediately
notices that the Abelian diagram is suppressed by a factor of $N_c^2$
as compared to the non-Abelian one. Several 
authors \cite{Bender:2002as,Bhagwat:2004hn,Bhagwat:2004kj,Matevosyan:2006bk}
have recently employed truncations of the quark-gluon vertex based on the
$N_c$-subleading Abelian one-loop diagram. This is necessary in the context
of the Munczek-Nemirovsky model \cite{Munczek:1983dx} where one substitutes
the gluon propagator by a Dirac delta function. Naturally, this operation is
not possible in the non-Abelian diagram due to the product of two-delta functions. 
However those authors are aware that the Abelian construction is subleading
by a factor of $N_c^2$ and has the wrong sign with respect to the dominant,
non-Abelian construction. Therefore as an ad-hoc fix to alleviate the
problem, they have changed the sign of the Abelian diagram and adapted the overall 
strength to the order of magnitude expected from the non-Abelian graph. 
The hope is then that the momentum dependence of both one-loop graphs is similar and
therefore a calculation of the non-Abelian part can be sidestepped. 

We have checked this assumption; see Appendix \ref{section:semipert} for details. 
\begin{figure}[h]
\includegraphics[width=11cm]{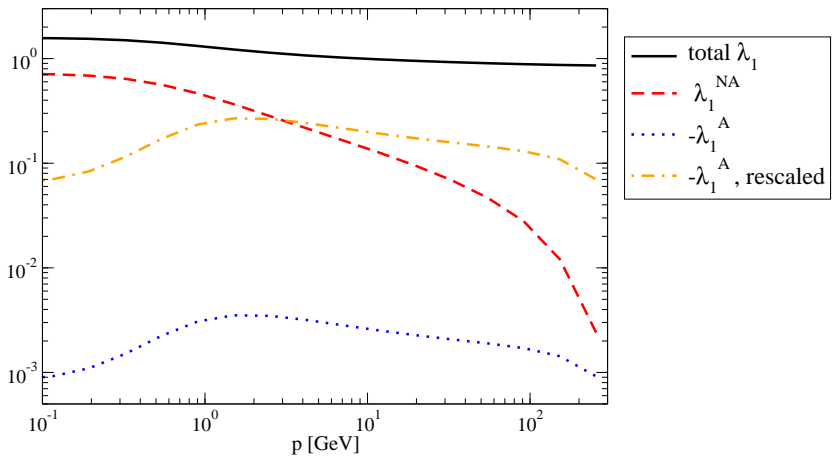}
\caption{\label{fig:compareAnotA}
Dressing factor $\lambda_1$ of the $\gamma_\mu$ vertex in the semi-perturbative
construction with dressed propagators and bare internal vertices. We show 
the total dressing function $\lambda_1 = Z_{1F} + \lambda_1^{NA} + 
\lambda_1^{A}$
(solid line, renormalized to 1 at $\mu=13 \ GeV$ in the soft-gluon kinematic section)
as well as the separate contributions $\lambda_1^{NA}$ and $\lambda_1^{A}$
from the non-Abelian and Abelian triangle loops. Finally, we show the
non-Abelian contribution rescaled by an arbitrary factor to compare the
momentum dependence.}
\end{figure}
Here we only discuss the result for a typical momentum configuration given in fig.~
\ref{fig:compareAnotA}. It can be clearly seen that differences remain even 
after arbitrary rescaling, so this rescaling factor must be dependent on 
the kinematics and adds to the systematic uncertainties.
In general we find from our numerics that the $N_c^2$ suppression of the 
Abelian diagram w.r.t.\ to the non-Abelian one is even amplified by the dynamics.

Another important difference between the Abelian and non-Abelian 
one-loop
diagram (inherited by the semi-perturbative construction) is the quark mass
dependence due to the quark propagators. The Abelian diagram contains two
quark propagators, the non-Abelian only one. Therefore, upon increasing 
the quark mass, the diagram's leading pieces fall as $1/M^2$ and $1/M$
respectively. This behavior can not be mocked by a constant factor unless
this factor depends itself on the mass; but since different Dirac amplitudes
depend differently on the quark mass, not all of them can be mocked
simultaneously.

Finally, we comment on the potential of this semi-perturbative construction  to
induce dynamical chiral symmetry breaking in the quark-DSE. To this end we
inserted our semi-perturbatively calculated vertex (Abelian plus  non-Abelian 
diagram plus tree-level contribution) into the DSE for the quark propagator,
fig.\ \ref{quarkdse0}. As a result we obtained only chirally symmetric
solutions with $B(p^2) \equiv 0$. This leads us to the conclusion that 
semi-perturbative truncations of the vertex-DSE are in general not suited 
to reproduce the patterns of dynamical chiral symmetry breaking.
Consequently we now focus on a more selfconsistent approach.

%%%%%%%%%%%%%%%%%%%%%%%%%%%%%%%%%%%%%%%%%%%%%%%%%%%%%%%%%%%%%%%%%%%%
\section{Infrared analysis of the quark  
sector \label{sec:self}}
%%%%%%%%%%%%%%%%%%%%%%%%%%%%%%%%%%%%%%%%%%%%%%%%%%%%%%%%%%%%%%%%%%%%

After the general structure of the quark-gluon vertex DSE and its diagrammatic 
contributions have been studied in the last section via semi-perturbative methods, 
we will study its interplay in the full set of DSEs of QCD. This is done within
a selfconsistent, purely analytical, infrared analysis of the coupled system of 
quark and vertex DSEs, which is later confirmed by our numerical solutions. 
 
\begin{comment}
In the previous section we have studied the semi-perturbative quark-gluon 
vertex with a quark propagator taken from a truncated quark 
Dyson-Schwinger equation, based on a vertex ansatz introduced in 
\cite{Fischer:2003rp}. 
We now abandon this approximation and address the 
full feedback of the quark-gluon vertex calculated from the vertex-DSE
into the quark DSE. This improved truncation scheme enables us to perform 
a selfconsistent, purely analytical, infrared analysis of the coupled system of 
quark and vertex DSEs, which is reproduced by our numerical solutions. 
In the following we will first detail the analytical treatment and then 
present our numerical results. Possible consequences of our solutions 
and relations to quark confinement are discussed in detail in later
sections.
\end{comment}

%%%%%%%%%%%%%%%%%%%%%%%%%%%%%%%%%%%%%%%%%%%%%%%%%%%%%%%%%%%%%%%%%%%%%%%
\subsection{Yang-Mills sector}
%%%%%%%%%%%%%%%%%%%%%%%%%%%%%%%%%%%%%%%%%%%%%%%%%%%%%%%%%%%%%%%%%%%%%%%

To begin, we shortly summarize previous results for the infrared
behavior of the Green's function in Landau gauge Yang-Mills theory. 
We discuss the analytical behavior of these functions when all
external momenta go to zero. 
Here, we choose all momenta proportional to 
each other and require the largest one to be much smaller than the 
typical scales of the theory, i.e. $p^2 \ll \Lambda^2_{\tt QCD}$. 
Here $\Lambda_{\tt QCD}$ is generated dynamically in the gluon sector
of the theory via dimensional transmutation. Then a self-consistent and 
unique power-law solution of the whole (untruncated) tower of DSEs is 
given by 
\cite{Alkofer:2004it,Fischer:2006vf} 
\be 
\Gamma^{n,m}(p^2) \sim \left({p^2}\right)^{(n-m)\kappa},
\label{IRsolution} 
\ee
where $\Gamma^{n,m}$ stands for the infrared leading dressing function of a 
Green's functions with $n$ external ghost and antighost
legs and $m$ external gluon legs. 
The exponent $\kappa$ is known to be positive \cite{Watson:2001yv,Lerche:2002ep} 
but more restrictive limits have been suggested recently \cite{Alkofer:2008jy}. 

We stress that this IR behavior of gauge correlation functions is not 
influenced by the dynamics of massive quarks as will be confirmed subsequently. 
We note, however, that due to the possibility of strong divergences in the 
vertices this does not follow directly from their IR decoupling
\cite{Appelquist:1974tg,Fischer:2003rp}. 
In the heavy quark limit even 
the value of the scale $\Lambda_{\tt QCD}$ is determined completely by 
the gauge dynamics and obeys $\Lambda^2_{\tt QCD} \ll M^2$. In contrast, 
in the chiral limit, where $\Lambda_{\tt QCD}$ is the unique scale of 
the system, there could in principle be a different conformal scaling 
solution in the Yang-Mills sector triggered by quark-loop insertions.
This is discussed in more detail in \cite{parallelpaper}. Here we 
concentrate on quenched QCD, where (\ref{IRsolution}) is certainly
unaffected by the quark sector.

Examples of the general solution (\ref{IRsolution}) are given by
the inverse ghost and gluon dressing functions, $\Gamma^{1,0}(p^2) =
G^{-1}(p^2)$ and  $\Gamma^{0,2}(p^2) = Z^{-1}(p^2)$, respectively. 
They are related to the ghost and gluon propagators via eqs. (\ref{gg},\ref{ghgh}).
The corresponding power laws in the infrared are
\beq
G(p^2) \sim \left({p^2}\right)^{-\kappa}, 
\hspace*{1cm} Z(p^2) \sim \left({p^2}\right)^{2\kappa}\,.
\label{kappa}
\eeq
Since $\kappa$ is positive one obtains an infrared diverging ghost
propagator, a behavior which is sufficient to ensure a well defined, 
{\it i.e.} unbroken, global color charge \cite{Kugo:1995km}. 
\footnote{We emphasize that the eq.~(\ref{IRsolution}) solves the 
untruncated system of DSEs and the corresponding equations from the 
functional renormalization group. There is, however, an ongoing discussion, 
fostered by recent lattice results \cite{Cucchieri:2007md,Sternbeck:2007ug} 
and also continuum investigations \cite{Aguilar:2008xm,Boucaud:2008ji} 
whether there is an alternative solution to the DSEs with an infrared 
finite or only weakly diverging ghost propagator and a massive gluon. 
This discussion involves subtile points of effects of Gribov copies 
and other artefacts on the lattice as well as in the continuum gauge 
theory. Pending further clarifications we will ignore this discussion 
for the moment and proceed by exploring the consequences of 
eq.~(\ref{IRsolution}) for quenched QCD. We are confident that the 
results of our investigation, presented below also provides direct 
support for the validity of the solution eq.~(\ref{IRsolution}).} 

Note that in Landau gauge an explicit value for $\kappa$ can be 
derived from the observation that the dressed ghost-gluon vertex remains 
(almost) bare in the infrared \cite{Taylor:1971ff}.
One then obtains $\kappa = (93 - \sqrt{1201})/98 \approx 0.595$ 
\cite{Lerche:2002ep,Pawlowski:2003hq,Zwanziger:2001kw}, 
which implies that the gluon propagator vanishes in the infrared. A direct 
consequence of this behavior are positivity violations in the gluon
propagator and therefore the confinement of transverse gluons 
\cite{vonSmekal:1997is,Alkofer:2003jj}. Another consequence of the
solution (\ref{IRsolution}) is the qualitative universality of the
running coupling of Yang-Mills theory, see \cite{Alkofer:2004it} 
and subsection \ref{coupling}.

Another important example of the solution (\ref{IRsolution}) is the 
three-gluon vertex, which later on will serve as input for our analytical 
and numerical calculations. Restricting ourselves here to dressings of  
the tree-level structure $\Gamma^0_\mu(p_1,p_2,p_3)$
(for a more detailed treatment see \cite{Huber}), this vertex goes like
\beq
\Gamma_\mu(p_1,p_2,p_3) \sim (p^2)^{-3\kappa} \Gamma^0_\mu(p_1,p_2,p_3), \label{IR3g}
\eeq
in all kinematical sections where there is only one external scale $p^2$ i.e.
for $p^2 \sim p_1^2 \sim p_2^2 \sim p_3^2$.

As has been noted recently \cite{Alkofer:2008jy,Huber}, there are additional 
kinematic divergences in the Yang-Mills 3-point functions that arise when 
only a single gluon momentum becomes small which we will refer to as the "soft-gluon limit". 
The IR solution is parameterized by IR-exponents $\delta$ for the leading 
correlation functions in the gauge sector. A lower index denotes the Greens 
function and the upper indices $\delta^{u}$, $\delta^{gl}$ and $\delta^{gh}$ 
denote the uniform, soft-gluon and soft-ghost limits.
The unique scaling solution is given in Table \ref{tab:IR-scaling-YM}.
\begin{table}[h]
\begin{tabular}{|c|c|c|c|c|c|c|c|c|}
\hline 
$\delta_{gh}$ & $\delta_{gl}$ & $\delta_{gg}^{u}$ & $\delta_{3g}^{u}$ & 
$\delta_{4g}^{u}$ & $\delta_{gg}^{gh}$ & $\delta_{gg}^{gl}$ & 
$\delta_{3g}^{gl}$ & $\delta_{4g}^{gl}$ \tabularnewline
\hline 
$-\kappa$ & $2\kappa$ & $0$ & $-3\kappa$ & $-4\kappa$ & $0$ & 
$1\!-\!2\kappa$ & $1\!-\!2\kappa$ & $1\!-\!2\kappa$ \tabularnewline
\hline
\end{tabular}
\caption{\label{tab:IR-scaling-YM}The IR exponents for the primitively divergent
Green's
functions of the unique IR scaling fixed point of Landau gauge Yang-Mills theory.}
\end{table}
It involves mild kinematic singularities in the three-gluon vertex 
$\delta_{3g}^{gl}$ when only one external gluon becomes soft. \\

%%%%%%%%%%%%%%%%%%%%%%%%%%%%%%%%%%%%%%%%%%%%%%%%%%%%%%%%%%%%%
\subsection{Including quarks: basics \label{irbasics}}
%%%%%%%%%%%%%%%%%%%%%%%%%%%%%%%%%%%%%%%%%%%%%%%%%%%%%%%%%%%%%

Based on the infrared solution (\ref{IRsolution}) summarized in the last
subsection we now extend our analysis to also include the quark sector of 
QCD. In the following we only consider quenched QCD and only 
briefly discuss possible effects of unquenching in section \ref{stringbreaking}.

As a starting point we analyze the Dyson-Schwinger equation of the
quark propagator, given diagrammatically in fig.~\ref{quarkdse0}.
Stripping the color factor, the quark propagator eq.~(\ref{qq}) can 
alternatively be parameterized as
\beq
S(p) = \frac{i \pslash + M(p^2)}{p^2 + M^2(p^2)}Z_f(p^2)\,,
\eeq
with the quark mass function $M(p^2)$ and the wave function $Z_f(p^2)$. In
this notation the quark-DSE can be written as
\beqa
\frac{M(p^2)}{Z_f(p^2)} &=& Z_2 Z_m m + g^{2}C_{F}Z_{1F} \, \mathrm{tr} \int
\frac{d^4q}{(2 \pi)^4}
\gamma_\mu \frac{M(q^2) Z_f(q^2)}{q^2 + M^2(q^2)} \Gamma_\nu(k,q) P_{\mu 
\nu}
\frac{Z(k^2)}{k^2} \\
\frac{1}{Z_f(p^2)} &=& Z_2 + g^{2}C_{F}Z_{1F} \, \mathrm{tr} \int \frac{d^4q}{(2
\pi)^4}
\pslash \gamma_\mu \frac{\qslash Z_f(q^2)}{q^2 + M^2(q^2)} 
\Gamma_\nu(k,q)
P_{\mu \nu} \frac{Z(k^2)}{p^2 k^2} 
\eeqa  
with $k=p-q$, the Casimir $C_{F}=(N_{c}^{2}-1)/(2N_{c})$ and the
renormalization 
factors $Z_2$ of the quark fields and $Z_{1F}$ of the quark gluon vertex.
The symbol $tr$ denotes the Dirac-trace.

Now consider the loop integral on the right hand side of the DSE. When
chiral symmetry is broken dynamically we have $M(p^2) \ne 0$ and the denominator of the
internal quark 
propagator cannot be smaller than the generated quark mass $M(p^2)$. As an
effect the integral is dominated by momenta larger than the quark mass and the
resulting 
quark mass on the left hand side of the equation freezes out to a constant
in the infrared, in selfconsistent agreement with $M(p^2) \ne 0$. 
This 
infrared screening
mechanism of the quark, observed in a broad range of works so far (see
\cite{Fischer:2003rp}
and refs. therein), can only be overruled by an extremely singular
quark-gluon vertex.
We have not found a selfconsistent solution for the vertex-DSE that supports
such an extreme case and therefore discard this possibility. 
In general this screening mechanism will also make the quark wave function 
constant in the infrared, as again observed repeatedly in the literature.

The only other possibility we see is given by the \lq scaling solution' 
$Z_f(p^2)/M(p^2) \sim (p^2)^{-\kappa-1/2}$. Since this solution is refuted
by lattice calculations of the quark propagator \cite{Bowman:2005vx} 
we do not discuss it here, 
but refer the interested reader to \cite{parallelpaper} for more details.

We thus arrive at an IR finite solution of the quark propagator with
\beq
M\equiv M(p^2 \rightarrow 0)\ {\rm and } 
\ Z_f\equiv Z_f(p^2 \rightarrow 0) \label{const}
\eeq
where we define the constant and non-vanishing limits by $M,Z_f$ and 
always refer to these limits in the following when no momentum 
arguments are given. The quark propagator can then be written as
\beq
S(p) = \frac{i \pslash + M(p^2)}{p^2 + M^2(p^2)}Z_f(p^2) \rightarrow 
\frac{i \pslash Z_f}{M^2}+\frac{Z_f}{M} \label{quark}
\eeq
for momenta $p^2 \ll \Lambda_{\tt QCD}^2$.

\begin{figure}[t]
\centerline{\epsfig{file=DSE.eps,width=150mm}}
\centerline{\epsfig{file=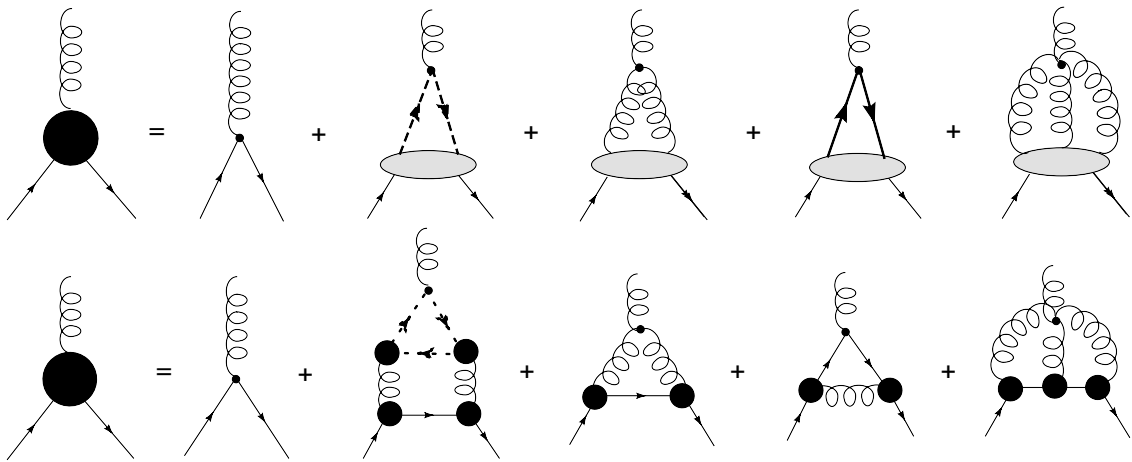,width=150mm}}
\caption{Dyson-Schwinger equation for the quark-gluon vertex. In the first
line we show the full equation \cite{Marciano:1978su}. The second line 
shows important terms of the lowest order of a skeleton expansion, that 
already reveal the correct power counting in the infrared.
All internal lines in the diagrams represent fully dressed 
propagators. Symmetry factors and signs have been 
omitted.
The infrared counting analysis can be followed by referring to the second
diagram in the right-hand-side of the second line, where one can
abstract the ghost loop as if forcing a dressed three-gluon vertex. This
is the same counting derived from the last diagram of figure \ref{DS3}.
}\label{vertex-DSE}
\end{figure}
In the following we show that a selfconsistent solution of the quark-DSE and
the DSE 
for the quark-gluon vertex can be found based on eq. (\ref{quark}). 
To this end we now apply the power counting method of 
\cite{Alkofer:2004it,Fischer:2006vf} to the Dyson-Schwinger 
equation for the full quark-gluon vertex, given diagrammatically in the
first line of fig.~\ref{vertex-DSE}. In the second line the higher 
$n$-point
functions of this equation have been expanded to lowest order in a skeleton expansion
in terms of full propagators and vertices. The dressed quark-gluon vertex
$\Gamma_\mu$ 
can be decomposed in a basis of twelve tensor structures \cite{Ball:1980ay} given 
explicitly in eq.~(\ref{newdec}) in the appendix.

For the internal loops of the skeleton expansion in 
fig.~\ref{vertex-DSE}
there are two possibilities: 

\begin{itemize}
\item[(i)] they are dominated by loop momenta $q_\mu$ of the order of the 
small external scale $p^2$; this then requires infrared singularities in the
internal
quark-gluon vertices, 
\item[(ii)] the same screening mechanism applies as in the quark-DSE; for a
small external
scale $p^2$ the loop diagram is then dominated by loop momenta in the
mid-momentum region
$q^2 \sim \Lambda_{\tt QCD}^2$ and consequently the vertex freezes out in the
infrared.
\end{itemize}

We emphasize that both possibilities may lead to selfconsistent solutions of
the quark- and vertex-DSE, although numerically we only found solutions 
with the property (i). Whether this observation is truncation dependent or
not needs to be investigated further in future work. Noting that the second 
possibility requires no further infrared analysis anyway we proceed to show
that indeed selfconsistent solutions can be found when (i) is realized. 

In this scenario the dressing functions of internal Yang-Mills propagators 
and vertices can be approximated 
by the power laws given in eq.~(\ref{IRsolution}). We therefore use
eq.~(\ref{kappa}) for the scaling 
of the ghost and gluon propagators. The ghost-gluon vertex scales as a
constant 
in the infrared \cite{Taylor:1971ff}. The internal quark propagator lines
are replaced by (\ref{quark}). 

We contracted the right hand side of the DSE with appropriate tensor
structures to project out the various components of the vertex, c.f. eq.~(\ref{newdec}).
Thus we eventually have to deal with scalar integrals only, which
depend on powers
of internal and external momenta. The evaluation of such diagrams has been
described in \cite{Anastasiou:1999ui}. Most important for our purpose is the fact that
all powers
of internal momenta transform to powers of external momenta after
integration. This
is also clear from simple dimensional consideration, provided there is only
one external scale present. In our case this is ensured by the condition 
$p^2 \ll \Lambda_{\tt QCD}^2 \le M^2$, which leaves only one scale $p^2$ in
the deep
infrared. The following analysis rests on the assumption that there is no
second small scale present. In principle this assumption may not be true 
for heavy quarks, since then one has a potential additional small scale 
$\Lambda_{\tt QCD}^2/M$. We have investigated this possibility in our
numerical calculations of the quark-gluon vertex and did not see any
influence of such a quantity. The corresponding results of our quark mass
study are presented in subsection \ref{massstudy}, where we discuss this 
issue further.

%%%%%%%%%%%%%%%%%%%%%%%%%%%%%%%%%%%%%%%%%%%%%%%%%%%%%%%%%%%%%%%%%%%%%
\subsection{How not to include quarks}
%%%%%%%%%%%%%%%%%%%%%%%%%%%%%%%%%%%%%%%%%%%%%%%%%%%%%%%%%%%%%%%%%%%%%

Before we come to our actual solution we would like to point out the 
complication that the Dirac structure induces.
Suppose we only consider a single form factor $\lambda_1(p^2)$ with 
a single IR exponent $\delta_{qg}$ for the quark gluon vertex corresponding to
the tree-level tensor structure $\gamma_\mu$. Due to the
Dirac-algebra the integral then picks up only contributions from the 
vector part of the internal quark 
propagator. 
The resulting vertex-DSE (IR-leading diagram only), that can be followed 
from figure \ref{vertex-DSE} and its caption, is then given
by
\beqa
\lambda_1(p_1,p_2) &=& 
\left( \frac{N_c}{2} g^2 \right) \int \frac{d^4q}{(2\pi)^4}
\frac{Z(k_1^2)}{k_1^2} \frac{Z(k_2^2)}{k_2^2}
\frac{Z_f(q^2)}{q^2 + M^2(q^2)}  \Gamma^{0,3}(k_1,k_2) \times \nonumber\\
&&\lambda_1(p,q) \lambda_1(q,2p)
\left( \frac{3 p^2}{2} - \frac{5 p^4}{4 k_1^2} - \frac{p^4}{4 k_2^2} +
\frac{p^6}{4 k_1^2 k_2^2}
\right)
\eeqa
with the momenta $k_1=p_1-q$, $k_2 = q-p_2$ and the dressing $\Gamma^{0,3}$
of the three gluon
vertex. Here we used tree-level tensor structure in the three-gluon vertex
without loss of generality. After integration, all powers of internal momenta are
converted into
powers of external momenta for dimensional reasons (potential conversions
into powers of the
quark mass are subleading in the infrared and are therefore discarded here;
we further justify this treatment {\it a posteriori} in our numerical 
section \ref{massstudy}).

Power counting in terms of the external momentum $p^2$ then results in
\beqa
(p^2)^{\delta_{qg}} &\sim& p^4 \times
\left[\frac{(p^2)^{2\kappa}}{p^2}\right]^2
\times\frac{Z_f}{M^2} \times (p^2)^{-3\kappa} \times
\left[\lambda_1(p^2)\right]^2 \times p^2\\
	        &\sim& Z_f \, \frac{(p^2)^{1+\kappa+2\delta_{qg}}}{M^2} 
\eeqa
Thus a selfconsistent power-law solution of this equation would entail that
\beq
\delta_{qg} = -1-\kappa\,, \label{wrongsol}
\eeq
i.e. a singular quark-gluon vertex in the infrared. This solution, however,
only serves illustrational purposes, since it does not survive when the 
full Dirac structure of the vertex included in the other
tensor structures is considered. This is shown in the next subsection.

%%%%%%%%%%%%%%%%%%%%%%%%%%%%%%%%%%%%%%%%%%%%%%%%%%%%%%%%%%%%%%%%%%%%%
\subsection{Including quarks: the correct way \label{irquarks}}
%%%%%%%%%%%%%%%%%%%%%%%%%%%%%%%%%%%%%%%%%%%%%%%%%%%%%%%%%%%%%%%%%%%%%

To see how the Dirac structure of the vertex can introduce additional 
complications we note that the tensor structures of the vertex group 
into two different sets of Dirac-scalar and vector parts, as detailed
in appendix \ref{app:tensor}. We have to take into account that within 
the DSEs the two parts receive qualitatively different contributions 
from the various loop corrections and introduce distinct IR exponents 
$\delta_{qgv}$ and $\delta_{qgs}$ for the vector (qgv) and scalar (qgs)
Dirac structures. 
One then arrives at a coupled set of equations with various contributions:
\beqa
(p^2)^{\delta_{qgv}} &\sim& \mathrm{max} \left\{ Z_f \, \frac{(p^2)^{1+\kappa+2\delta_{qgv}}}{M^2},
                             Z_f \, \frac{(p^2)^{1+\kappa+2\delta_{qgs}}}{M^2},
			     Z_f \, \frac{(p^2)^{1/2+\kappa+\delta_{qgv}+\delta_{qgs}}}{M}\right\}\nonumber\\ 
(p^2)^{\delta_{qgs}} &\sim& \mathrm{max} \left\{ Z_f \, \frac{(p^2)^{1/2+\kappa+2\delta_{qgv}}}{M},
                             Z_f \, \frac{(p^2)^{1/2+\kappa+2\delta_{qgs}}}{M},
			     Z_f \, \frac{(p^2)^{1+\kappa+\delta_{qgv}+\delta_{qgs}}}{M^2}\right\} 
			     \label{IRsimple2} 
\eeqa
where the \lq max' indicates that the most singular term in the infrared is
the leading one on the right hand
sides of these equations. This equation can be solved by a range of
solutions given by
\beq
\delta_{qgv} \in [-1/2-\kappa , -\kappa], \ \ \ \delta_{qgs} = -1/2-\kappa \label{IRsimple1}
\eeq
As is demonstrated explicitly in appendix \ref{IRanalysis}, this infrared solution
for $\delta_{qgv}$ and $\delta_{qgs}$ does not change when also the tensor components 
of the quark-gluon vertex are taken into account. The above encountered 
behavior 
eq.~(\ref{wrongsol}) is therefore an artefact due to the ommission of the scalar Dirac 
components.

It is furthermore instructive to look closer into the mechanism that generates the 
solution (\ref{IRsimple1}) and recall the 
properties of the different tensor structures with respect to chiral symmetry. 
The corresponding transformations act on fermion fields via
$
e^{i\alpha \gamma^5} \psi
$
with $\alpha$ labeling the transformation. A chirally symmetric quark propagator or
quark-gluon vertex would therefore correspond to 
\beq
\{\gamma^5,\Gamma^\mu\}=0 \ \ \mbox{and} \ \ \{\gamma^5,S\}=0.
\eeq
These relations are satisfied by all tensor structures with an \emph{odd} number of 
gamma matrices. Those parts with an \emph{even} number of gamma matrices can only be
present when chiral symmetry is violated. From 
eqs.~(\ref{IRsimple2}) we find that the driving pieces for the selfconsistent solution 
(\ref{IRsimple1}) of the vertex DSE are the ones with broken chiral symmetry. If these
structures are left out in both the quark propagator and the vertex, one finds a different
(infrared weaker) selfconsistent solution, as discussed in subsection \ref{chiralsymm} 
below\footnote{Note in this respect that the (inconsistent) behavior eq.~(\ref{wrongsol})
has been obtained from the chiral symmetry breaking part of the propagator while the
vertex had only the symmetry preserving term $\gamma_\mu$.}. The fact that the presence or absence
of chiral symmetry breaking terms generates different solutions of the vertex-DSE
seems a remarkable and novel result to us. It underlines that dynamical chiral symmetry 
breaking is not restricted to the quark propagator alone but happens (consistently) in 
every Green's function with fermionic content. Since the underlying mechanism is not 
well appreciated in the literature we found it useful to illustrate it with a simple 
mechanical model, discussed in appendix \ref{model}.

Before we discuss some aspects of the solution eq.~(\ref{IRsimple1}), it is important 
to check that it persists to all orders in the skeleton expansion used in 
fig.~\ref{vertex-DSE}. Higher order terms in this expansion can
be generated by inserting diagrammatical pieces like 
\beq
\epsfig{file=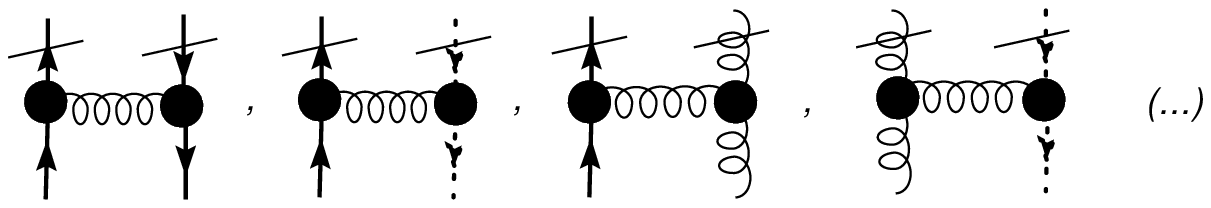,width=80mm} \nonumber
\eeq
into lower order diagrams (all propagators are fully dressed). In ref.\
\cite{Alkofer:2004it} some of us 
showed that insertions involving
pure Yang-Mills propagators and vertices do not change the overall infrared
divergence
of a given diagram. This is also true for the three pieces involving quark
lines: the first one introduces two quark-gluon vertices, two quark propagators 
and one gluon propagator plus a 
new integration into a given diagram. This amounts to an additional power of
\beq
Z_f \, \frac{(p^2)^{-1-2\kappa + 2\kappa-1 + 2}}{M^2} = \frac{Z_f}{M^2} (p^2)^0
\eeq
in the external momentum $(p^2)$. A similar result is obtained for the other
pieces. Thus diagrammatic insertions of quark lines do not change the infrared 
behavior of a given diagram but
introduce additional powers of $Z_f/M^2$. This implies that the solution eqs.
(\ref{IRsimple1}) is valid 
to all orders in the skeleton expansion. It is therefore also an infrared
solution of the full vertex-DSE. 

Note that similar to the DSEs in the Yang-Mills sector it is the diagram containing 
the ghost triangle loop that dominates the right hand side of the equation in 
fig.~\ref{vertex-DSE}. Thus the infrared divergent behavior of the 
quark-gluon 
vertex is an induced effect from singularities in the pure Yang-Mills theory. 

\begin{figure}[h]
\begin{center}
\begin{picture}(280,90)(30,0)
    \SetWidth{1.5}
    \SetScale{0.6}
    \ArrowLine(12,1)(51,29)
    \ArrowLine(52,29)(114,30)
    \ArrowLine(114,29)(145,1)
    \Gluon(50,31)(50,71){4.5}{2.57}
    \Gluon(110,71)(110,31){4.5}{2.57}
    \DashArrowLine(51,74)(110,74){5}
    \DashArrowLine(110,73)(79,107){5}
    \DashArrowLine(79,107)(50,72){5}
    \Gluon(81,104)(82,144){4.5}{2.57}
    \Vertex(110,30){4.47}
    \Vertex(49,28){4.47}
    \GCirc(110,74){6}{0}
    \GCirc(50,74){6}{0}
    \GCirc(110,30){6}{0}
    \GCirc(49,28){6}{0}
    \put(95,35){\Large{$\mathbf \simeq $}}
    \ArrowLine(162,1)(201,29)
    \ArrowLine(202,29)(264,30)
    \ArrowLine(264,29)(295,1)
    \Gluon(200,31)(232,84){4.5}{3}
    \Gluon(232,84)(260,31){4.5}{3}
    \Gluon(232,84)(232,144){4.5}{3}
    \Vertex(260,30){4.47}
    \Vertex(199,28){4.47}
    \GCirc(232,84){6}{0}
    \GCirc(260,30){6}{0}
    \GCirc(199,28){6}{0}
  \end{picture}
\end{center}
\begin{center}
  \begin{picture}(284,70)
    \SetWidth{1.5}
    \ArrowLine(196,0)(222,18)
    \ArrowLine(263,19)(284,0)
    \ArrowLine(224,19)(263,19)
    \Gluon(240,55)(220,20){3}{3}
    \Gluon(240,55)(264,20){3}{3}
    \Gluon(242,55)(242,80){3}{2}
    \ArrowLine(112,0)(138,23)
    \Gluon(139,23)(137,62){3}{3}
    \ArrowLine(138,23)(159,0)
    \put(175,36){\Large{$\mathbf+$}}
    \Vertex(25,20){4.47}
    \Vertex(138,22){2}
    \ArrowLine(0,0)(26,21)
    \Gluon(27,21)(27,60){3}{3}
    \put(72,35){\Large{$\mathbf =$}}
    \ArrowLine(26,21)(47,0)
    \Vertex(263,19){5.1}
    \Vertex(224,20){5.1}
    \Vertex(242,55){5.1}
  \end{picture}
\end{center}
\caption{The infrared leading diagram of the quark-gluon vertex DSE is shown in the first line. 
In the second line we show an approximation of the full quark-gluon vertex DSE that preserves all
infrared features of the equation, while also capturing the leading $1/N_c$-parts of the 
ultraviolet properties.
\label{irdiag}}
\end{figure}
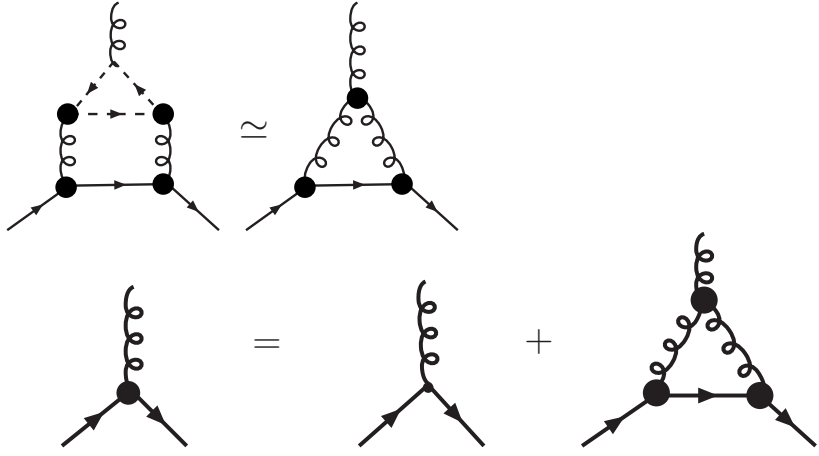

Finally, we emphasize that the ghost triangle loop in the infrared leading diagram 
is also the leading infrared contribution to the fully dressed three-gluon vertex 
\cite{Alkofer:2004it}. This offers an interesting opportunity for an approximation 
which enables our numerical analysis in section \ref{sec:numsec}.
Since we are mainly interested in the infrared behavior of the quark-gluon vertex 
we may replace the ghost triangle sub-diagram by the dressed three-gluon vertex as 
shown in fig.~\ref{irdiag}.
The resulting diagram greatly resembles the \lq \lq non-Abelian'' diagram 
already discussed above
in our semi-perturbative analysis, but with all bare vertices replaced by dressed ones. It is
interesting to note that such a completely dressed non-Abelian diagram is also present
in the equations of motion for the quark-gluon vertex from the functional renormalization group
(FRG) and nPI formulations 
(with $n > 2$). This suggests that our infrared solution eqs. (\ref{IRsimple1}) also solves the 
corresponding FRG and nPI equations self-consistently. We have checked that this is 
indeed the case.

%%%%%%%%%%%%%%%%%%%%%%%%%%%%%%%%%%%%%%%%%%%%%%%%%%%%%%%%%%%%%%%%%%%%%%%%%
\subsection{Soft-gluon singularity in the quark-gluon vertex\label{softgluondiv}} 
%%%%%%%%%%%%%%%%%%%%%%%%%%%%%%%%%%%%%%%%%%%%%%%%%%%%%%%%%%%%%%%%%%%%%%%%%

In addition to the main infrared singularity that appears when all scales 
in a given Green function are sent to zero, there can be kinematical 
singularities that appear in specific kinematic sections. The present counting 
rules relied on the implicit assumption that all external momenta $p_i$ scale 
as some common scaling variable $p$. However, in case any of the momenta 
vanishes identically this is not fulfilled. In the parameterization with a 
common scaling variable $p$ and dependencies on the other momenta given in 
terms of dimensionless momentum ratios $p_i/p \in (0,\infty)$ or angle cosines 
between momentum vectors $x_i\in (-1,1)$, the IR expression reads
$$
F(p_i/p, x_i) (p^2)^\alpha \; ,  \quad i=1\dots n-1
$$
and these kinematic singularities appear as singularities of the function $F$.
We have identified such a kinematic singularity for the quark-gluon vertex 
in the soft-gluon kinematic section where the gluon momentum $p_3=p_2-p_1=0$, 
whereas $p_1^2$ and $p_2^2$ may be large. In this configuration the 
singularity is even self-consistently enhanced and thereby is likely to 
constitute a stable solution of the DSE system. This singularity directly 
extends the uniform IR singularity discussed above and reported previously 
in \cite{Alkofer:2006gz} to this larger kinematic section.

\begin{figure}
\centerline{\epsfig{file=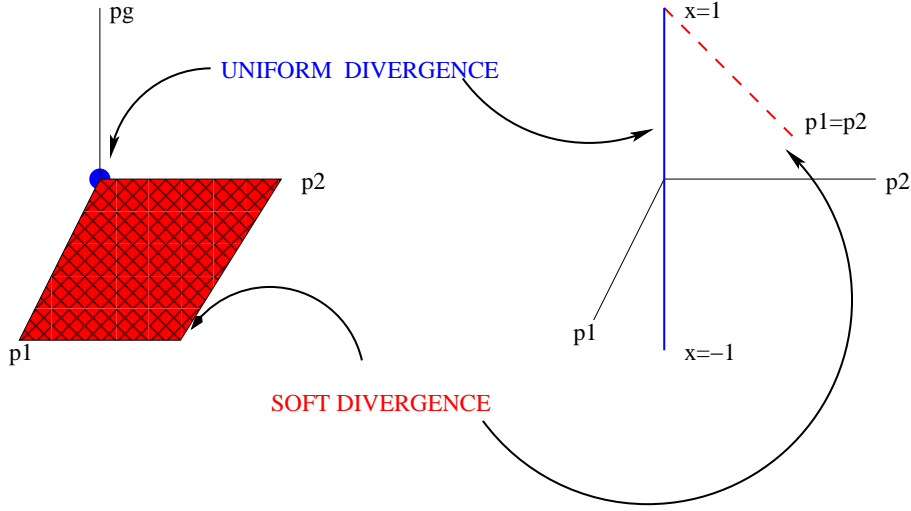,width=12cm}}
\caption{Support of uniform versus soft IR divergences in the 
quark-gluon vertex.
Left: if the quark-gluon vertex is expressed in terms of the three
Lorentz invariants $p_1^2$,  $p_2^2$,  $p_g^2=(p_1-p_2)^2$, the uniform
divergence corresponds to the point where all momenta simultaneously
vanish, whereas the soft divergence lies on a plane in this
three-dimensional space, since only the gluon momentum is required to
vanish. This shows the larger phase-space (and therefore larger
contribution to loop integrals) associated with soft divergences.
Right: if one uses instead of $p_g$ a Euclidean spherical angle between
$p_1$ and $p_2$, whose cosine $x\in(-1,1)$ is on the $OZ$-axis,
the uniform divergence corresponds precisely to this segment. However
the soft divergence is now also a segment, at fixed $x=1$, running for
$p_1=p_2$. Thus a grid optimized to capture only the uniform-divergence
in a numerical code will have difficulties to capture
the soft-divergence too.  }
\end{figure}

The mechanism behind this singularity is based on the strong
$\left( p^2\right)^{-3\kappa}$ singularity in the three-gluon
vertex when all its momenta vanish simultaneously. To see this
consider the infrared leading diagram (\ref{irdiag})
of the quark-gluon vertex DSE. 
\begin{center}
  \begin{picture}(200,140)(0,-30)
    \SetWidth{1.5}
    \SetScale{0.7}
    \Gluon(100,140)(100,84){5}{3}
    \Gluon(150,40)(100,84){5}{3}
    \ArrowLine(50,40)(150,40)
    \Gluon(50,40)(100,84){5}{3}
    \ArrowLine(25,-15)(50,40)
    \ArrowLine(150,40)(175,-15)
    \Vertex(100,84){4}
    \Vertex(50,40){4}
    \Vertex(150,40){4}
    \put(30,-10){\large{$\mathbf{p_1}$}}
    \put(130,-10){\large{$\mathbf{p_2}$}}
    \put(80,90){\large{$\mathbf{p_3=\epsilon}$}}
    \put(50,13){\large{$\mathbf{p_1-\delta}$}}
    \put(40,50){\large{$\mathbf{\delta}$}}
    \put(95,50){\large{$\mathbf{\delta+\epsilon}$}}
  \end{picture}
\end{center}
We choose a momentum configuration, 
where the external momenta $p_1^2$ and $p_2^2$ may be large, but
the external gluon momentum $p_3 \equiv \epsilon$ is small. This 
allows to route the loop momentum $\delta$  to expose the 
relevant phase-space where both
 internal gluon momenta, $\delta$ and $\delta+\epsilon$, are small.
The incoming quark momentum is fixed at $p_1$ and the outgoing is 
$p_2=p_1+\epsilon$:
Now, if there are soft-gluon kinematic singularities in the quark-gluon
vertex, these also appear in the internal quark-gluon vertices at this
particular kinematical configuration and if these singularities are
strong enough loop momenta around $\delta$ will dominate the integral.

Then, leaving out finite factors, self-consistency of the soft-gluon 
divergence demands that the corresponding exponent $\delta_{qg}^{gl}$ fulfills
\begin{eqnarray}
(\epsilon^2)^{\delta_{qg}^{gl}} \propto \int d^4\delta \ (\delta^2)^{2\kappa-1} \ 
((\delta+\epsilon)^2)^{2\kappa-1}\
(\delta^2)^{\delta_{qg}^{gl}} \ ((\delta+\epsilon)^2)^{\delta_{qg}^{gl}} 
\ 
(\delta^2+\epsilon^2+(\delta+\epsilon)^2)^{-3\kappa} \ \ \Gamma_{3g}^0
\end{eqnarray}
where 
$(\delta^2+\epsilon^2+(\delta+\epsilon)^2)^{-3\kappa}$
stems from the dressing of the three gluon vertex. Note that the bare
three-gluon vertex $\Gamma^0_{3g}$ also contributes a power of momentum 
$(\epsilon^2)^{1/2}$
after integration. 

Now, if $\epsilon\to 0$, the integral is inconsistently divergent in
$\delta\to 0$. Therefore, $\epsilon$ needs to be kept finite, in which case
the integral is regulated to the RHS of
\begin{equation}
(\epsilon^2)^{\delta_{qg}^{gl}} \propto (\epsilon^2)^{\kappa+2\delta_{qg}^{gl} +1/2}
\end{equation}
and therefore it must be that
\begin{equation}
\delta_{qg}^{gl}= -1/2 - \kappa \label{ss}
\end{equation}
which indeed corresponds to a soft singularity with an equally large
exponent in the vertex.

Below, we check explicitly that the counting of soft infrared
divergencies is not affected by the remaining diagrams in the quark-gluon
vertex DSE. In fact the situation is similar to the kinematic section where 
all external momenta go to zero: In a Dyson-Schwinger formulation the
non-Abelian diagram is the infrared leading one with all other diagrams
being subleading. In an nPI formulation or within the FRG framework, where
all vertices are dressed in general, we encounter a uniform scaling of
the whole equation, i.e. all diagrams have the same infrared divergencies.

The importance of the self-consistently enhanced soft-gluon divergence comes 
from its role for the confinement mechanism of quarks discussed below. A 
mechanism based purely on the uniform IR divergence as discussed above and 
proposed in \cite{Alkofer:2006gz} would require the external quarks to have 
vanishing Euclidean momenta - corresponding to light-like momenta 
in Minkowski space to trigger the divergence. This is only possible for chiral 
quarks, whereas any mass would act as a cutoff that shields the divergence. 
As known from the structure functions measured in deep inelastic scattering 
experiments \cite{Yao:2006px}, the quarks in hadrons have a broad momentum 
distribution which even vanishes at zero momentum for valence quarks. 
Similarly, constituent quarks inside charmonia or bottomonia have momenta 
of the order of or larger than $\Lambda_{\tt QCD}$ (typically of order 
$\alpha_s M_Q$). The 
soft-gluon divergence is realized independent of the quark kinematics and 
requires only the gluon momentum to become small. Therefore this type of singularities has the 
potential to provide a static linear potential between heavy quarks since the 
momentum of an exchanged gluon is always small as far as the quarks are 
sufficiently spatially separated. This will be discussed in more detail in 
sections \ref{qqkernel} and \ref{sec:stringtension}.

Let us now analyze the IR power counting for the full system of DSEs for 
the quark sector of quenched QCD. In contrast to the propagator which has only one external scale and is parameterized by a single IR exponent $\delta_q$, the quark-gluon vertex has two independent external momenta. There are three different limits that can be IR sensitive, namely the {\em uniform} limit when all momenta vanish and the {\em soft-gluon} limit respectively the {\em soft-quark} limit when only one corresponding momentum vanishes. The corresponding IR exponents are distinguished by upper indices $\delta_{qg}^{u}$, $\delta_{qg}^{gl}$ and $\delta_{qg}^{q}$, respectively. Including all these possibilities in our analysis we can show that the uniform and soft 
divergences studied before indeed provide a solution of the full non-perturbative system of DSEs (for 
simplicity we display results only for the one-loop skeleton expansion, but the 
generalization to an arbitrary order should now be evident, and a 
skeleton-free proof of the dimensional counting along the lines of 
\cite{Alkofer:2006xz} possible).

In a case where scales of different order are present the IR scaling
law can in principle involve all of them \cite{Alkofer:2008jy}. Since it results from the
conformal pure gauge sector, the part in the scaling law that involves
powers of the anomalous scaling exponent $\kappa$ always involves
the characteristic scale $\Lambda_{\tt QCD}$. When there are soft $s$
and hard $h$ external momenta\footnote{Note one could also speak perhaps 
more appropriately of ``ultrasoft'' and ``soft'' momenta but this might 
lead to confusion with NRQCD naming.}, 
the scaling law can also involve ratios
of them and finally the quark propagators introduce the quark mass
$M$. We write all scaling laws in terms of the soft external momentum over 
$\Lambda_{\tt QCD}$ and skip any finite ratios of the different hard scales.
From the point of view of power counting the integral
can be dominated by modes in the regions around all the external scales.
Therefore, it has to be separated into two parts, namely an 'infrared' part
where the loop momenta are of the order of the soft external momentum $s\ll\Lambda_{\tt QCD}$ 
that scales to zero and an 'ultraviolet' part where the loop momenta are 
either of the order of the quark mass $M>\Lambda_{\tt QCD}$ or of the order of the
hard momentum $h<\Lambda_{\tt QCD}$ that stays finite in the limit.

The explicit power counting analysis is presented in appendix   
 \ref{sec:qpc} and yields the following limits: 
 \begin{eqnarray}\label{neweq}
 \lim_{p\rightarrow 0} S(p) &\sim& (p^2)^{\delta_{q}} 
 \hspace*{3.8cm} \mbox{with} \hspace*{3mm} \delta_{q} = 0 \nonumber\\
 \lim_{p,q,k\rightarrow 0} \Gamma_\mu(p,q,k) &\sim& 
 (k^2+p^2+q^2)^{\delta_{qg}^{u}} \,\,\Gamma^0_\mu(p,q,k)  
 \hspace*{4.5mm} \mbox{with} \hspace*{3mm} 
 \delta_{qg}^{u} = \left\{\begin{array}{r}-\frac{1}{2}-\kappa 
 \\ 1\!-\!2\kappa \end{array}\right. \nonumber\\
 \lim_{p\rightarrow 0} \Gamma_\mu(p,q,k) 
 &\sim& (p^2)^{\delta_{qg}^{q}} \,\, \Gamma^0_\mu(p,q,k)  \nonumber\\
 \lim_{q\rightarrow 0} \Gamma_\mu(p,q,k) 
 &\sim& (q^2)^{\delta_{qg}^{q}} \,\, \Gamma^0_\mu(p,q,k)  
 \hspace*{2cm} \mbox{with} \hspace*{3mm} \delta_{qg}^{q} = 0 \nonumber\\
 \lim_{k\rightarrow 0} \Gamma_\mu(p,q,k) 
 &\sim& (k^2)^{\delta_{qg}^{gl}} \,\, \Gamma^0_\mu(p,q,k)  
 \hspace*{2cm} \mbox{with} \hspace*{3mm} 
 \delta_{qg}^{gl} = \left\{\begin{array}{r}-\frac{1}{2}-\kappa 
 \\ 1\!-\!2\kappa \end{array}\right. 
 \end{eqnarray}
 where $k$ is the gluon momentum and $p,q$ are the quark momenta.
 Here $\Gamma_\mu$ is the fully dressed quark-gluon vertex and 
 $\Gamma_\mu^0$ are the undressed tensor structures given in 
 eq.~(\ref{newdec}). Eq.~(\ref{neweq}) is the central result of our
 infrared analysis. It shows indeed that the strong dynamics in the 
 quark sector can self-consistently induce both a uniform and a 
 soft-gluon singularity in the quark-gluon vertex, that are of equal
 strength and which are fully compatible with each other.

%%%%%%%%%%%%%%%%%%%%%%%%%%%%%%%%%%%%%%%%%%%%%%%%%%%%%%%%%%%%%%%%
\subsection{Restoring chiral symmetry \label{chiralsymm}}
%%%%%%%%%%%%%%%%%%%%%%%%%%%%%%%%%%%%%%%%%%%%%%%%%%%%%%%%%%%%%%%

The infrared power law solution found in the last subsection is valid in the 
chirally broken phase of quenched QCD. In the following we will study the chirally
symmetric case, i.e. we will omit all those Dirac structures in the propagator
and quark-gluon vertex that break chiral symmetry. We still restrict ourselves
to the case of quenched QCD, i.e. we do not take into account the backreaction
of the chiral quarks onto the Yang-Mills sector. The solution (\ref{IRsolution}) 
is then still valid. In general, however, unquenching effects are certainly important
for chiral quarks and can lead to a different solution in the gauge sector.
This possibility will be discussed in \cite{parallelpaper}. 

With chirally symmetric Dirac structures only, the quark propagator
and quark-gluon vertex are given by
\be
S_\chi(p) = \frac{i \pslash}{p^2}Z_f(p^2) \,, \ \ \ \Gamma_\chi^\mu (p) = 
i g \left( \sum_{i=1,2} \lambda_i (p^2) L^{\mu}_i + \sum_{i=2,3,6,8} \tau_i(p^2) T^\mu_i \right) 
\ee
in the presence of only one external scale $p^2$; the chirally symmetric 
vertex structures in the general case are given
explicitly in the numerical section \ref{num:chiralsymm}.

For the infrared analysis we project the vertex-DSE on the remaining tensor components. 
We then arrive at the reduced system for the IR exponents $\delta_{q\chi}$ and 
$\delta_{qg\chi}$ in the chiral limit
\beqa
(p^2)^{-\delta_{q\chi}+1/2} &\sim& \mathrm{max} \left\{ (p^2)^{1/2}, (p^2)^{2 \kappa+\delta_{qg\chi}+\delta_{q\chi}+1/2} \right\} \\
(p^2)^{\delta_{qg\chi}} &\sim& \mathrm{max} \left\{ (p^2)^0, (p^2)^{\kappa+2\delta_{qg\chi}+\delta_{q\chi}} \right\}
\eeqa
where $Z_f(p^2) \sim (p^2)^{\delta_{q\chi}}$ denotes a potential power law for the quark wave function
in the infrared. Assuming for the moment that $Z(p^2=0)=Z_f$ stays constant in the infrared,
similar to the chirally broken case discussed above, we then find the
solution
\be
\delta_{q\chi} = 0; \ \ \ \delta_{qg\chi} = - \kappa\,, \label{irweak}
\ee

i.e. an infrared finite quark wave functions and an infrared divergent vertex.
Note, however, that the vertex is considerably less divergent than in the case
of broken chiral symmetry above.

The behavior eq.~(\ref{irweak}) is also realized in our numerical 
solutions, as shown 
in subsection \ref{numsec:chiral}. We have also investigated the chiral quark-DSE 
for a range of possible behaviors of the quark-gluon vertex and found that this 
equation is always controlled by intermediate momenta. In the infrared this 
naturally leads to a constant $Z_f(p^2=0)=Z_f$ in agreement with our assumption from above. 

In addition to the solution presented above, a pure power counting analysis cannot exclude the possibility that both the propagator and the vertex scale trivially in the IR limit
\be
\delta_{q\chi} = 0; \ \ \ \delta_{qg\chi} = 0 \,.
\ee
This possible alternative fixed point is interesting since it would yield an IR vanishing running coupling according to the discussion in the next subsection. A similar behavior has been found previously for the scaling regime of the ungapped magnetic sector at high density \cite{Schafer:2005mc}. We defer a complete analysis of the fixed point structure for chiral quarks to a future work \cite{parallelpaper}.

%%%%%%%%%%%%%%%%%%%%%%%%%%%%%%%%%%%%%%%%%%%%%%%%%%%%%%%%%%%%%%%%
\subsection{The running coupling \label{coupling}}
%%%%%%%%%%%%%%%%%%%%%%%%%%%%%%%%%%%%%%%%%%%%%%%%%%%%%%%%%%%%%%%

We now wish to investigate the running coupling $\alpha_{qg}$ that can be defined from
the quark-gluon vertex. To this end we first recall the corresponding definitions for the
couplings from the Yang-Mills sector. Here, renormalization group invariant
couplings have been defined from either of the primitively divergent vertices 
of Yang-Mills-theory, {\it i.e.} from the ghost-gluon vertex ($gg$), 
the three-gluon vertex ($3g$) or the four-gluon vertex ($4g$) via
\beqa
\alpha_{gg}(p^2) &=& \frac{g^2}{4 \pi} \, G^2(p^2) \, Z(p^2) 
     \hspace*{9mm} \stackrel{p^2 \rightarrow 0}{\sim} \hspace*{2mm} 
     {\bf const}/N_c \,, \label{gh-gl}\\
\alpha_{3g}(p^2) &=& \frac{g^2}{4 \pi} \, [\Gamma^{0,3}(p^2)]^2 \, Z^3(p^2) 
    \hspace*{2mm} \stackrel{p^2 \rightarrow 0}{\sim}
     \hspace*{2mm} {\bf const}/N_c \,,\\
\alpha_{4g}(p^2) &=& \frac{g^2}{4 \pi} \, \Gamma^{0,4}(p^2) \, Z^2(p^2) 
    \hspace*{2mm} 
    \stackrel{p^2 \rightarrow 0}{\sim} \hspace*{2mm} {\bf const}/N_c \,.
     \label{alpha}
\eeqa
Using the DSE-solution (\ref{IRsolution}) it is easy to see that all three 
couplings approach a fixed point in the infrared \cite{Alkofer:2004it}. This fixed point can be
explicitly calculated for the coupling eq. (\ref{gh-gl}). Employing a bare
ghost-gluon vertex one obtains $\alpha_{gg}(0) \approx 8.92/N_c$
\cite{Lerche:2002ep}. A first step in the evaluation of the coupling of the
four gluon vertex has been taken in \cite{Kellermann:2008iw}, where the value 
$\alpha_{4g}(0) \approx 0.0083/N_c$ has been found.

A nonperturbative and renormalization group invariant definition of the
corresponding coupling from the quark gluon vertex is given by
\be
\alpha_{qg}(p^2) = \frac{g^2}{4\pi} \, \lambda_1^2(p^2) \, Z_f^2(p^2) \,
Z(p^2) \hspace*{2mm} \stackrel{p^2 \rightarrow 0}{\sim} \hspace*{2mm} \frac{1}{p^2} \,,
\ee
where $\lambda_1$ dresses the $\gamma_\mu$-part of the vertex. With the 
exponents $\delta_{qgv}=\delta_{qgs}=-1/2-\kappa$, as obtained in our numerical
solutions, we find by power counting that the coupling is proportional 
to $1/p^2$. Thus, contrary to the couplings from the Yang-Mills vertices, 
we find this coupling to be singular in the infrared, which corresponds to
the old notion of infrared slavery 
\cite{Weinberg:1973un,Pagels:1977xv,vonSmekal:1991fp}, that has been 
deemphasized for a decade or two.

If, however, quenched QCD is forced to the chirally symmetric phase as done
in the previous subsection, we recover the fixed point behavior also in the
coupling from the quark-gluon vertex. Then the weaker infrared 
singularity (\ref{irweak}) of the vertex leads to 
\be
\alpha_{qg\chi}(p^2) = \frac{g^2}{4\pi} \, \lambda_1^2(p^2) \, Z_f^2(p^2) \,
Z(p^2) \hspace*{2mm} \stackrel{p^2 \rightarrow 0}{\sim} \hspace*{2mm} {\bf const} \,.
\ee
We thus uncovered a close connection between dynamical chiral symmetry breaking
and two different possible behaviors of the coupling from the quark-gluon
vertex. One obtains infrared slavery when chiral symmetry is broken and a fixed
point when chiral symmetry is restored. We will see later on in section \ref{qqkernel}
that this connection also extends to the quark-antiquark potential and thus
suggests a novel mechanism linking confinement and dynamical chiral symmetry 
breaking in quenched QCD.

%%%%%%%%%%%%%%%%%%%%%%%%%%%%%%%%%%%%%%%%%%%%%%%%%%%%%%%%%%%%%%%%
\subsection{Heavy-quark mass dependence of the quark-gluon vertex \label{IRmass}}
%%%%%%%%%%%%%%%%%%%%%%%%%%%%%%%%%%%%%%%%%%%%%%%%%%%%%%%%%%%%%%%

In the previous subsections we discussed in some detail the emergence
of selfconsistent infrared power laws in the quark-gluon vertex. From our
analytical infrared analysis we derived an infrared behavior of 
$(p^2)^{-1/2-\kappa}$ in case of only one external scale $p$ that goes 
to zero and a soft-gluon infrared divergence of $(p_g^2)^{-1/2-\kappa}$
in case only the external gluon momentum $p_g$ goes to zero. We now investigate
the behavior of the corresponding coefficients and in particular their
dependence on the quark mass.

%%%%%%%%%%%%%%%%%%%%%%%%%%%%%%%%%%%%%%%%%%%%%%%%%%%%%%%%%%%%%%%%
\subsubsection{Naive counting: $M(0) \sim m$ \label{subsec:naive}}
%%%%%%%%%%%%%%%%%%%%%%%%%%%%%%%%%%%%%%%%%%%%%%%%%%%%%%%%%%%%%%%
To this end we start with the vertex constructed in one-loop perturbation 
theory. Then the internal vertices are not dressed and only feature
a color charge $g$ that is universal and independent of the quark mass,
with scale-running effects not yet manifest.
In first-order perturbation theory, all dependence on the mass comes from
the quark propagator, that can be written as
\begin{equation}
S=\frac{1}{\not k - M} = \frac{\not k+M}{k^2+M^2}  \xrightarrow[M \rightarrow \infty] \; 
\frac{1}{M} + \frac{\not k}{M^2} \ .
\end{equation}
Therefore, in the infinite mass limit, the contribution with an even number
of Dirac $\gamma$ matrices (the scalar piece) is suppressed by one power of
the heavy quark mass, the contribution with an odd number (the vector
piece), by two powers.

The behavior of the quark-gluon vertex in perturbation theory, given by fig. \ref{DS3} 
where the  dressed vertices are replaced by bare ones, is then easy to obtain. 
The vector piece is dominant thanks to the bare (mass
independent) diagram, and subsequently takes $1/M^2$ corrections from both
the Dirac structure $(\gamma \not \! k \gamma)/M^2$ of the non-Abelian (second to last)
diagram and the Dirac structure $(\gamma M \gamma M \gamma)/M^4$ of the Abelian diagram. 
Likewise the scalar piece has no zero order term, and takes $1/M$ corrections 
from the structure $(\gamma M \gamma)/M^2$ of the non-Abelian diagram at one-loop. 
For brevity we will from now on use the symbol $V_1$ for the vector part of
the vertex (odd number of gamma matrices) and $S_1$ for the scalar part 
(even number of gamma matrices), where the subscript indicates the number of gluon legs.
Consequently we will use in the following $V_0$ to denote the vector part of the
quark propagator and $S_0$ for its scalar part, i.e. we have 
\begin{equation}
V_0 \sim 1/M^2 \quad , \quad S_0 \sim 1/M \; . 
\end{equation}
We will read Feynman diagrams following the spin 
line from left to right. In this notation we can state for the mass dependence at 
one-loop order
\begin{equation}
V_1 \sim O(M^0) + g^2 O(1/M^2) \quad , \quad S_1 \sim  g^2 O(1/M) \; , \label{leading} 
\end{equation}
where the leading correction is in the scalar piece and arises from the 
$V_1S_0V_1$-combination in the non-Abelian diagram.

Notice that the Abelian diagram also gives subleading corrections through the
$V_1V_0V_1V_0V_1$-combination and the $V_1S_0V_1V_0V_1$-combination. In terms 
of mass dependence these are further suppressed compared to the leading terms 
(\ref{leading}) and can be ignored. Thus the conclusion would be that in 
perturbation theory, heavy quarks
increasingly see a Coulombic potential from perturbative
one-gluon-exchange.

We will now proceed beyond perturbation theory and discuss the leading 
$N_c$, leading skeleton, self-consistent equation for the vertex as 
given in  fig. \ref{irdiag}.
\begin{comment}
\begin{center}
  \begin{picture}(284,93) (0,-1)
    \SetWidth{1.5}
    \ArrowLine(196,-2)(222,18)
    \ArrowLine(263,19)(284,0)
    \ArrowLine(224,19)(263,20)
    \Gluon(238,55)(220,21){5}{2 }
    \Gluon(238,55)(264,20){5}{2}
    \Gluon(238,55)(238,91){5}{2}
    \ArrowLine(112,3)(138,23)
    \Gluon(139,23)(137,62){7.5}{2}
    \ArrowLine(138,23)(159,4)
    \put(163,36){\huge{$\mathbf+$}}
    \Vertex(25,20){4.47}
    \ArrowLine(0,1)(26,21)
    \Gluon(27,21)(25,60){7.5}{2}
    \put(72,35){\huge{$\mathbf =$}}
    \ArrowLine(26,21)(47,2)
    \Vertex(263,19){5.1}
    \Vertex(224,20){5.1}
    \Vertex(240,55){5.1}
  \end{picture}
\end{center}
\end{comment}
Let us first consider the Dirac amplitudes with an odd number of gamma matrices. 
The bare term is of order $O(M^0)$ whereas the one-loop skeleton graph on the 
right gives three types of contributions, namely $V_1V_0V_1$, $S_1S_0V_1$ 
and $S_1V_0S_1$. 
For the even amplitudes, as $S_1$, the bare term does not contribute and the 
one-loop skeleton graph yields structures $S_1V_0V_1$, $V_1S_0V_1$ and $S_1S_0S_1$.

The entire system is still consistent with the perturbative result, that is,
with vertex counting
\begin{equation}
V_1 \sim M^0 , \ \ S_1 \sim M^{-1} \ . \label{pos1}
\end{equation}
But now in addition  self-consistent solutions arise with
\begin{eqnarray} \label{mcountingsols}
V_1 \sim M^{0}, \ \ S_1 \sim M^{1} \label{pos2}\\ 
V_1 \sim M^{1}, \ \ S_1 \sim M^{1} \label{pos3}
\end{eqnarray}
for example, the first stemming from the $S_1=S_1S_0S_1$ term for the scalar
piece and from the $S_1S_0V_1$, $S_1V_0S_1$ terms for the vector piece.
Note that the system has no fixed-point solution with $S_1 \sim M^{n}$ with $n>1$ 
(the $S_1S_0S_1$ term then runs away upon iteration) nor with $V_1 \sim M^{n}$ with $n>1$. 
This still leaves us with the situation that, for a self-consistent solution
both the scalar and vector pieces grow with the quark mass. As will be seen later
on in section \ref{sec:stringtension} this yields an enhanced string-tension between 
heavy-quarks that grows quadratically with $M$. 
\footnote{We presently ignore how the
lattice data would rule out such a scenario. It is far from intuitive since 
lattice practitioners often fix the scale to the string tension, 
now scaling with the quark mass, and in the heavy mass limit all other 
splittings they could measure in heavy quarkonium also scale with the mass.
Happily experimental data, as seen later, seems to rule this solution out.}

Let us now back-feed these solutions to the quark-propagator DSE to ascertain
whether any change in the perturbative $1/M$ behavior of the propagator is
to be expected. 
Since the inverse propagator is the amputated Green's
function with two quarks and no gluons, by analogy it is obvious that the $1/M$ 
counting for higher-point functions (with two quarks, $n$ gluons) is the same. Consider first the 
vector part of the propagator. We have $V_0 \geq O(1)$ because of the presence of the
un-suppressed vector term of the inverse bare propagator $\not k$ on the right 
hand side of the DSE. Substituting now $V_1=1,\ M$ in the $\gamma V_0 V_1$ 
structure of the loop we obtain $M^{-2}$ or $M^{-2} M=M^{-1}$. Thus the vector 
part of the propagator remains bare in the heavy-quark 
limit. Consider now the scalar part of the propagator equation. This gives 
$S_0 \geq O(M)$ from the bare mass term of the propagator. If we substitute 
$S_1=M$ and $V_0=1$ into the self-energy, the relevant structures $\gamma S_0 V_1$ 
and $\gamma V_0 S_1$ are $O(M^{-1})$ and $O(M^0)$ respectively, and therefore 
subleading compared to the bare mass term. Also, the solution with $V_1=1$ 
yields a power $M^0$. We are thus left with the purely perturbative counting 
in the presence of a heavy quark mass scale, that is,
\begin{equation}
V_0 \sim M^0,\ \ S_0 \sim M^1 \ ,
\end{equation}
consistent with our assumption above. 

For the quark-gluon vertex we are thus left with all the three possibilities 
(\ref{pos1}), (\ref{pos2}) and (\ref{pos3}). It is then a purely dynamical 
question which of these three possibilities are realized. As a matter of fact 
there is even a fourth possibility which we discuss now.

%%%%%%%%%%%%%%%%%%%%%%%%%%%%%%%%%%%%%%%%%%%%%%%%%%%%%%%%%%%%%%%%%%%%%%%%%5
\subsubsection{Modified counting: $M(0) \not\sim m$ 
\label{subsec:newcounting} }
%%%%%%%%%%%%%%%%%%%%%%%%%%%%%%%%%%%%%%%%%%%%%%%%%%%%%%%%%%%%%%%%%%%%%%%%%%%

There is yet another logical possibility that solves the
coupled system of propagator and vertex equations, and this is a propagator
that is not suppressed by a power of the heavy quark current mass in the
infrared, that is, with some degree of saturation in 
$M(p^2)=B/A$, a mass-independent constant. Substituting this in the
vertex equation, it is obvious that self-consistency is reached by a
quark-gluon vertex that is also mass-independent, i.e. we have
\begin{equation}
V_1\sim M^0\ , \ S_1 \sim M^0 . \label{pos4}
\end{equation}

This case has interesting consequences. As will be discussed later on in
section \ref{sec:stringtension},
the string tension is then mass-independent. However, in the Schr\"odinger equation
one cannot take the infinite-quark mass limit and simultaneously study the
two quarks at infinite distance. The infinitely heavy-quark mass limit only
makes sense at intermediate momentum where the propagator is still affected
(suppressed) by the heavy quark mass.

As mentioned before, the four cases (\ref{pos1}), (\ref{pos2}), (\ref{pos3}) 
and (\ref{pos4}) are all selfconsistent and it is a question of dynamics
which one of them is realized. It may even be, that in different regions
of the quark mass different scenarios are realized, i.e. one may e.g.
encounter the scaling law (\ref{pos4}) for physical masses up to the
bottom quark mass, whereas the naive perturbative scaling (\ref{pos1})
is realized for quark masses as large as the top quark.

In fact, as we will see in section \ref{sec:numsec}, the scenario (\ref{pos4})
is favored by our numerics. The vertex is only mildly dependent on the
current quark mass for masses at least up to the bottom quark. 

However, this behavior does not persist for larger masses. Our numerical
calculations show that for masses in the range of $20-40$ GeV the scaling 
goes over into the perturbative one given by eq.~(\ref{pos1}).

%%%%%%%%%%%%%%%%%%%%%%%%%%%%%%%%%%%%%%%%%%%%%%%%%%%
\subsection{The Slavnov-Taylor identity}
%%%%%%%%%%%%%%%%%%%%%%%%%%%%%%%%%%%%%%%%%%%%%%%%%%%
In this subsection we check that the infrared power-counting that we 
have devised satisfies the pertinent Slavnov-Taylor identity for the 
quark and gluon vertex. Of course, since our analysis largely hinged 
on dimensional reasoning one does not expect to find inconsistencies 
here. 

 In the following we assume
all momenta to be small and proportional to each other, such that our
infrared counting rules can be applied.

The Slavnov-Taylor identity for the quark-gluon vertex is explicitly given and 
derived in \cite{Eichten:1974et}. One needs to introduce a non-standard 
vertex $H^a(k,q)$ that corresponds to a ghost-to quark off-diagonal 
self-energy. This of course is an auxiliary device that violates all of
fermion number, color and momentum conservation. One can think of it as 
a quark-ghost source.
This vertex can be obtained in terms of an equivalent \lq \lq bare vertex'' 
$gT^a$ via an inhomogeneous Bethe-Salpeter equation
\begin{equation}
H^a(k,q) = gT^a + B^a(k,q)
\end{equation}
with the loop integral coupling it to the ghost-quark scattering kernel.
 \begin{center}
  \begin{picture}(217,44) (0,-37)
    \SetWidth{1}
    \GBox(30,-34)(62,-1){0.882}
    \ArrowLine(1,-33)(29,-25)
    \DashArrowLine(0,0)(29,-8){5}
    \DashArrowArcn(88.56,-28.67)(34.28,140.79,23.5){5}
    \ArrowArc(89.63,-2.23)(33.77,-142.05,-25.93)
    \Vertex(119,-16){3.61}
    \put(129,-16){\Large{$g T^a$}}
    \put(175,-16){\Large{$=-B^a(k,q)\ .$}}
  \end{picture}
\end{center}
The momenta $k$ and $q$ correspond respectively to the incoming ghost 
and quark legs. 

The function $H$ is infrared finite within our power counting. To see 
this, expand the quark-ghost scattering kernel in terms of a skeleton 
graph and count the anomalous dimensions of the lowest order using the 
general counting rule (\ref{IRsolution}) and the infrared leading
$1/M$-terms in the quark-propagator. One arrives then at a total 
infrared anomalous dimension of $\kappa-1$. This power is precisely 
canceled by the $-\kappa-1$ from the ghost propagator in the diagram 
above and the $+2$ from the loop integration measure. Hence the total 
exponent is $0$ and the function $H$ takes a finite constant value 
in the infrared.

The Slavnov-Taylor identity for the quark-gluon vertex reads
\begin{equation}
p_3^\mu \Gamma^a(p_1,p_2) \times G^{-1}(p_3^2)=
H^a(p_1,p_2)S^{-1}(p_2) - S^{-1}(p_1) H^a(p_1,p_2)
\end{equation}
in terms of the inverse quark propagator $S^{-1}$ and inverse ghost 
self-energy $G^{-1}$ (note this is written down as $1+b(p_3^2)$ in the 
original reference \cite{Eichten:1974et} but there it was assumed to be 
infrared finite as opposed to infrared suppressed).
Since the quark propagator in the chirally symmetric broken phase is a 
constant and we just argued that $H$ is also an infrared constant, so is 
the right hand side of the STI. 

The left hand side power-law is easy to count: $-1/2-\kappa$ from the 
quark-gluon vertex, $1/2$ from the explicit gluon momentum, and
$+\kappa$ from the inverse ghost dressing $G^{-1}$ precisely cancel out to 
yield $0$. Thus our infrared power-counting is consistent with the 
requisite gauge invariance. Note also that the argument is independent 
of whether the quark momentum is small or not, so it applies to either 
soft or uniform divergence counting, as long as the same counting is 
used for the quark-gluon vertex and the quark-ghost scattering kernel. 

Finally we discuss the case of the chirally symmetric solution. 
Here the H-function is still constant since to lowest order in the 
skeleton expansion there are two powers of $(p^2)^{-1/2}$ less in 
the quark-gluon vertices but also two powers of $1/p$ more in the 
quark propagators. The inverse propagator $S^{-1}$ on the right hand 
side of the STI is now given by $i \not p$, whereas the vertex 
dressing on the left hand side cancels the contribution of the ghost
dressing function. So both  the left and the right hand side of the 
equation are proportional to $p$ and the matching still works.

It is also easy to see how at least the perturbative and modified mass-
countings for the quark and gluon vertex, for heavy-quark masses, are
consistent with the Slavnov-Taylor identity. In the modified counting,
neither propagator nor quark-gluon vertex are dependent on the quark
mass, therefore there is no running in either LHS or RHS of the STI. The
perturbative counting can be understood from the Abelian WTI in QED
\be
i(k_\mu-k'_\mu) \Gamma_\mu(k,k') = S^{-1}(k) - S^{-1}(k')
\ee
where for fixed $k$, $k'$ and $M\to \infty$ the RHS appears to have a
vector part of the fermion propagator $V_0 \propto M^0$, and a scalar 
part $S_0\propto M^1$; however, neglecting momentum dependence,
the scalar part on the RHS cancels ($M-M=0$) and since the LHS stays
mass-independent, because $\gamma_\mu$ is mass-independent, the mass
independence of both sides coincides.
Coming now to the QCD STI for the qqg vertex in perturbation theory, in
this counting the $H^a$ kernel becomes a mass-independent constant at
large mass (only the $gT^a$ bare term survives) and the $M-M$
cancellation happens again on the RHS. Thus at least the perturbative
and modified large-mass countings are seen to be consistent with the
STI.
For other mass countings the situation is more complicated and we will
not discuss them further.

%%%%%%%%%%%%%%%%%%%%%%%%%%%%%%%%%%%%%%%%%%%%%%%%%%%%%%%%%%%%%%%%%%%%%%%%
\section{Numerical results for the quark-gluon vertex \label{sec:numsec}}
%%%%%%%%%%%%%%%%%%%%%%%%%%%%%%%%%%%%%%%%%%%%%%%%%%%%%%%%%%%%%%%%%%%%%%%%

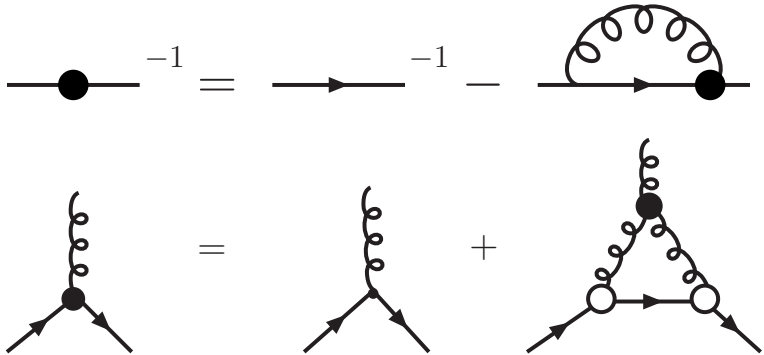
\begin{figure}[b]
\begin{center}
   \begin{picture}(284,50)
     \SetWidth{1.5}
     \ArrowLine(0,0)(50,0)
     \ArrowLine(100,0)(150,0)
     \ArrowLine(200,0)(280,0)
     \put(72,-5){\huge{$\mathbf=$}}
     \put(172,-5){\huge{$\mathbf-$}}
     \put(52,7){\large{$\mathbf-1$}}
     \put(152,7){\large{$\mathbf-1$}}
     \CCirc(25,0){5}{1}{1}
     \GlueArc(240,0)(25,0,180){5}{5}
     \CCirc(265,0){5}{1}{1}
   \end{picture}
\end{center}
\begin{center}
  \begin{picture}(284,90)
    \SetWidth{1.5}
    \ArrowLine(196,0)(222,18)
    \ArrowLine(263,19)(284,0)
    \ArrowLine(224,19)(263,19)
    \Gluon(240,55)(220,20){3}{3}
    \Gluon(240,55)(264,20){3}{3}
    \Gluon(242,55)(242,80){3}{2}
    \ArrowLine(112,0)(138,23)
    \Gluon(139,23)(137,62){3}{3}
    \ArrowLine(138,23)(159,0)
    \put(175,36){\Large{$\mathbf+$}}
    \Vertex(25,20){4.47}
    \Vertex(138,22){2}
    \ArrowLine(0,0)(26,21)
    \Gluon(27,21)(27,60){3}{3}
    \put(72,35){\Large{$\mathbf =$}}
    \ArrowLine(26,21)(47,0)
    \Vertex(242,55){5.1}
    \BCirc(263,20){5}
    \BCirc(224,20){5}
  \end{picture}
\end{center}
\caption{The DSE for the quark propagator and the quark-gluon vertex as used in 
our numerical investigation. The open circles in the diagram on the right hand side of
the vertex-DSE indicate that we use fit functions for these internal vertices, 
whereas the quark-gluon vertex in the quark-DSE is the one calculated from the 
vertex-DSE. \label{numDSE}}
\end{figure}

Having discussed the infrared structure of the propagator and vertex DSEs
analytically, it is now time to underpin our findings with numerical studies.
To this end we solve the coupled system of the DSE for the quark propagator
and the one for the quark-gluon vertex in the infrared preserving approximation 
discussed around fig.~\ref{irdiag}. For the convenience of the
reader we display the resulting system again diagrammatically in fig.~\ref{numDSE}. 
In the following we first discuss our numerical techniques and then present
our results.

%%%%%%%%%%%%%%%%%%%%%%%%%%%%%%%%%%%%%%%%%%%%%%%%%%%%%%%%%%%%%%%%%%%%%
\subsection{Renormalization and numerical methods \label{numerics}}
%%%%%%%%%%%%%%%%%%%%%%%%%%%%%%%%%%%%%%%%%%%%%%%%%%%%%%%%%%%%%%%%%%%%%

In our numerical calculations we use solutions of the
coupled ghost and gluon propagator DSEs as input, cf. appendix \ref{propinput}.
These have been renormalized at the scale $\mu^2=170 \mbox{GeV}^2$, which
we also use to renormalize the quark-gluon vertex and the quark propagator.
The corresponding renormalization conditions for the quark wave function
$Z_f(p^2)$ and the vertex $\Gamma_\nu(p_1^2,p_2^2,p_3^2)$ are
\beqa
Z_f(\mu^2) &=& 1 \nonumber\\
\lambda_1(\mu^2,2\mu^2,3\mu^3) &=& 1. \label{rencond}
\eeqa
Note that only $\lambda_1$ appears in this condition, since the accompanying
tensor $\gamma_\nu$ is the only divergent part of the quark-gluon vertex;
all other dressing functions $\lambda_{2..4}$ and $\tau_{1..8}$ are UV-finite. 
Note also that we renormalize the vertex at the totally asymmetric momentum slice
(c.f. appendix \ref{kin}), which is convenient for our numerical setup.
The conditions (\ref{rencond}) then unambiguously determine the renormalization
constants $Z_2$ and $Z_{1F}$ for the quark propagator and the quark-gluon
vertex.

The loop integral on the right hand side of the DSE is treated employing
standard techniques. We perform all four integrals numerically in 
hyper-spherical coordinates using Gauss-Legendre integration;
no angular approximations are done. The radial integral is treated on a 
logarithmic integration grid with explicit infrared and ultraviolet momentum 
cutoffs. We verified that our results are independent of these cutoffs. 

The methods to project the right hand side of the vertex equation onto the
various tensor structures of the vertex are described 
in some detail in appendix \ref{app:dirac}. We evaluate the vertex from the 
vertex-DSE on a three-dimensional grid, which is adapted to the quark DSE.
To this end we use the propagator fits (\ref{eq:fitsigs}) and (\ref{eq:fitsigv}) 
as our start guess for the quark propagator.
Upon backfeeding the vertex into the quark DSE, this equation is iterated
until convergence. The resulting quark propagator is then again back-feeded
into the vertex-DSE and so on until complete convergence of the coupled
system of equations is achieved. 

Due to the considerable numerical complexity we have not yet found a way to
completely back-feed the vertex into its own DSE. We therefore employ only
parts of its tensor structure for the two internal vertices in the
non-Abelian diagram. These are represented by fit-functions which are
matched against the calculated vertex in the asymmetric kinematical 
momentum slice. From our infrared analysis in subsection (\ref{irquarks})
we saw that in principle it is sufficient to back-feed the most important 
vector and scalar components into the equation to generate selfconsistent
infrared physics (at least on a qualitative basis). We therefore employ 
the two structures
\beq
\Gamma_\mu (p_1,p_2) = \gamma_\mu \lambda_1(p_1,p_2) -i (p_1+p_2)_\mu \lambda_3(p_1,p_2) 
\eeq
for the internal quark-gluon vertices in the non-Abelian diagram. Here we will use the fit
forms 
\beq
\lambda_1(p_1,p_2) = f^{\tt IR}(x) g_1^{\tt UV}(x) \quad , \quad 
\lambda_3(p_1,p_2) = \frac{1}{\sqrt{(p_1+p_2)^2}} 
                     f^{\tt IR}(x) g_3^{\tt UV}(x) \label{vertexfit}
\eeq
with $x=p_1^2+p_2^2+p_3^2$. These include the proper asymptotics in the IR and UV
\beq
f^{\tt IR}(x) \equiv \left(\frac{x}{d_1+x}\right)^{-\kappa-1/2} , \; 
g_1^{\tt UV}(x) \equiv \left(\frac{d_1}{d_1+x} + d_2 
\log{\left[\frac{x}{d_1}+1\right]}\right)^{-9/44} , \; 
g_3^{\tt UV}(x) \equiv \left(\frac{d_2}{d_2+x^{n_1}}\right)^{n_2}  \left(\frac{d_3}{d_3+x}\right)^2 \; .
\eeq
For the three-gluon vertex we use the tree-level tensor structure and represent
the corresponding dressing function by
\beq   
\Gamma^{0,3}(p_1,p_2) = \left(\frac{x}{d_1+x}\right)^{-3\kappa} 
                \left(d_3 \frac{d_1}{d_1+x} + d_2 \log{\left[\frac{x}{d_1}+1\right]}\right)^{17/44}
		\label{3gfit}
\eeq
which reflects the strong $(p^2)^{-3\kappa}$ singularity known from
Yang-Mills theory, eqs.(\ref{IRsolution}),(\ref{IR3g}) and a logarithmic ultraviolet tail 
with coefficients from resummed perturbation theory. In all fits the momentum
argument is given by the sum of the three squared
momenta from the external legs. The fit forms thus represent overall divergences
only and do not take care of possible soft gluon divergencies. We will modify
the fits accordingly when we investigate this issue in subsection \ref{num:softglue}.

In general, backfeeding only parts of the vertex into the vertex equation
is somewhat problematic. Since most of the tensor structures of the 
vertex
contribute considerably to the loop integral of the non-Abelian diagram,
omitting some of them leads to reduced interaction strength in the integral.
As can be seen from our power counting in subsection \ref{irquarks}
this fact does not affect the qualitative behavior of the resulting
vertex dressing functions (i.e. their infrared exponents), but does 
influence the coefficients of the resulting power laws. 
%\begin{comment}
As a result one faces a dilemma: 
\begin{itemize}
\item[(i)] One can insist that the input fit functions match both the
output infrared exponents AND their coefficients. We found that this is only
possible if our input for the three-gluon vertex is very strong, i.e. much
stronger than one would expect from corresponding DSEs for the three-gluon 
vertex \cite{Huber}. The missing interaction strength, presumably from 
the neglected $ggg$ vertex pieces is then simulated by the strong 
three-gluon interaction.
This setup allows one to systematically compare the resulting quark-gluon
vertices for different quark masses. However, as we will see, it leads to 
unnaturally large scales, which cannot be true.
\item[(ii)] If we instead insist on correct physical scales, measured e.g.
by the chiral condensate from the quark propagator in the chiral limit, the
missing interaction strength in the non-Abelian diagram does not permit
a match to the scales from the input fits with the resulting vector and scalar
vertex pieces. The missing interaction strength from the neglected vertex 
pieces in the loop diagram is then taken over by the input dressing
functions, which are stronger than the calculated ones. 
\end{itemize}
%\end{comment}
Naturally this quantitative problem would not exist if we were able to back-feed the 
complete
vertex into the vertex-DSE. Postponing this considerable numerical effort
for future work, we will use a combination of the possibilities (i) and (ii) here.
We first investigate the chiral limit using the procedure (ii) and thus
producing phenomenologically acceptable scales in the resulting vertex and quark
propagator. For our quark mass study, however, we have to resort to the
procedure (i). Since we are dealing with qualitative statements only, this
compromise seems acceptable to us and we believe that the results we obtain
are meaningful.

%%%%%%%%%%%%%%%%%%%%%%%%%%%%%%%%%%%%%%%%%%%%%%%%%%%%%%%%%%%%%%%%%%%%%%
\subsection{The chiral limit \label{numsec:chiral}}
%%%%%%%%%%%%%%%%%%%%%%%%%%%%%%%%%%%%%%%%%%%%%%%%%%%%%%%%%%%%%%%%%%%%%%

Let us first have a look at the chiral limit and the corresponding structure
of the quark-gluon vertex. The input parameters in our calculation for the 
fits (\ref{vertexfit}) and (\ref{3gfit}) are given in table \ref{param0}.
Here the ultraviolet behavior of the three-gluon vertex is controlled
by the parameter $d_2$, which is chosen such that the vertex is 
normalized
to the tree-level expression at the renormalization point $\mu^2=170 \mbox{GeV}^2$.
The corresponding ultraviolet coefficients $d_2$, $d_3$, $n_1$ and $n_2$ of the input functions
$\lambda_1$ and $\lambda_3$ are chosen such that the calculated vertex dressings
are matched by the input. The remaining infrared coefficients $d_1$ in 
$\lambda_{1,3}$ are chosen such that the relative strength of the 
corresponding calculated dressings is matched. Finally, the parameters $d_1$ and $d_3$ 
in $\Gamma^{0,3}$ are chosen such that the chiral condensate of the quark 
propagator reproduces the phenomenological value.  

\begin{table}[h]
\begin{tabular}{ c|c|c|c|c|c}
               & $d_1$   &  $d_2$   &  $d_3$ &  $n_1$ & $n_2$    \\\hline\hline
 $\lambda_1$   & 2.00 & 0.5   &  -  &  -  & -     \\
 $\lambda_3$   & 4.00 & 0.5   & 0.5 &  1  & -0.5  \\\hline
 $\Gamma^{0,3}$& 1.00 & 0.40  & 10  &     &            
\end{tabular}
\caption{Input parameters for the fit functions (\ref{vertexfit}) and 
(\ref{3gfit}) in the chiral limit.\label{param0}
}
\end{table}

\begin{figure}[t]
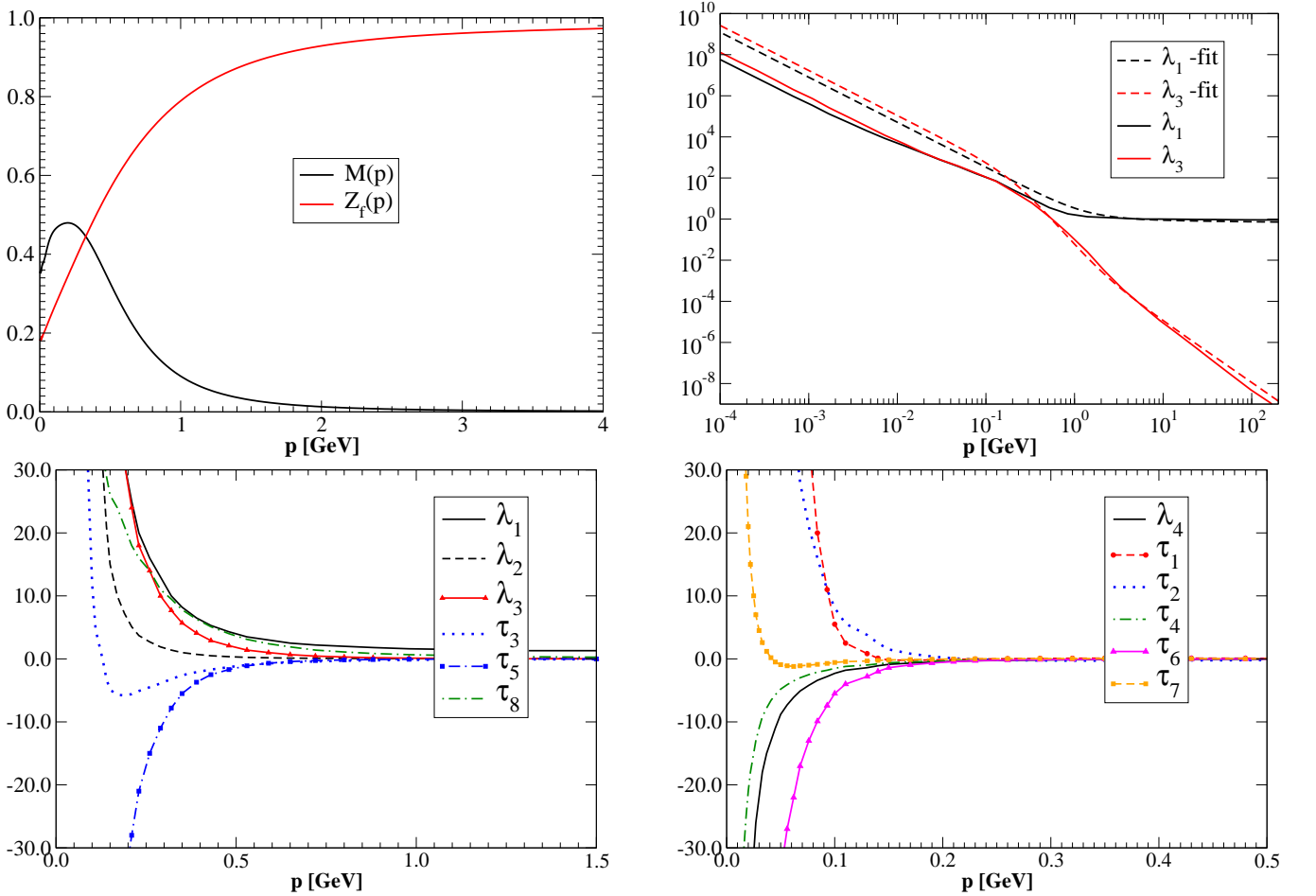

\includegraphics[width=8.5cm]{res.quark.X.eps}\hfill
\includegraphics[width=8.5cm]{res.vertex.X1.eps}
\vspace*{5mm}
\includegraphics[width=8.5cm]{res.vertex.X2.eps}\hfill
\includegraphics[width=8.5cm]{res.vertex.X3.eps}
\caption{\label{fig:chiral}
Top left corner: the chiral quark propagator on a linear momentum scale. Note that
the maximum seen in the quark mass function is not a strict prediction of our truncation
and may or may not be a truncation artefact. Top
right corner: internal representations of the vector and scalar vertex dressings 
(dashed lines) compared to the corresponding calculated vertex pieces (full lines).
Bottom left corner: The six large tensor components of the vertex. Bottom right
corner: the six small tensor components of the vertex. Note the different
momentum-scale as compared to the plot of the large components. All vertex dressings have been multiplied by appropriate powers of momenta, given in table II, such that they correspond to
a normalized version of the basis eq. (2) and can be compared
quantitatively. The dressings are plotted versus $p \equiv p_1$ and
and evaluated in the asymmetric momentum slice with $p_3^2=3p_1^2$ and $p_2^2=2p_1^2$.}
\end{figure}

The resulting input and output vertex dressings for $\lambda_1$ and $\lambda_3$
are shown in fig.~\ref{fig:chiral} together with the resulting quark
propagator and further vertex dressing functions. 
%%%%%%%%%%%%%%%%%%%%%%%%%%%%%%%%%%%%%%%%%%%%%%%%%%%%%%%%%%%%%%%%%%%%%%%%%%%%%%%%%
%%%%%%%%%%%%%%%%%%%%%%%%%%%%%%%%%%%%%%%%%%%%%%%%%%%%%%%%%%%%%%%%%%%%%%%%%%%%%%%%%

In the top right corner we compare the internal representations of the vector
($\gamma_\mu$) and the scalar ($p_\mu$) piece of the quark-gluon vertex with
the calculated components. The most important observation here is the perfect
matching of the generated infrared exponent $(p^2)^{-1/2-\kappa}$ of both
the $\lambda_1$ and $\lambda_3$ components with the infrared exponents from 
the input fits. In fact the very same power law is developed by all twelve 
tensor structures $\lambda_{1..4}$ and $\tau_{1..8}$. One then expects that 
backfeeding additional structures besides $\lambda_1$ and $\lambda_3$ 
would just reproduce the same infrared exponents as seen from fig.~
\ref{fig:chiral} with backfeeding $\lambda_1$ and $\lambda_3$ only. 
We have explicitly checked that this is indeed the case. Thus the numerical 
result confirms that {\it all dimensionless vertex dressings share the same 
infrared divergence}\footnote{In the notation of appendix \ref{IRanalysis}
and ref. \cite{Alkofer:2006gz} this amounts to 
$\beta_1=\beta_2=\beta_3=\beta_4=-1/2-\kappa$.}. 

For the quark mass function $M(p)$ we observe the characteristic rise 
towards small momenta that is typical for dynamical chiral symmetry 
breaking. Quite unexpectedly
we also observe a turnover at very small momenta, which to our knowledge has 
not been seen yet in any other truncation scheme. Whether this turnover is a 
genuine effect independent of the truncation of the vertex-DSE remains to be 
investigated. Also for the quark wave function $Z_f(p)$ we observe a stronger
decrease in the infrared than has been observed in previous truncation schemes
(see e.g. \cite{Fischer:2003rp}).  

As concerns the scale we did not match the input and output scales here but 
pursued scenario (ii) and
seeked to generate the quark-gluon vertex at a \lq physical' scale such that the 
chiral condensate has the correct magnitude. This condensate can be extracted 
from the quark propagator in the chiral limit according to
\beqa
-\langle \bar{\Psi}\Psi\rangle(\mu) := Z_2 \, Z_m \, N_c \,\mathrm{tr} \int
\frac{d^4q}{(2\pi)^4} S_{0}(q^2) \,,
\label{ch-cond}
\eeqa
where the trace is over Dirac indices, $S_{0}$ is the quark propagator in the 
chiral limit and $Z_m$ is the quark mass renormalization factor that can also 
be determined from the quark-DSE. We obtain
\beq
\langle \overline{\Psi} \Psi \rangle^{0,N_f=0}_{\overline{MS}}(2 \, \mbox{GeV}) 
= (-260)^{3} \,\mbox{MeV},
\eeq
where we have used $\Lambda_{\overline{MS}} = 225$ MeV. This value is in agreement
with the values reported in the literature.

\begin{table}[b]
\begin{tabular}{ c|c|c|c|c|c|c|c|c|c|c|c}
$\lambda_1$ & $\lambda_2$ & $\lambda_3$ & $\lambda_4$ & $\tau_1$ & $\tau_2$ &
$\tau_3$ & $\tau_4$ & $\tau_5$ & $\tau_6$ & $\tau_7$ & $\tau_8$ \\\hline
 $1$       &
 $4 p_1^2$ &
 $2 p_1$   &
 $2 p_1$   &
 $  p_1^3$ &
 $2 p_1^4$ &
 $  p_1^2$ &
 $4 p_1^3$ &
 $  p_1$   &
 $2 p_1^2$ &
 $2 p_1^3$ &
 $2 p_1^2$ 
\end{tabular}
\caption{\label{table:factors}
Dimensionful factors with which the vertex dressings are multiplied
to obtain dimensionless quantities.}
\end{table}
In the bottom line of fig.~\ref{fig:chiral} we compare the strengths of the
different structures of the vertex. These are multiplied by appropriate factors
to make them dimensionless; the factors are given in table 
\ref{table:factors}.
Clearly some structures are more dominant in
terms of infrared strength (i.e. coefficients of the universal power law 
$(p^2)^{-1/2-\kappa}$), the most important pieces being $\lambda_1$, $\lambda_3$ and
$\tau_5$ \footnote{This has already been noted in \cite{Bhagwat:2004hn,Bhagwat:2004kj}.}. 
There is no uniform behavior of the various vertex dressings but some
are strictly positive and monotonic, some are strictly negative and some change
sign in the infrared. This behavior may or may not be an artifact of the chosen truncation 
scheme of the quark-gluon vertex DSE and will be investigated further in future work. 
Of all structures perhaps the most interesting one is the $\lambda_3$ piece,
since it represents the leading chiral symmetry breaking component of the vertex.
Its considerable infrared strength signals a large amount of dynamical chiral symmetry
breaking not only in the quark propagator but also in the vertex structures. This
is consistent with and in fact corroborates the strategy and results of our infrared 
analysis in subsection \ref{irquarks} and appendix \ref{IRanalysis}.

\begin{figure}[t]
\includegraphics[width=12cm]{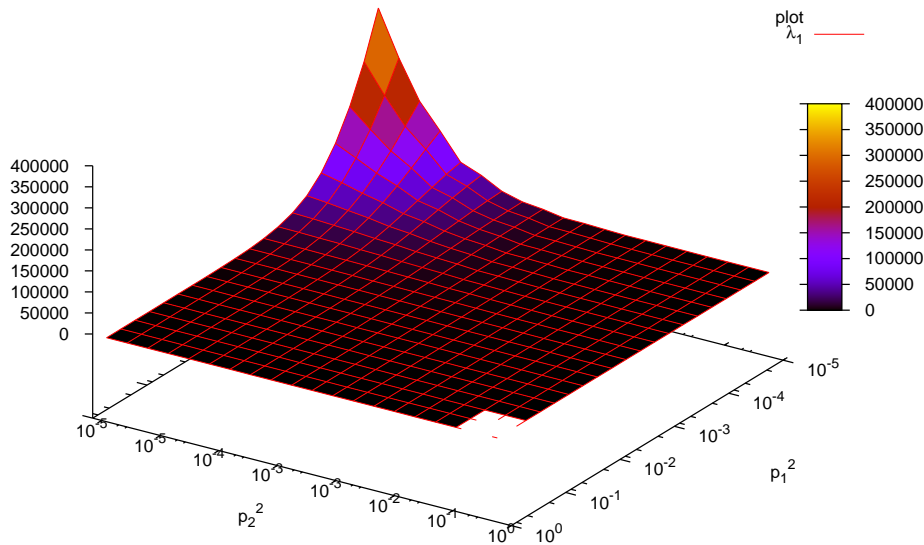}
\caption{\label{fig:3dplot}
The vertex dressing function $\lambda_1(p_1,p_2)$ at fixed angle 
$\arccos(\hat{p}_1 \cdot \hat{p}_2)=0.1$.}
\end{figure}

A two-dimensional plot of the dressing function $\lambda_1(p_1,p_2)$ at fixed angle 
$\arccos(\hat{p}_1 \cdot \hat{p}_2)=0.1$ is shown on in fig.~\ref{fig:3dplot}.
%%%%%%%%%%%%%%%%%%%%%%%%%%%%%%%%%%%%%%%%%%%%%%%%%%%%%%%%%%%%%%%%%%%%%%%%%%%%%%%%%
%%%%%%%%%%%%%%%%%%%%%%%%%%%%%%%%%%%%%%%%%%%%%%%%%%%%%%%%%%%%%%%%%%%%%%%%%%%%%%%%%
One clearly observes, that the infrared divergence of $\lambda_1$ occurs almost 
uniformly in the $p_1^2 - p_2^2$-plane. This is in qualitative agreement with lattice 
calculations of the vertex \cite{skullerud:2002ge} at those momenta available on the 
lattice so far. Unfortunately the volumes of these lattice simulations are by far not 
large enough to be able to confirm or refute the divergence of the
vertex in the deep infrared.

\begin{figure}[t]
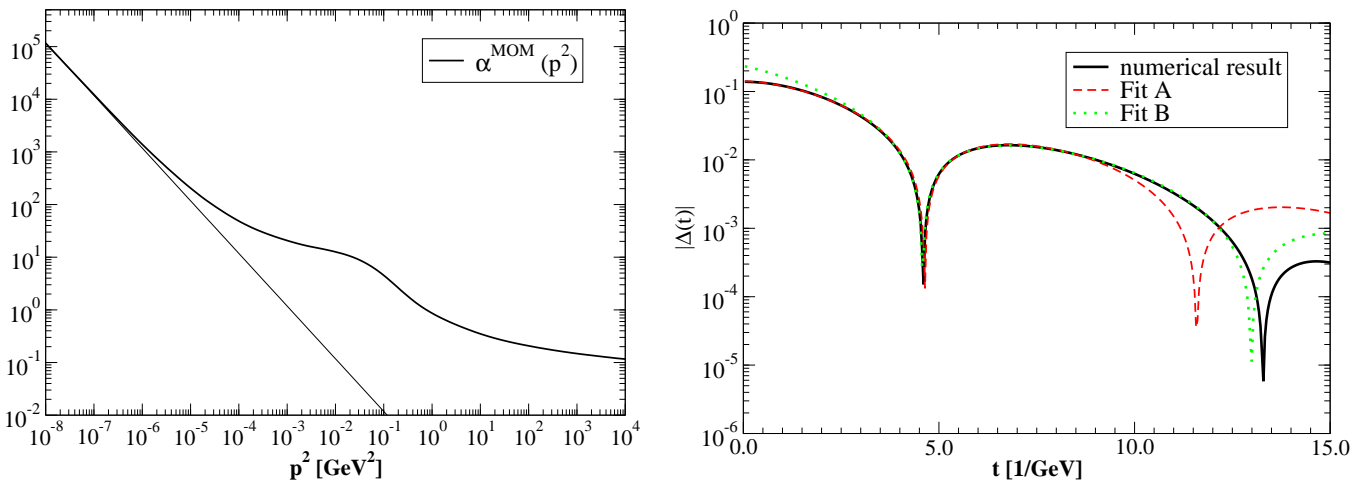

\includegraphics[width=8.5cm]{res.coupling.eps}\hfill
\includegraphics[width=9cm]{res.FFT.eps}
\caption{\label{fig:coupling}
Left diagram: The running coupling from the quark-gluon vertex in the 
asymmetric MOM-scheme with
$\mu^2=p_1^2=p_2^2/2=p_3^2/3$. {Infrared slavery in the form of a 
power-law is seen at extremely low-momenta, whereas the coupling is 
quite flat at intermediate to low-momenta, just as needed for conformal 
arguments \cite{Brodsky:2006uq}.} Right diagram: absolute value of the 
Schwinger function 
of the scalar part of the quark propagator in the chiral limit. Clearly
visible are positivity violations at a scale of about $5/GeV \sim 1 fm$.}
\end{figure}

Next we determine the running coupling from the quark-gluon vertex. For
the convenience of the reader we repeat its definition here, (c.f. subsection \ref{coupling}):
\beq
\alpha_{qg}(p^2) = \frac{g^2}{4\pi} \, \lambda_1^2(p^2) \, Z_f^2(p^2) \,
Z(p^2) \,.
\eeq
Although the right hand side of this definition is renormalization point invariant,
it still depends on the renormalization scheme. The scheme we are using, eq.~(\ref{rencond}), 
is adapted to the asymmetric kinematical momentum slice.
We call this an \lq asymmetric MOM-scheme'. The resulting running coupling
as numerically calculated in this scheme is displayed in fig.~\ref{fig:coupling}.
Clearly one sees the usual logarithmic running of the coupling in the ultraviolet,
with a value $\alpha_{qg}^{MOM-asymm}(M_Z^2) = 0.118$ at the mass of the Z-boson.
(Note that the exact agreement of this result with the experimentally determined
coupling in the $\overline{MS}-scheme$ is incidental.)
Towards lower momenta we see a gradual increase of the coupling, where the  
nonperturbative region $\alpha_{qg}^{MOM-asymm} > 1$ is reached at about
$p^2= 0.75 \mbox{GeV}^2$. The coupling then continually grows, seems to reach a plateau
(corresponding to a similar plateau in $\lambda_1$) and then enters the scaling region
in the infrared. In this region, below $p^2= (1 \mbox{MeV})^2$ we see a $1/p^2$ singularity
in agreement with our analytical analysis in subsection \ref{coupling}. 
%%%%%%%%%%%%%%%%%%%%%%%%%%%%%%%%%%%%%%%%%%%%%%%%%%%%%%%%%%%%%%%%%%%%%%%%%%%%%%%%%
%%%%%%%%%%%%%%%%%%%%%%%%%%%%%%%%%%%%%%%%%%%%%%%%%%%%%%%%%%%%%%%%%%%%%%%%%%%%%%%%%

Finally we investigate the possibility of positivity violations in the quark
propagator. These are related to the analytic properties of the quark propagator 
which can in part be read off from the corresponding Schwinger function
\beq
\sigma(t)\; =\; \int d^3x \int \frac{d^4p}{(2\pi)^4} \exp(i p \cdot x) \sigma_{S,V}(p^2),
\label{FT}
\eeq
where $\sigma_{S,V}$ are the scalar and the vector parts, respectively, of the 
dressed quark propagator. (This method has a long history, see {\it e.g.} 
\cite{Alkofer:2003jj,Krein:1990sf,Oehme:1994pv,Burden:1997ja} 
and references therein). According to the Osterwalder-Schrader axioms of 
Euclidean field theory \cite{Osterwalder:1973dx}, this function has to be positive 
to allow for asymptotic states in the physical sector 
of the state space of QCD. Conversely, positivity violations in the Schwinger function
show that the corresponding asymptotic states (if present) belong to the
unphysical part of the state space. Thus positivity violations constitute a 
sufficient condition for confinement. Our results for the Schwinger function of the
chiral limit quark propagator are shown in the right diagram of fig.~\ref{fig:coupling}.
Shown is the absolute value of the Schwinger function on a logarithmic scale. The
thick black curve denotes our numerical result for the Fourier transform (\ref{FT}).
Clearly visible is a cusp in the Schwinger function at about 
$t \approx 5$ GeV$^{-1} \approx 1$ fm. Certainly, this is a typical scale where quark 
confinement is expected to occur. 

A possible analytical structure for such a positivity violating quark propagator
has been suggested in \cite{Stingl:1996nk}. It reads
\beq
\sigma(t)\; =\; |\, b_0 \exp(-b_1 t) \cos(b_2 t+b_3)\,| \quad , \label{cc}
\eeq
which corresponds to a pair of complex conjugate poles of the propagator 
in the timelike momentum plane. These poles correspond to a \lq quark mass' given by
$m = b_1 \pm i b_2$. In our case we can fit this form either to the small time behavior of
the numerical result or to the region between the first two cusps, as shown in the
diagram. The results are then $m = 305(5) \pm i \, 453(10)$ MeV for \lq Fit A' and
$m = 351(5) \pm i \, 374(10)$ MeV for \lq Fit B'. Here the errors reflect variations
in the bounds of the fitted regions.\footnote{This result demands a comment. 
In \cite{Alkofer:2003jj} two of us showed that the quark propagator is positive 
definite if an ansatz for the quark-gluon vertex is used which contains an Abelian 
part, satisfying the Abelian Ward-Takahashi identity, and an overall non-Abelian 
dressing function multiplying all tensors uniformly. The resulting DSE for the
quark propagator looks similar in structure to the corresponding one for the
fermion propagator in QED. Consequently one obtains an analytical structure of
a singularity at the fermion mass with a branch cut attached, i.e. exactly the
structure expected for a physical electron. It was then left open in  
ref.~\cite{Alkofer:2003jj}  whether the exact quark-gluon interaction of QCD has 
such a structure, leading to a positive definite quark confined by another 
mechanism, or whether the quark-gluon interaction is sufficiently different
from this ansatz so as to allow for positivity violations. Our results here
point in exactly this direction. However, since the relative strength of the
vertex components are somewhat truncation dependent further research is necessary
to confirm this result.}

%%%%%%%%%%%%%%%%%%%%%%%%%%%%%%%%%%%%%%%%%%%%%%%%%%%%%%%%%%%%%%%%%%%%%%%%%%%
\subsection{Mass dependence of the coupled system \label{massstudy}}
%%%%%%%%%%%%%%%%%%%%%%%%%%%%%%%%%%%%%%%%%%%%%%%%%%%%%%%%%%%%%%%%%%%%%%%%%%%

We now investigate the dependence of the quark-gluon vertex on the 
current
quark mass. To compare the results for different quark masses in a meaningful way
we match the input and output scales of the vector $\lambda_1$ and scalar
$\lambda_3$ components of the vertex in a given momentum configuration; 
since the results are qualitatively similar for different kinematical 
slices,
they are most easily presented using the asymmetric momentum slice 
(c.f. subsection \ref{kin}). This choice
is not unique, since we saw in a previous subsection that the vertex
is not uniform with the angle $\hat{p}_1 \cdot \hat{p}_2$ in the infrared, 
whereas our input fits are. However, we checked that different choices would 
not affect the qualitative statements of this subsection, so the asymmetric
slice can be chosen without loss of generality.
 
The input parameters in our calculation for the fits (\ref{vertexfit}) and 
(\ref{3gfit}) and the current quark mass are given in table \ref{param}.
\begin{table}[h]
\begin{tabular}{ c c|c|c|c|c|c|c }
   &            & $d_1$   &  $d_2$   &  $d_3$ &  $n_1$ & $n_2$     & m(170 GeV$^2$)	\\\hline\hline
 u & $\lambda_1$& 3.51 & 0.5   &  -  &  -  & -      &  0.003		\\
   & $\lambda_3$& 7.20 & 500   & 1.2 &  1  & -1.25  & 			\\\hline
 s & $\lambda_1$& 3.22 & 0.5   &  -  &  -  & -      &  0.050		\\
   & $\lambda_3$& 6.65 &  90   & 1.8 &  1  & -1.25  &			\\\hline 
 c & $\lambda_1$& 1.88 & 0.5   &  -  &  -  & -      &  1.200		\\
   & $\lambda_3$& 3.90 &  18   & 3.0 &  1  & -1.25  &			\\\hline
 b & $\lambda_1$& 2.50 & 0.5   &  -  &  -  & -      &  4.200		\\
   & $\lambda_3$& 5.20 & 0.006 & 0.4 &  4  & -0.3125&			\\\hline\hline
$\Gamma^{0,3}$& & 3.00 & 0.40  & 1200&     &        &    
\end{tabular}
\caption{Input parameters for the fit functions (\ref{vertexfit}) and 
(\ref{3gfit}) and the current quark mass. The parameters are chosen such that the input
fits to $\lambda_1$ and $\lambda_3$ match the calculated dressing as closely as possible
in the asymmetric momentum slice.\label{param}}
\end{table}
Again the ultraviolet scale of the three-gluon vertex is matched such that the
vertex goes bare at the renormalization point $\mu^2=170 \mbox{GeV}^2$. The infrared 
strength of this vertex is now considerably stronger than in the 
previous sections: 
the vertex now effectively takes over the missing strength from the vertex structures 
which are not back-feed into the loop integral. The remaining parameters in the 
$\lambda_{1,3}$ input fits are then chosen to match their calculated counterparts.
Since we are matching in a particular kinematic slice only, it does not make sense to
seek for perfect matching here; we therefore concentrated on the infrared power law
coefficients and the ultraviolet tails and left some discrepancies at the intermediate 
momenta. These discrepancies are irrelevant for the points we wish to make here.

\begin{figure}[t]
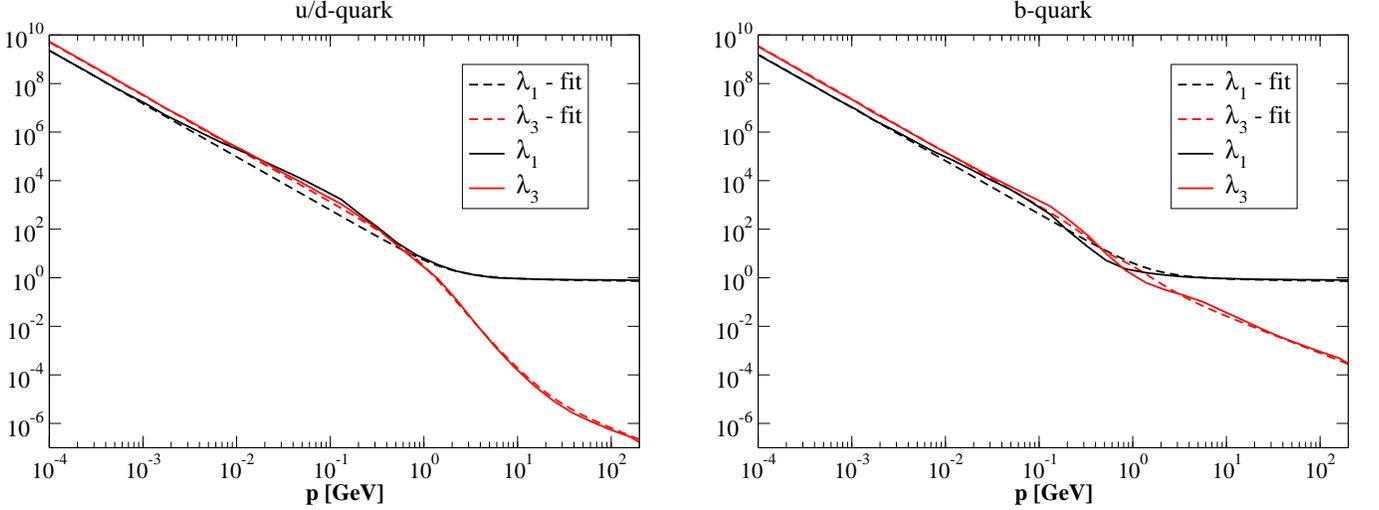

\includegraphics[width=8.5cm]{res.vertexu.eps}\hfill
\includegraphics[width=8.5cm]{res.vertexb.eps}
\caption{\label{fig:vertmass}
Internal representations of the vector and scalar vertex dressings (dashed lines)
compared to the corresponding calculated vertex pieces (full lines). Shown is
the asymmetric momentum slice with $p_3^2=3p_1^2$ and $p_2^2=2p_1^2$.}
\end{figure}

The resulting input and output vertex dressings for the two extreme cases 
(u/d and bottom quark) are shown in fig.~\ref{fig:vertmass}. 
%%%%%%%%%%%%%%%%%%%%%%%%%%%%%%%%%%%%%%%%%%%%%%%%%%%%%%%%%%%%%%%%%%%%%%%%%%%%%%%%%
%%%%%%%%%%%%%%%%%%%%%%%%%%%%%%%%%%%%%%%%%%%%%%%%%%%%%%%%%%%%%%%%%%%%%%%%%%%%%%%%%
%%%%%%%%%%%%%%%%%%%%%%%%%%%%%%%%%%%%%%%%%%%%%%%%%%%%%%%%%%%%%%%%%%%%%%%%%%%%%%%%%
\begin{figure}[t]
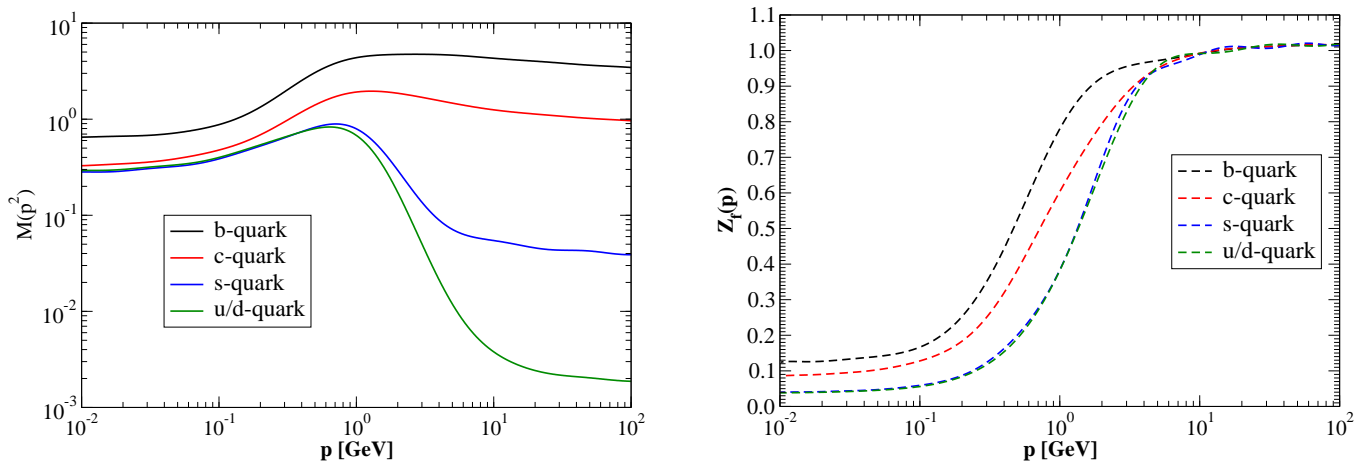

\includegraphics[width=8.5cm]{res.quarkM.eps}\hfill
\includegraphics[width=8.5cm]{res.quarkZ.eps}
\caption{\label{fig:propmass} (Color online)
The fully converged quark propagator from the coupled system of vertex 
and propagator DSEs for the different current quark masses (from bottom 
to top)
$m_{u/d}(\mu^2)=3 \,\mbox{MeV}$, 
$m_{s}(\mu^2)=50 \,\mbox{MeV}$, $m_{c}(\mu^2)=1200 \,\mbox{MeV}$ and 
$m_{b}(\mu^2)=4200 \,\mbox{MeV}$ at the renormalization point 
$\mu^2=170 \,\mbox{GeV}$.}
\end{figure}
%%%%%%%%%%%%%%%%%%%%%%%%%%%%%%%%%%%%%%%%%%%%%%%%%%%%%%%%%%%%%%%%%%%%%%%%%%%%%%%%%
One observes the following:
\begin{itemize}
\item In the infrared the calculated vertex dressings reproduce the
power laws from the internal fits regardless of the quark mass. This observation
nicely confirms our infrared analysis from subsection \ref{irquarks} and underlines
the presence of universal power laws with mass independent exponents in the infrared.
\item We now come back to a point already discussed at the end of section \ref{irbasics}:
The presence of these mass independent infrared anomalous dimensions also shows
that the potentially small scale $\Lambda_{\tt QCD}/M$ is not small 
enough (at least for 
physical quark masses) to disturb the power counting in the infrared. In fact we do not see 
any effects from this scale in our numerical results at all. This again justifies {\it a 
posteriori} our infrared analysis in sections \ref{irquarks} and 
\ref{IRanalysis}.
\item Comparing the coefficients of these power laws, given in table \ref{table}, we see
a much smaller dependence on the quark mass than could have been anticipated
from the naive analysis in subsection \ref{IRmass}. 
\begin{table}[b]
\begin{tabular}{c|c|c|c|c|}\label{stringt}
                                & $u/d$ & $s$  & $c$  & $b$   \\\hline
$m(\mu^2) [MeV]              $  &   3   &  50  & 1200 & 4200  \\\hline 
$M(0)     [MeV]              $  &  270  & 270  &  320 & 650   \\\hline 
$\lambda_1 [GeV^{1/2+\kappa}]$  &  3.95 & 3.60 & 2.00 & 2.73  \\\hline 
$\lambda_3 [GeV^{1/2+\kappa}]$  &  8.70 & 7.97 & 4.44 & 6.08  
\end{tabular}
\caption{The infrared coefficients for the dressing functions $\lambda_{1,3}$ for 
different current quark masses $m(\mu^2)$ at the renormalization point $\mu^2=170 
\,\mbox{GeV}$. Also given is the (approximate) value of the mass function at zero 
momentum.\label{table}}
\end{table}
Assuming some systematic errors one could interpret the results for both, the
vector $\lambda_1$ and the scalar $\lambda_3$ dressing function as constant if not
slightly decreasing with the quark mass. The value of the quark mass function at
zero momentum is only slightly increasing with the current quark mass. Clearly, 
this pattern does not match any of the three possibilities found in our analytical 
analysis with \lq naive' counting in subsection 
\ref{subsec:naive}.
Where does the naive counting of quark masses go wrong? The basic 
assumption of the naive counting is an increase of the infrared mass 
$M=M(p^2 \ll \Lambda_{\tt QCD}^2)$
roughly proportional to the current quark mass. This is, however, not 
the case in our solutions as can be seen from fig.~\ref{fig:propmass}. 
This could be in agreement with subsection \ref{subsec:newcounting}.
Whereas the ultraviolet 
behavior of the quark mass functions indeed is proportional to the current quark 
mass (as strictly demanded from perturbation theory), the quark mass in the
infrared is not. Whereas the current quark mass varies by a factor of $O(10^3)$
from the up- to the bottom-quark mass, the infrared mass $M$ varies only by
roughly a factor of two. Together with the variation of $Z_f(p^2)$ in the infrared
this mass dependence explains the result for the mass (in-)dependence of the 
quark-gluon vertex observed above.
\item We also did calculations for very heavy quark masses in the range of 
$20-40$ GeV. In this region, and presumably also for heavier quarks, the 
naive scaling again sets in, i.e. $M(0) \propto m$ and we see perturbative
counting behavior for the vertex. It therefore seems that the region of 
physical quark masses is clearly distinct.
\end{itemize}

These two last observations are somewhat unexpected, since they deviate from the usual
behavior found in the literature. It remains to be seen whether this effect
persists if the quark-gluon vertex is back-coupled self-consistently into
its own DSE. Nevertheless it is interesting to speculate of the possible 
consequences of this result. This will be done in section 
\ref{sec:stringtension}, where we derive some qualitative aspects of the 
string tension related to an infrared divergent quark-gluon vertex.

%%%%%%%%%%%%%%%%%%%%%%%%%%%%%%%%%%%%%%%%%%%%%%%%%%%%%%%%%%%%%%%%%%%%%%
\subsection{Soft-gluon divergence \label{num:softglue}}
%%%%%%%%%%%%%%%%%%%%%%%%%%%%%%%%%%%%%%%%%%%%%%%%%%%%%%%%%%%%%%%%%%%%%%

In the last subsection we showed that indeed the quark-gluon vertex as 
calculated
from our vertex-DSE has a strong divergence like $(p^2)^{-1/2-\kappa}$ in the
infrared once all external scales go to zero, i.e. 
$p^2 \sim p_1^2 \sim p_2^2 \sim p_3^3 \rightarrow 0$. However, in our infrared 
analysis, subsection \ref{softgluondiv} we also discovered an additional 
selfconsistent infrared singularity when only the external gluon momentum goes 
to zero, i.e $p^2 \sim p_3^2 \rightarrow 0$, but $p_1^2$ and $p_2^2$ are kept 
constant. We now confirm this analytical result also numerically. To this end 
we change the internal dressing functions (\ref{vertexfit}) to the following 
forms:
\beq
\lambda_1(p_1,p_2) = f^{\tt IR}(p_3^2) g_1^{\tt UV}(x) \quad , \quad 
\lambda_3(p_1,p_2) = \frac{1}{\sqrt{(p_1+p_2)^2}} f^{\tt IR}(p_3^2) g_3^{\tt UV}(x)
\eeq
which have additional strength in the soft gluon point, i.e. they go like
$\left((p_3)^2\right)^{-1/2-\kappa}$, where $p_3=p_2-p_1$ is the gluon momentum. Similar
as in eqs. (\ref{vertexfit}) we have $x=p_1^2+p_2^2+p_3^2$ for the ultraviolet part of our fits.
The parameters are unchanged, i.e. given in table \ref{param0}.
For the quark propagator we use the chiral result of the last subsection. 
\begin{figure}[t]
\includegraphics[width=9cm]{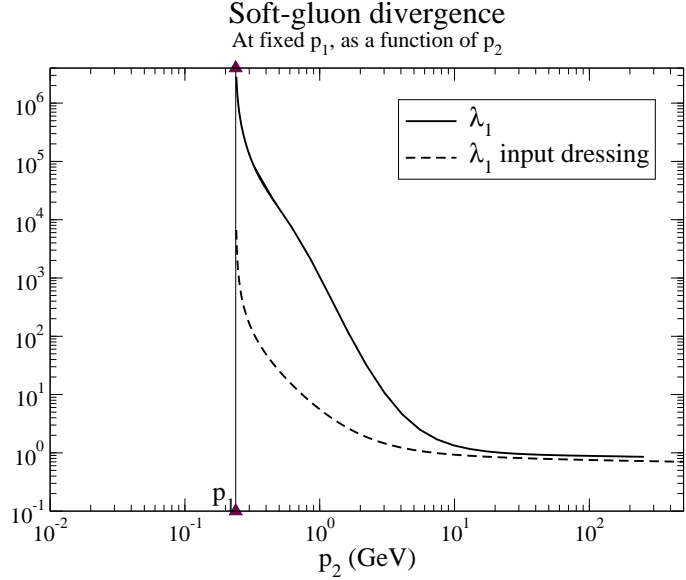}
\caption{\label{fig:softdivinL1}
Input and output of a coupled quark-propagator (totally self-consistent)
to quark-gluon vertex (one loop iteration with internal dressings) showing
that the soft-singularity is indeed a property of the self-consistent vertex
equation with the non-Abelian diagram.}
\end{figure}
In fig.~\ref{fig:softdivinL1} we plot a calculation based on these input forms.
Shown is the output dressing function $\lambda_1$ in the kinematical section with
$p_1 \cdot p_3 = 0$. We fix $p_1 = 0.24$ GeV and vary $p_2^2$. The soft gluon
point $p_3^2=0$ is then reached when $p_1^2$=$p_2^2$. As can be seen from the plot
we clearly encounter a singularity in this limit. From a corresponding log-log
plot we extracted the exponent of the power law and found 
$\left((p_1-p_2)^2\right)^{-1/2-\kappa}$ to good accuracy in agreement with our
infrared analysis.

This observation has several important consequences. First, the limit 
$p_3^2 \rightarrow 0$ is important for the confining potential, as 
detailed 
later in section \ref{sec:stringtension}. Second, the soft-gluon endpoint 
$p^2_3=0$ is a preferred point chosen in several lattice calculations 
\cite{Lin:2005zd,skullerud:2002ge}. However, we just saw that the vertex 
is singular at precisely this kinematical section. This then begs the
question: how can lattice 
calculations obtain results at this point. 

If it is true that a soft-gluon power law develops, then at fixed gluon momentum
$p_3=0$ one should obtain an infinite result no matter what quark momentum
is chosen. Thus, a computation in the soft-gluon kinematic section is bound
to be infinite. On the lattice, however, this infinity is regulated and one
obtains a finite answer for all momenta $p^2=p_1^2=p_2^2$ at $p_3^2=0$.
This finite answer should then vary with the volume and lattice spacing, 
provided there is indeed a soft-gluon singularity in the infinite-volume/continuum 
limit. We therefore propose to study the scaling properties of the vertex
in the soft-gluon kinematical section. There is also one further subtile point 
here: all scaling effects may be wiped out if the vertex is also renormalized
in this section (as partly done in \cite{Lin:2005zd,skullerud:2002ge}).
Therefore it is recommended to change the renormalization point to a non-divergent
section, as for example our totally asymmetric kinematic section 
$p_1^2=1/2p_2^2=1/3p_3^2$. Then the graph of $\lambda_1$ in the
soft-gluon kinematic section is not forced to pass by one, and the
sensitivity to varying the grid is exposed. 

\begin{figure}[h]
\includegraphics[width=8cm,angle=-90]{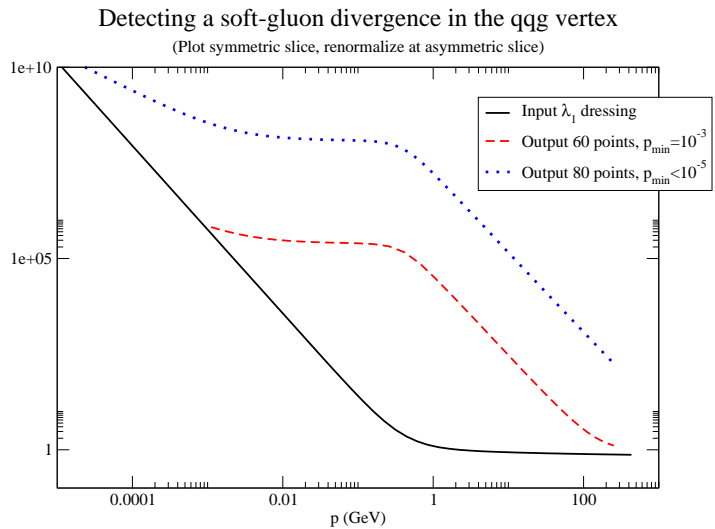}
\caption{\label{fig:checkyourgrid}
Black solid line: input $\lambda_1$ dressing for the internal vertex with a soft-gluon
singularity, plotted here against the momentum of the gluon in a safe
kinematic section. The calculation also employs a $\lambda_3$ dressing, not
shown.\
Red dashed line: the output $\lambda_1$ dressing after one iteration of the
non-Abelian loop with the Curtis-Pennington quark propagator as input. The
grid has 60 points and is cut-off in the infrared at about 1 MeV (therefore
the loop integral too). The plot is in the soft-gluon kinematic section
where the analytical result is divergent. Therefore, the curve is an
artifact of the grid regularization.\
Blue dotted line: same but with an 80-point grid and a lower cutoff of 0.01
MeV. As seen, the grid sensitivity is large when the renormalization point
is not imposed on this kinematic section, forcing the curves to pass by 1.
This spurious effect might be at work in existing lattice calculations.}
\end{figure}

This exercise is carried out in fig.~\ref{fig:checkyourgrid}, where we indeed find a 
large sensitivity of the results to the grid in the presence of our soft-gluon divergence.

%%%%%%%%%%%%%%%%%%%%%%%%%%%%%%%%%%%%%%%%%%%%%%%%%%%%%%%%%%%%%%%%%%%%%%%%%%%
\subsection{Chiral symmetric solution \label{num:chiralsymm}}
%%%%%%%%%%%%%%%%%%%%%%%%%%%%%%%%%%%%%%%%%%%%%%%%%%%%%%%%%%%%%%%%%%%%%%%%%%%
%%%%%%%%%%%%%%%%%%%%%%%%%%%%%%%%%%%%%%%%%%%%%%%%%%%%%%%%%%%%%%%%%%%%%%%%%%%%%%%%%
\begin{figure}[t]
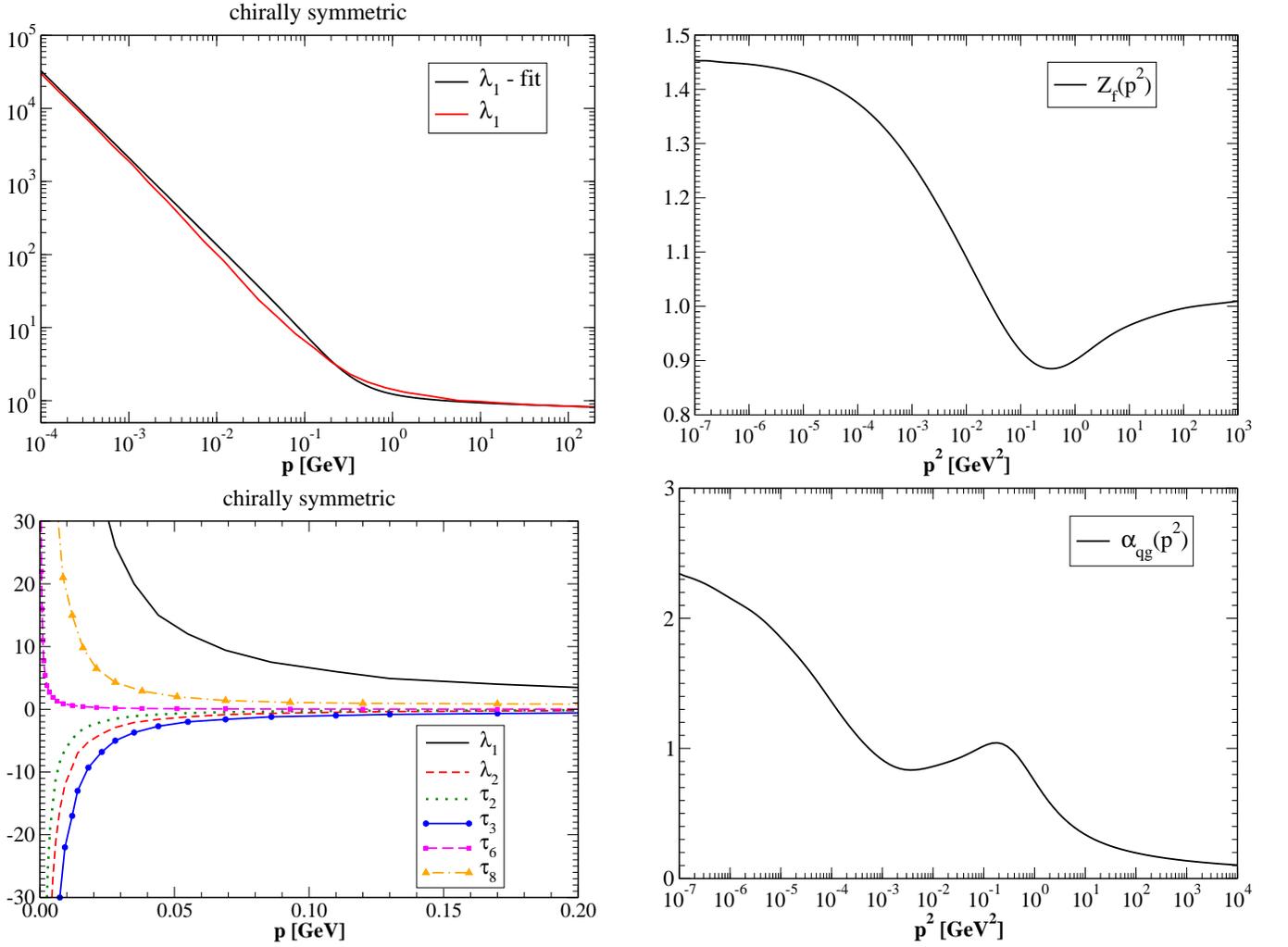

\includegraphics[width=8.5cm]{res.vertexsymm.X1.eps}\hfill
\includegraphics[width=8.5cm]{res.vertexsymm.X2.eps}
\vspace*{5mm}
\includegraphics[width=8.5cm]{res.vertexsymm.X3.eps}\hfill
\includegraphics[width=8.5cm]{res.vertexsymm.X4.eps}
\caption{\label{fig:symmetricresults}
All diagrams show results for the quark-gluon vertex, the quark propagator and the
running coupling in the case where chiral symmetry is forced to be unbroken.
Top left corner: Internal and external (calculated) dressing function $\lambda_1$ in
the asymmetric momentum slice with $p_3^2=3p_1^2$ and $p_2^2=2p_1^2$.
Top right corner: The quark propagator function $Z_f(p^2)$.
Bottom left corner: The six chirally symmetric tensor components of the quark-gluon vertex. 
Bottom right corner: The running coupling for the chirally symmetric 
case. Note that within numerical errors the coupling approaches a fixed point in the
infrared in agreement with our analytical analysis. The bump at mid-momentum is 
model dependent and potentially an artifact due to scale mismatch 
between the gluon 
propagator and the vertex functions. } \end{figure}
%%%%%%%%%%%%%%%%%%%%%%%%%%%%%%%%%%%%%%%%%%%%%%%%%%%%%%%%%%%%%%%%%%%%%%%%%%%%%%%%%

Finally we present numerical solutions for the case of unbroken chiral 
symmetry. This means we solve the coupled system of quark propagator DSE and
vertex-DSE employing chirally symmetric tensor structures only. As detailed in
subsection \ref{chiralsymm} the quark propagator is then given by
\be
S_\chi(p) = \frac{i \pslash}{p^2}Z_f(p^2) .
\ee
In general kinematics the corresponding chirally symmetric quark-gluon vertex
has the tensor structures $L_1$, $L_2$, $T_2$, $T_3$, $T_6$ and $T_8$ given in eqs. (\ref{newdec}) 
with dressing functions $\lambda_{1,2}$ and $\tau_{2,3,6,8}$. All other 
dressing functions vanish identically. This is reproduced by our numerical
solutions.

\begin{table}[b]\label{tab:symm}
\begin{tabular}{ c|c|c|c}
               & $d_1$   &  $d_2$   &  $d_3$       \\\hline\hline
 $\lambda_1$   & 0.10 & 0.20  & 0.02      \\
 $\Gamma^{0,3}$& 1.00 & 0.40  & 1.00             
\end{tabular}
\caption{Input parameters for the fit functions (\ref{vertexfit}) and 
(\ref{3gfit}) in the chiral limit.}
\end{table}
Since $\lambda_3$ is among the vanishing dressing functions we only work
with $\lambda_1$ in the internal quark-gluon vertices. The corresponding 
fit function is given by
\beqa
\lambda_1(p_1,p_2) &=& \left(\frac{x}{d_1+x}\right)^{-\kappa}
                    \left(d_3\frac{d_1}{d_1+x} + d_2  
\log{\left[\frac{x}{d_1}+1\right]}\right)^{-9/44} \nonumber\\
\eeqa
Note that the infrared exponent is changed from $-1/2-\kappa$, as in eq.~(\ref{vertexfit}), to
$-\kappa$ and we have introduced an additional parameter $d_3$. The 
input for the 
three-gluon vertex is unchanged compared to (\ref{3gfit}). Again we matched the input and
output values for $\lambda_1$ in the asymmetric momentum slice $p^2=p_1^2=p_2^2/2=p_3^2/3$.
The resulting values for the parameters are given in table \ref{tab:symm}.
Here the parameters for the three-gluon vertex allowed for the 'natural' 
values $d_1=d_3=1$. These set the scale of the results. Certainly this 
scale is
arbitrary, since we have no physical situation to compare with. 

Our results for the vertex and the quark propagator are displayed in 
fig.~\ref{fig:symmetricresults}.
In the top left corner we show the matching between the internal fit form for
$\lambda_1$ and our calculated result for  the asymmetric momentum slice. The
overall matching is very good, the slight deviations in the mid-momentum regime
do not affect the point we would like to make: in the infrared there is a clear 
matching between the internal and external power laws once again justifying 
our analytical analysis from subsection \ref{chiralsymm}. The
vertex indeed diverges in this case with $(p^2)^{-\kappa}$. The resulting
quark dressing function $Z_f(p^2)$, displayed in the top right corner, shows
a somewhat unexpected behavior. It starts with the expected perturbative running
in the ultraviolet momentum region. Then it bends down towards smaller momenta in
the mid-momentum region similar to the chirally broken case. It then, however,
starts rising again and settles for a finite infrared value larger than one.
Taken at face value such a behavior forbids a positive definite spectral function
\cite{Itzykson:1980rh}, i.e. such a quark propagator does not describe a physical 
quark. Since this behavior may or may not be an artifact of our truncation it 
creates no problem {\it per se}, since the situation we are investigating in this 
subsection has no counterpart in the real world. For our purposes it is 
sufficient to note
that the quark dressing function  $Z_f(p^2)$ indeed goes to a constant in the infrared,
thus confirming our central assumption of the infrared analysis in subsection 
\ref{chiralsymm}. 

In the bottom left corner of fig.~\ref{fig:symmetricresults} we show 
the
dressing of the quark-gluon vertex as a function of the momentum. Similar to the chirally
broken case we find $\lambda_1$, $\tau_3$ and $\tau_8$ among the large vertex 
components, whereas $\lambda_2$, $\tau_2$ and $\tau_6$ are somewhat
smaller. The running coupling in the chirally symmetric case is shown in the
bottom right diagram of fig.~\ref{fig:symmetricresults}. In agreement with
our infrared analysis we find something like a fixed point in the infrared. 
The scaling behavior is not as convincing as in the chirally broken case. To
a large extent this may be to numerical uncertainties, which are much larger
than in the chirally broken setup. We are working to improve this situation in 
the future. Note that the precise value of this fixed point depends on the input 
strength of the three-gluon vertex and is therefore not a prediction of our 
calculation. It is the mere existence of this fixed point that will be 
important for our analysis of the quark-antiquark potential later on.

%%%%%%%%%%%%%%%%%%%%%%%%%%%%%%%%%%%%%%%%%%%%%%%%%%%%%%%%%%%%%%%%%%%%%%
\section{The quark-quark scattering kernel \label{qqkernel}}
%%%%%%%%%%%%%%%%%%%%%%%%%%%%%%%%%%%%%%%%%%%%%%%%%%%%%%%%%%%%%%%%%%%%%%

In the previous two sections we found an infrared divergent, selfconsistent
solution for the quark gluon vertex using both analytical and numerical
methods. We now wish to investigate the consequences of such a behavior
for the quark-quark scattering kernel, i.e. the Green's function that
should contain the confining quark potential in the heavy quark limit of 
quenched QCD. We will first briefly discuss the relevance of chiral 
Ward-identities, then discuss the properties of four-quark Green's function
in a skeleton expansion (cf. ref.~\cite{Alkofer:2006gz}) and finally show 
that the results derived from this expansion can also be directly 
justified from the Dyson-Schwinger equation of the four-quark function. 

%%%%%%%%%%%%%%%%%%%%%%%%%%%%%%%%%%%%%%%%%%%%%%%%%%%%%%%%%%%%%%%%%%%%%%
\subsection{Quark-quark scattering kernel and chiral Ward identities}
%%%%%%%%%%%%%%%%%%%%%%%%%%%%%%%%%%%%%%%%%%%%%%%%%%%%%%%%%%%%%%%%%%%%%

In the heavy quark limit chiral symmetry breaking is dominated by the
effects of the large bare quark mass. Thus chiral Ward Identities, 
associated with effects from the dynamical breaking of chiral symmetry, 
play only a minor role. Certainly, the opposite is true when quarks with
small masses are considered. Then, the axial vector Ward-Takahashi
identity enforces an intimate relation  between the scattering kernel 
and the gap equation. 

A systematic prescription to construct a scattering kernel that is 
consistent with a given quark gluon interaction has been given by 
Munczek \cite{Munczek:1994zz}. 
Beyond-rainbow-ladder approaches have been demonstrated, for example in
\cite{Bender:1996bb}, with semiperturbative-type vertex constructions.
The basic idea is to take functional 
derivatives of the quark self-energy with respect to the quark propagator. 
In appendix \ref{app:scattering} we detail a truncation scheme of the 
coupled system of Dyson-Schwinger equations for the quark-gluon vertex,
the quark-gluon scattering kernel and the quark propagator DSE that is
amenable to Munczek's prescription. We then give a formal expression
for a resulting Bethe-Salpeter kernel that is both chirally symmetric and
confining. The construction is formally straightforward but leads to a
complicated system of equations whose explicit treatment is far beyond
the scope of this work. Instead we proceed discussing the heavy quark 
limit of the scattering kernel, where we expect a skeleton expansion to
deliver meaningful results.

%%%%%%%%%%%%%%%%%%%%%%%%%%%%%%%%%%%%%%%%%%%%%%%%%%%%%%%%%%%%%%%%%%%%%%
\subsection{Quark-quark scattering kernel in a skeleton expansion \label{kernel-skel}}
%%%%%%%%%%%%%%%%%%%%%%%%%%%%%%%%%%%%%%%%%%%%%%%%%%%%%%%%%%%%%%%%%%%%%

\begin{figure}[t]
\includegraphics[width=11cm]{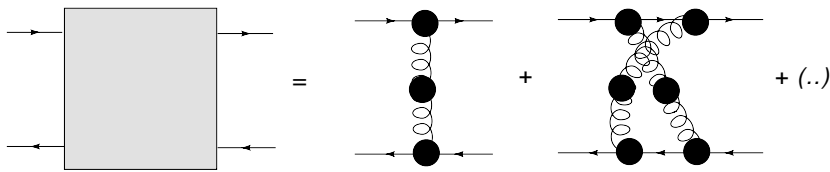}
\caption{The four-quark 1PI Green's function and the first terms of its
skeleton expansion}
\label{qq-1PI}
\end{figure}

We investigate the infrared behavior of the full quark-quark four-point 
function in the heavy quark limit and in the presence of only one
dynamical external scale (the distance between a quark and an antiquark). 
This function together with the first two terms of a skeleton expansion
are given in fig.~\ref{qq-1PI}. As in subsection \ref{IRanalysis}
we examine the infrared behavior of these graphs and perform a scaling 
analysis in terms of infrared anomalous dimensions.

The first term of fig.~\ref{qq-1PI} is nonperturbative one-gluon exchange,
i.e. one-gluon exchange with a dressed propagator and two dressed quark-gluon
vertices. Since we are working in the heavy quark limit with only one
small external scale, i.e. the gluon momentum, we are probing the soft 
singularity of the quark gluon vertex $(k^2)^{-1/2-\kappa}$. We have seen in
subsection \ref{softgluondiv} that this singularity appears in all tensor
structures of the vertex. Thus we do not need to distinguish between vector
and scalar components here. 

Together with the gluon propagator scaling $(k^2)^{2\kappa-1}$ we thus arrive 
at a total scaling 
\be
{\rm term1} \sim (k^2)^{-2}
\ee
of the first term in the skeleton expansion of the four-quark kernel.
Now observe how adding the second rung modifies the first term. For the quark
propagator we have $S = \frac{i \pslash Z_f}{M^2} + \frac{Z_f}{M} \rightarrow \frac{Z_f}{M}$
in the infrared which leads to 
\beq
term2 \sim term 1 \times \left(Z_f^2 \, \frac{(k^2)^{2+2\kappa-1+2(-1/2-\kappa)}}{M^2} + \cdots
\right) 
\sim \frac{Z_f^2}{M^2} (k^2)^{-2}
\eeq
for the second term. Here the dots represent subleading contributions. The first factor of
$(k^2)^2$ comes from the four-momentum integral, then we have $(k^2)^{2\kappa-1}$ 
from the extra dressed gluon propagator and two scaling factors $(k^2)^{-1/2-\kappa}$ 
from the quark-gluon vertices. We thus obtain the same scaling behavior of the second
term than for the first term. With the methods of section \ref{sec:self} it can be shown 
that this behavior persists for all other terms of the skeleton expansion. 

As a result we obtain for the four-quark kernel the power law $H(k^2) \sim (k^2)^{-2}$.
This is in perfect agreement with the $1/k^4$ behavior in the old infrared-slavery scenario.
The crucial difference, however, is that in the former scenario the vertices have been assumed 
constant in the infrared, whereas the gluon was believed to be singular as $1/k^4$. Here we have an
infrared vanishing gluon compensated for by the strength of two infrared divergent quark-gluon
vertices. 

Then, the well-known relation
\be
V({\bf r}) = \frac{1}{(2\pi)^3} \int d^3k \,H(k^0=0,{\bf k})\, e^{i {\bf k r}} 
= \frac{1}{(2\pi)^3} \int d^3k \,\,\frac{e^{i {\bf k r}}}{{\bf k}^4}   
\ \ \sim \ \ |{\bf r} |
\ee
between the static four-quark function $H(k^0=0,{\bf k})$ and the quark
potential $V({\bf r})$ therefore gives a linearly rising potential by naive 
dimensional arguments. We therefore find confinement in the chirally broken
phase of quenched QCD.
A more refined treatment, as described in \cite{Gromes:1981cb}, 
leads to the same result. By construction already the first term in the skeleton 
expansion, {\it i.e.\/} nonperturbative one-gluon exchange displayed in 
fig.~\ref{qq-1PI}, generates this result. Since the following terms in 
the 
expansion are equally enhanced in the infrared, the string tension will be 
built up by summing over an infinite number of diagrams. We will come back
to this point below.

It is instructive to also have a look at the chirally symmetric case. 
The corresponding infrared solutions for the quark propagator and quark-gluon
vertex have been derived in subsection \ref{chiralsymm} and confirmed
numerically in subsection \ref{num:chiralsymm}. They correspond to
a vertex scaling of $(k^2)^{-\kappa}$ and a gluon propagator which still goes
like $(k^2)^{2\kappa-1}$. From the first term of the skeleton expansion
for the quark-quark scattering kernel one obtains $H(k^2) \sim 1/k^2$.
If one would derive a potential for such an interaction one finds  
\be
V({\bf r}) = \frac{1}{(2\pi)^3} \int d^3 pH(k^0=0,{\bf k}) e^{i {\bf k r}}  
= \frac{1}{(2\pi)^3} \int d^3k \,\,\frac{e^{i {\bf k r}}}{{\bf k}^2}   
\ \ \sim \ \ \frac{1}{|{\bf r}|}\,, 
\ee
i.e. the well-known Coulomb potential. (Note, however, that the notion of
a potential in chirally symmetric QCD is certainly not appropriate.) 
As we have seen in subsection \ref{coupling}, also the resulting running 
coupling from the  quark-gluon vertex is no longer diverging but goes to 
a fixed point in the infrared similar to the couplings from the Yang-Mills 
vertices. In a sense, the restoration of chiral symmetry is therefore directly 
linked with the disappearance of infrared slavery. This is one of the main 
results of this work.

%%%%%%%%%%%%%%%%%%%%%%%%%%%%%%%%%%%%%%%%%%%%%%%%%%%%%%%%%%%%%%%%%%%%%%
\subsection{The four-quark Dyson-Schwinger equation \label{four-quark-vertex}}
%%%%%%%%%%%%%%%%%%%%%%%%%%%%%%%%%%%%%%%%%%%%%%%%%%%%%%%%%%%%%%%%%%%%%

\begin{figure}[t]
\centerline{\epsfig{file=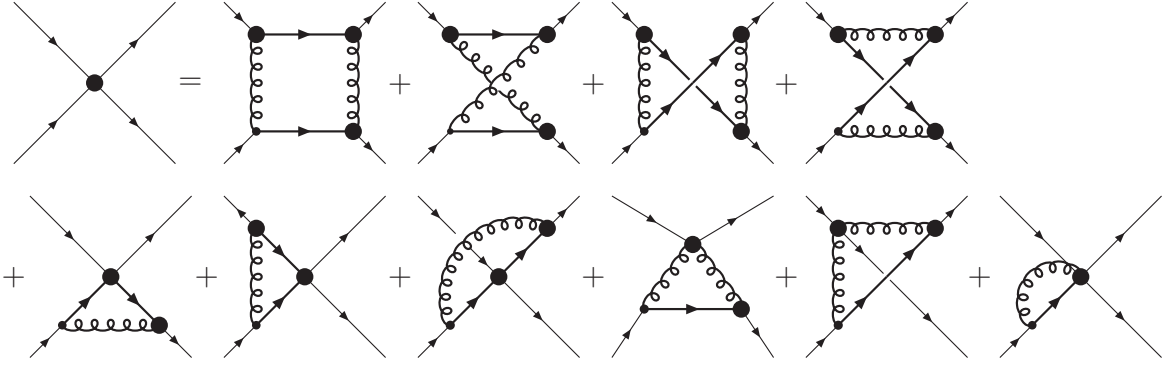,width=16cm}}
\caption{The DSE for the four-quark 1PI Green's function. \label{four-quark-DSE}}
\end{figure}

We now proceed by justifying the above results directly from the 
Dyson-Schwinger equation for the four-quark 1PI Green's function, which
is shown in fig.~\ref{four-quark-DSE}. This equation describes the full 
amplitude that includes possible bound states, scattering as well as 
off-shell effects. In contrast to the skeleton expansion fig.~\ref{qq-1PI},
the diagrams on the right hand side of fig.~\ref{four-quark-DSE} all
involve a bare quark-gluon vertex. Since this vertex is infrared divergent
it is therefore not obvious how both equations can give a similar infrared
behavior for the four-quark function. Indeed, when considering only the 
graphs in the first line of fig.~\ref{four-quark-DSE} an infrared power 
counting similar to that in section \ref{sec:self} yields scaling laws 
\begin{equation}
\left( p^2 \right)^2 \left( \left( p^2 \right)^{-\kappa-1/2} \right)^3 
\left( \left( p^2 \right)^{2\kappa -1} \right)^2= \left( p^2 \right)^{-3/2+\kappa} \; , 
\label{reduced-scaling}
\end{equation}
independently of the quark kinematics.
Here, the second term arises from the three dressed quark-gluon vertices and 
the third one from the two gluon propagators, whereas the quark propagators 
proved to be scale invariant. These diagrams therefore cannot be responsible
to generate a confining quark-antiquark potential.

\begin{figure}[t]
\centerline{\epsfig{file=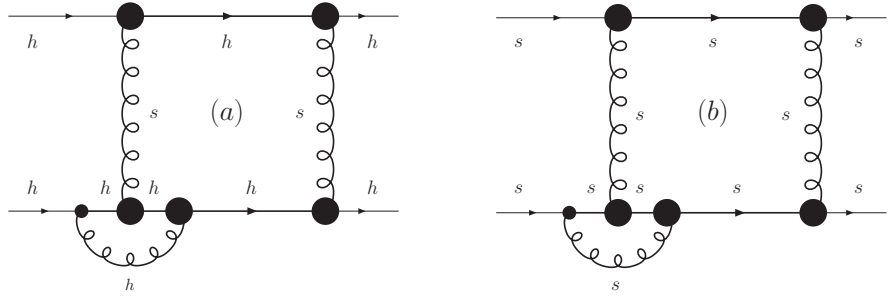,width=12cm}}
\caption{Contribution to the skeleton expansion of the last term in the 
four-quark DSE fig. \ref{four-quark-DSE}. Soft and hard momenta are 
denoted by an $s$ respectively $h$. \label{four-quark-graph}}
\end{figure}

However, these diagrams are not the leading ones in the infrared.
To see this, consider the last graph in the second line of the DSE, 
fig.~\ref{four-quark-DSE}. A skeleton expansion 
of this graph involves, among others, the two-loop graphs shown in 
fig.~\ref{four-quark-graph}. 
In complete analogy to the analysis of the second ('t-channel') version 
of the DSE discussed in appendix \ref{app:secondDSE} this two-loop 
integral has a kinematic region, shown on the left side (a) of 
fig.~\ref{four-quark-graph}, where the vertex correction given by 
the 'small' loop is dominated by hard loop momenta of the order of the 
quark mass and not by the infrared regime. This graph (a)
represents the proper kinematics to study the confining properties 
of bound states. Here the momenta of the massive quarks in a meson are 
finite and do not scale to zero. However, the exchanged gluon momentum would 
become soft when the two quarks are sufficiently spatially separated. In 
this infrared limit the counting for the whole graph involves four dressed 
quark-gluon vertices in the soft-gluon limit $p\to 0$ where these vertices
are divergent according to our analysis in section \ref{softgluondiv}. 
This then yields
\begin{equation}
\left( p^2 \right)^2 \left( \left( p^2 \right)^{-\kappa-1/2} \right)^4 
\left( \left( p^2 \right)^{2\kappa -1} \right)^2= \left( p^2 \right)^{-2} \; , 
\label{full-scaling}
\end{equation}
i.e. a strong singularity in agreement with the one found in the skeleton
expansion. As discussed above, this singularity has the potential to 
permanently confine quarks. 

Now consider the right graph (b) were the quark momenta are also soft and 
scale to zero. In Minkowski space this would correspond to the case of 
light quarks close to the on-shell limit, i.e. in some sense the opposite 
case to that where the quenched approximation we consider here is 
justified. Nevertheless let us study the power counting in this case. 
Naively the right graph shown in fig.~\ref{four-quark-graph} seems to 
yield an equally strong divergence than the preceding one. However, 
there is no large quark mass here, so that the integral is dominated 
by small modes. Consequently all dressing functions are probed in 
their scaling region and it is the uniform limit of the quark-gluon 
vertices that is relevant here. The resulting power law is similar to 
the one of eq.~(\ref{reduced-scaling}), i.e. not sufficiently divergent
to trigger a linear rising potential. Comparing to the analysis of diagram
(a) above, we see that (i) it is not the uniform infrared limit of the 
quark-gluon vertex that is responsible for quark confinement but the 
soft-gluon limit and (ii) it is impossible to confine light quarks close 
to the on-shell limit with this mechanism, whereas heavy quarks are confined. 

Unfortunately the four-quark DSE requires two-loop graphs and is far 
too involved to be useful for any explicit analysis of heavy bound states. 
Therefore in practical calculations one has to resort to the skeleton
expansion given in the last subsection. Nevertheless, we wish to
emphasize again that the strong interaction between heavy quarks seen
in this expansion is indeed also a feature of the underlying functional 
equation.

%%%%%%%%%%%%%%%%%%%%%%%%%%%%%%%%%%%%%%%%%%%%%%%%%%%%%%%%%%%%%%%%%%%
\section{Theoretical aspects of confinement \label{sec:conf}}
%%%%%%%%%%%%%%%%%%%%%%%%%%%%%%%%%%%%%%%%%%%%%%%%%%%%%%%%%%%%%%%%%%
%%%%%%%%%%%%%%%%%%%%%%%%%%%%%%%%%%%%%%%%%%%%%%%%%%%
\subsection{Absence of long range forces}
%%%%%%%%%%%%%%%%%%%%%%%%%%%%%%%%%%%%%%%%%%%%%%%%%%%

It is sometimes argued that a strong, confining, quark-quark scattering
kernel based on non-perturbative one-gluon exchange is not possible because it
gives rise to Van der Waals forces between color singlet hadrons. Confinement,
however, implies the absence of these forces. In nature this corresponds 
to nuclear forces described by an exponentially suppressed potential of the form 
\begin{equation}
V_{11}(R) \propto e^{-m_\pi R}\,,
\end{equation}
at medium to large distances. Prominent long range forces also are not observed  
in quenched lattice simulations \cite{Okiharu:2004ve}. We therefore have to 
address the question whether the strong quark-quark scattering kernel found in
this work necessarily implies power-law suppressed $R^{-\alpha_s}$ Van-der-Waals 
forces between hadrons. As will be argued in the following, this is not the case.

Let us begin with simple examples, building up to the more general case. First note
that a single gluon exchange between color singlet hadrons is forbidden by 
color algebra. Next consider a two-gluon exchange between the two singlets. A 
corresponding diagram is e.g. given by 
\begin{equation}
  \begin{picture}(224,100) (-40,-31)
    \SetScale{0.4}
    \SetWidth{2}
    \ArrowLine(30,131)(192,131)
    \ArrowLine(29,100)(191,100)
    \Vertex(131,133){5}
    \Vertex(86,101){5}
    \ArrowLine(31,-50)(193,-50)
    \ArrowLine(29,-22)(191,-22)
    \Vertex(82,-22){5}
    \Vertex(129,-21){5}
    \Gluon(86,102)(85,52){5}{2.14}
    \Gluon(82,27)(81,-23){5}{2.14}
    \Gluon(127,28)(126,-22){5}{2.14}
    \Gluon(130,56)(131,133){5}{4.07}
    \GBox(76,26)(151,55){0.882}
  \end{picture}\label{singletexchange1}
\end{equation}
where the two gluons combine to a color singlet exchange. Such diagrams are 
nonvanishing and indeed expected to drive pomeron exchange 
\cite{LlanesEstrada:2000jw}. However, they could only lead to Van der Waals 
forces in the absence of a mass gap, implying a massless glueball in 
contradiction with lattice gauge theory. Consequently such diagrams should
give rise to a Yukawa potential $V_{11} \sim e^{-M_G R}$ in the static limit,
which is suppressed at long distances $R$ by the glueball mass $M_G$. 

One can naturally employ a strong one-gluon exchange mechanism if
simultaneously quarks are exchanged to balance color, namely
\begin{equation}
  \begin{picture}(318,100) (-30,-30)
    \SetWidth{2}
    \SetScale{0.4}
    \ArrowLine(46,110)(149,111)
    \ArrowLine(149,111)(240,-11)
    \ArrowLine(240,-11)(363,-9)
    \ArrowLine(61,-9)(164,-8)
    \ArrowLine(236,111)(359,113)
    \ArrowLine(164,-6)(237,112)
    \Gluon(149,111)(152,-7){5}{7}
    \ArrowLine(60,-39)(363,-38)
    \ArrowLine(45,143)(348,144)
    \Vertex(151,-7){5.83}
    \Vertex(147,112){5.83}
    \GOval(51,-20)(28,-9)(0){0.882}
    \GOval(371,-22)(28,-9)(0){0.882}
    \GOval(356,130)(28,-9)(0){0.882}
    \GOval(42,131)(28,-9)(0){0.882}
  \end{picture}\label{singletexchange2}
\end{equation} 
where we have explicitly shown the Bethe-Salpeter amplitudes coupling quarks
to the physical hadron singlets. If we evaluate this diagram in the center of 
mass frame the two quark and gluon vertices and all propagators are evaluated 
in the small momentum scaling region. As a consequence, the infrared counting 
would give a power of $(p^2)^{-2}$ from the exchange kernel, which is
countered by powers of $(p^2)^{4+4BS}$ from the two momentum integrations and the
Bethe-Salpeter amplitudes. However, these amplitudes are finite in 
the infrared ($BS=0$), since the Bethe-Salpeter equation 
\begin{center}
  \begin{picture}(285,50) (-30,-35)
    \SetWidth{2}
    \SetScale{0.6}
    \SetColor{Black}
    \GOval(29,-18)(34,17)(0){0.882}
    \ArrowLine(47,1)(117,2)
    \ArrowLine(44,-34)(114,-33)
    \put(90,-10){\large{$\mathbf =$}}
    \GOval(210,-12)(34,17)(0){0.882}
    \ArrowLine(227,3)(297,4)
    \ArrowLine(225,-33)(295,-32)
    \Gluon(247,5)(272,-30){5}{3}
    \Vertex(269,-29){6.}
    \Vertex(248,3){6.}
  \end{picture}
\end{center}
displays a cancellation of the type $(p^2)^{2-2}$ between powers of momenta
from the enhanced quark-quark scattering kernel and the internal momentum 
integration. Finite Bethe-Salpeter amplitudes are also required from physical
considerations within the context of core nucleon-nucleon repulsion 
\cite{Ribeiro:1978gx}. We therefore conclude that also the amplitude 
(\ref{singletexchange2}) does not lead to colored Van-der-Waals forces.

This above analysis is quite general. For the purpose of analyzing long-range forces
we can limit ourselves to color singlet exchanges as the only non-vanishing
amplitudes. However, in these diagrams the absence of long-range residual forces 
can be guaranteed by the sufficient condition that
\emph{all color singlet, full scattering kernels in the theory are gapped,
that is, they have no pole with zero mass.}
This condition is natural in that no massless particle is present in the 
experimental spectrum. Of course its validity remains to be shown in full 
Quantum Chromodynamics. Our point here is only that enhanced 
one-gluon exchange 
potentials do not necessarily generate Van der Waals forces between 
singlet 
hadrons. The detailed behavior of the color singlet kernels that appear in
diagrams like (\ref{singletexchange1}) and (\ref{singletexchange2}) remain to 
be investigated in future work.

Finally, it is worth examining the original Hamiltonian argument for long ranged
colored forces given in \cite{Gavela:1979zu}. One divides the Hilbert space 
into two sectors, $P$ and $Q$, where $P$ projects over states in which two 
interacting quark and gluon clusters are in a color singlet configuration and 
$Q$ is the part in which the clusters are in non-singlet color configurations 
(the total state still being a singlet). Then the Hamiltonian can be exactly 
reduced to the color-singlet $P$ sector as
\begin{equation}
P H_{\rm eff}(E)P = PHP + PHQ\ \frac{1}{E-QHQ} \ QHP \ .
\end{equation}
The presence of an infrared enhanced confining potential $\sigma R$ in the
unconfined-color sector $QHQ$ appears to yield a power-suppressed $PH_{\rm
eff} P\propto 1/R$ Van der Waals potential in the singlet-singlet 
(confined) sector.
However, the correct procedure to obtain non-relativistic inter-hadron 
interactions from the Landau gauge Green's functions is to perform a 
non-relativistic reduction not of the quark-quark but of the hadron-hadron 
scattering kernel. Since the kernel is gapped in the infrared, such a procedure 
would lead to Yukawa interactions
$$
\frac{1}{k^2+M^2} \to \frac{1}{{\bf k}^2+M^2} \to e^{-MR} \ .
$$
It is certainly not correct to perform a non-relativistic reduction of a 
confining quark-quark scattering kernel and then attempt to construct a 
hadron-hadron interaction Hamiltonian from it. Thus the formal arguments
of Gavela \emph{et al.} do not apply to our approach.

%%%%%%%%%%%%%%%%%%%%%%%%%%%%%%%%%%%%%%%%%%%%%%%%%%%
\subsection{N-ality dependence of the string tension \label{stringbreaking}}
%%%%%%%%%%%%%%%%%%%%%%%%%%%%%%%%%%%%%%%%%%%%%%%%%%%
There are good arguments from both continuum and lattice QCD that 
the string tension $\sigma$ between static (anti-) quarks follows scaling laws 
that are dependent on the distance $R$ between the quark and antiquark
and the representation of the gauge group. At intermediate distances $R$
the string tension between quarks in the representation $r$ follow the 
so called Casimir scaling, i.e. $\sigma_r = C_r/C_F \sigma_F$, where 
the subscript $F$ denotes the fundamental representation and $C_r$,$C_F$
the respective Casimir operators. At asymptotically large distances, however,
the scaling law changes to $\sigma_r = f(k) \sigma_F$, where $f(k)$ is
a function of the N-ality $k$ of the representation $r$
(see \cite{Greensite:2003bk,Alkofer:2006fu} for reviews). 

Casimir scaling is exactly derivable from one-gluon-exchange type forces, thus
the first term of the skeleton expansion in fig.~\ref{qq-1PI} agrees
with this behavior. All other diagrams involve quark propagators. From our
discussions in sections \ref{IRmass} and \ref{massstudy} we recall that
for really heavy quarks  we find the
perturbative mass dependence of the quark-gluon vertex and the propagators
implying that all these diagrams are suppressed by powers of the quark mass 
$M$. Thus approaching the static limit we recover Casimir scaling in our 
quark-antiquark potential, as we should \cite{Bali:2000un}.

Next we discuss the N-ality dependence of the asymptotic string tension.
The N-ality of a given representation is the number of boxes in its Young 
tableau, modulo N. For SU(3) one has $k=1$ for the fundamental representation 
(one box for the quark) and $k=0$ for the adjoint representation (3(mod 3)=0) 
for the gluon or octet fermion. 
In string-type models inspired in lattice gauge theory such as the Flux tube
model \cite{Isgur:1984bm,Waidelich:2000,Buisseret:2007hf}, the string is postulated 
to snap if the hadron may decay to two color-singlet clusters. 
The energy stored in the flux tube has to exceed the threshold for the
hadron-hadron combined energy, and thus this happens only at large distances. 
It then follows from color algebra that the string between two fundamental 
sources in a triplet/anti-triplet representation can never snap
in quenched approximation, since sea quarks are absent. Only gluons in the
adjoint representation can be produced from the vacuum and these cannot combine
with fundamental quarks to yield a color singlet hadron. 
However the octet-octet string can break, via two-gluon production, one of
them combining with each source to make a color singlet. Thus, it is
apparent that the string tension indeed scales with the N-ality of the
representation. 

It is instructive to discuss how this works in the Green's functions approach 
to Landau gauge QCD. Certainly, the quark-quark scattering kernel with one gluon 
exchange cannot be responsible for this type of behavior, since it also confines
fermions in the adjoint representation, leading to Casimir scaling.
However this is only a four-point Green's function and by far not the only
contribution to the computation of the string tension, say, via the Wilson
loop.  The following two graphs represent the first skeleton diagram of the
6-point
Green's function that make N-ality scaling conceivable. 
The left diagram corresponds to
fermions in the adjoint representation, the right diagram in the
fundamental representation. 

\begin{center}
  \begin{picture}(215,80) (60,80)
    \SetWidth{2}
    \SetScale{0.6}
    \Photon(41,257)(176,257){4}{9}
    \Photon(41,148)(176,148){4}{9}
    \Gluon(83,149)(83,255){4}{6}
    \Gluon(84,229)(176,229){4}{6}
    \Gluon(83,180)(176,180){4}{6}
    \Vertex(83,148){7}
    \Vertex(83,181){7}
    \Vertex(83,229){7}
    \Vertex(83,256){7}
    
    \ArrowLine(360,257)(500,257)
    \ArrowLine(360,148)(500,148)
    \Vertex(404,148){7}
    \Vertex(404,181){7}
    \Vertex(404,229){7}
    \Vertex(404,256){7}
    \Gluon(404,149)(404,255){4}{6}
    \Gluon(407,229)(500,229){4}{6}
    \Gluon(406,180)(500,180){4}{6}
    
    \put(10,150){\bf{\large 8}}
    \put(10, 85){\bf{\large 8}}
    \put(110,140){\bf{\large 1}}
    \put(110, 95){\bf{\large 1}}
    
    \put(200,150){\bf{\large 3}}
    \put(200, 85){\bf{\large 3}}
    \put(310,140){\bf{\large $\mathbf{\not 1}$}}
    \put(310, 95){\bf{\large $\mathbf{ \not 1}$}}
  \end{picture}
\end{center}

If we send all scales simultaneously to zero as $P\to 0$, 
both diagrams are strongly infrared enhanced, with the help of our 
counting we establish a power law $(P^2)^{-3-2\kappa}$. However in this
case, all colored particles are simultaneously being separated, and we would
still expect confinement to be at work. If now we separate the two clusters,
keeping a finite momentum between the components of each cluster, most
vertices and propagators have finite momenta running through, and only the
middle gluon rung contributes to an infrared exponent of
$(p^2)^{-1+2\kappa}$, therefore being suppressed.

It is obvious that a full evaluation of the string tension from the Wilson
loop requires the interplay of many such higher Green's functions, with
complicated color structure. The interplay of these to yield the correct
N-ality is a complete other project in itself. However it is obvious from
these comments that demanding N-ality to arise from our quark-quark
scattering kernel alone is not appropriate. It should (as it correctly does)
yield Casimir scaling.

%%%%%%%%%%%%%%%%%%%%%%%%%%%%%%%%%%%%%%%%%%%%%%%%%%%%%%%%%%%%%%%%%
\section{Phenomenological aspects of confinement \label{sec:stringtension}}
%%%%%%%%%%%%%%%%%%%%%%%%%%%%%%%%%%%%%%%%%%%%%%%%%%%%%%%%%%%%%%%%%

%%%%%%%%%%%%%%%%%%%%%%%%%%%%%%%%%%%%%%%%%%%%%%%%%%%%%%%%%%%%%%%%%
\subsection{Kernel non-relativistic reduction \label{sec:nonrel}}
%%%%%%%%%%%%%%%%%%%%%%%%%%%%%%%%%%%%%%%%%%%%%%%%%%%%%%%%%%%%%%%%%

We would now like to examine the limit of our quark-quark scattering kernel
when the quark mass is large and see how close we can get to a Schroedinger 
description. Before we proceed, let us remark that within full QCD a 
description of the quark-antiquark interaction in terms of a time independent 
potential is not entirely adequate 
\cite{Bali:1997am,Leutwyler:1980tn,Voloshin:1978hc}. 
\footnote{One argument is: Within such a description the extension of a heavy 
meson at fixed radial quantum number $n_r$ would become vanishingly small 
in the static limit, i.e. $<r>_{n_r}\to 0$, and see exclusively the Coulomb 
part of the potential. This part being scaleless all Balmer splittings 
$m_{n_{r+1}}-m_{n_r}$ would scale with the quark mass $M$. However, the 
glueball masses $m_G$ would not. So the states neighboring any one $n_r$ heavy 
quarkonium state would not be the corresponding $n_{r\pm 1}$ state,
but instead a state with mass $m_{n_r}+ m_G$. 
We thank Gunnar Bali for this observation.}  
The time-independent Schroedinger equation following from Coulomb-gauge QCD
would be formulated in a Fock space with explicit transverse gluons. However,
considering the strict non-relativistic reduction of the quark-quark scattering 
kernel in Landau gauge, i.e. seeking for a Schroedinger equation for a two-body 
problem, we need to insist in the elimination of all gluonic degrees of freedom. 
This then leads to a strongly energy-dependent kernel, which is equivalent to 
a time-dependent potential in coordinate space.

Nevertheless it is instructive to go through the exercise, since there may be a 
window of quark masses where a potential could be of use (while certainly 
important effects from virtual states with transverse gluons are expected too). 
This allows us to make contact with time-honored phenomenological and lattice 
computations in the static limit.

%%%%%%%%%%%%%%%%%%%%%%%%%%%%%%%%%%%%%%%%%%%%%%%%%%%%%%%%%%%%%%%%%%%%%%%%
\subsection{Salpeter equation \label{salpeter}}
%%%%%%%%%%%%%%%%%%%%%%%%%%%%%%%%%%%%%%%%%%%%%%%%%%%%%%%%%%%%%%%%%%%%%%%%

The main problem in deriving a Salpeter equation for a bound meson state
are energy singularities in the loop momentum $k$ of the Bethe-Salpeter equation.
In QED, ladder and cross-ladder diagrams in the Bethe-Salpeter kernel
conspire to cancel these singularities, as detailed in appendix \ref{GY}.
However, due to factors from the color algebra such cancellations are
unlikely to also occur in QCD. If one insists in maintaining the simplicity 
of the ladder approximation one therefore faces singularities in $k_0$ from 
both the quark propagators and the kernel (gluon propagators and vertices). 
In non-relativistic reductions it has been common practice
\cite{Caswell:1978mt,Bicudo:1989si,LeYaouanc:1984dr,Alkofer:1988tc}, 
to \emph{ignore the $k_0$ dependence of the kernel and Bethe-Salpeter
wavefunctions} and perform the pole integrals by putting only the
intermediate quarks on-shell. The Bethe-Salpeter equation is thus inconsistently 
evaluated with $k_0=0$ for the kernel and $k_0 \neq 0$ for the propagating quarks. 
This procedure is justified empirically by demanding finite support for 
the BS-amplitude when the quarks are off-shell. 

The resulting Salpeter equation has many nice features such as color 
confinement (by sending colored objects to infinite mass outside the 
spectrum) in the presence of infrared enhancement kernel. It also 
respects chiral symmetry and it can be obtained as a collective excitation 
(Random Phase Approximation) in a
Hamiltonian framework \cite{LlanesEstrada:1999uh,LlanesEstrada:2001kr}.

In the following we go through the derivation of the Salpeter equation and 
discuss how it can be used to generate a static potential from our relativistic 
quark-quark interaction kernel. Here we follow Bicudo and Ribeiro 
\cite{Bicudo:1989si}; see also Itzykson and Zuber, \cite{Itzykson:1980rh}, 
for an analogous
derivation without propagator dressing functions), to define directly a
positive (negative) Salpeter function from the Bethe-Salpeter amplitudes 
to emit (absorb) a quark-antiquark pair from the meson, evaluated at zero
energy and integrated over the propagators
\begin{equation}
\phi^+({\bf k}) \equiv \int \frac{dk_0}{2\pi} S_q({\bf k},P_0/2+k_0)
S_{\bar{q}}(-{\bf k},P_0/2-k_0) \phi^+({\bf k},0)
\end{equation}
(and analogously for $\phi^-$).
The Bethe-Salpeter equations then reduce to the coupled Salpeter equations
for the two amplitudes $\phi^\pm$, the first of which becomes, now including
the spin index, 

\begin{eqnarray}
\phi^+({\bf k})_{\lambda_1,\lambda_2} &=& 
\left[\int\frac{dk_0}{2\pi} S_q({\bf k},P_0/2+k_0)
S_{\bar{q}}(-{\bf k},P_0/2-k_0)
\right]\times \\ \nonumber&&
\left( \int\frac{d{\bf k'}}{(2\pi)^3} u_{i}^\dagger({\bf k},\lambda_1)
v_{n}^\dagger(-{\bf k}',\lambda_4)
K_{ijmn}({\bf k}-{\bf k}')v_{j}(-{\bf k},\lambda_2) u_{m}({\bf
k}',\lambda_3) 
\phi^+({\bf k'})_{\lambda_3, \lambda_4}-
\right. \\ \nonumber &&\left.
\int\frac{d{\bf k'}}{(2\pi)^3} u_{i}^\dagger({\bf k},\lambda_1)
u_{n}^\dagger(-{\bf k}',\lambda_3)
K_{ijmn}({\bf k}-{\bf k}')v_{j}(-{\bf k},\lambda_2) v_{m}(-{\bf
k}',\lambda_3)
\phi^-({\bf k'})_{\lambda_3, \lambda_4}
\right)
\end{eqnarray}

where the interaction part can be represented graphically with an instantaneous Bethe-Goldstone
diagram (note fermions and anti-fermions are distinguished, anti-fermion always the
bottom line)
\begin{center}
  \begin{picture}(258,110) (7,-22)
    \SetScale{0.7}
    \SetWidth{1.5}
    \ArrowLine(8,101)(43,101)
    \ArrowLine(8,64)(43,64)
    \ArrowLine(84,64)(119,64)
    \ArrowLine(84,101)(118,101)
    \put(110,55){\large{$\mathbf -$}}
    \GBox(44,55)(85,110){0.882}
    \GOval(127,83)(35,9)(0){0.882}
    \put(85,55){\large{$\mathbf +$}}
    \GBox(224,55)(265,110){0.882}
    \GOval(178,13)(35,9)(0){0.882}
    \ArrowLine(187,64)(221,64)
    \ArrowLine(187,101)(222,101)
    \ArrowLine(265,64)(185,-7)
    \ArrowLine(265,101)(185,33)
    \put(120,5){\large{$\mathbf -$}}
  \end{picture}
\end{center}

The integral over the propagators in the first line yields
$$
\frac{1}{M-E_q-E_{\bar{q}}}
$$
and puts the quarks ``on-shell''. One should note however that in the 
presence
of a strong kernel, $E_q=\sqrt{A^2({\bf q}){\bf q}^2 + B^2({\bf q})}$ is
actually divergent \cite{Bicudo:1989si,Alkofer:2005ug}
(when the same approximations are applied to the
propagator Dyson-Schwinger equation) and these divergences in the
self-energies cancel only for color singlet states. 

To take the strict static limit one neglects $\phi^-<<\phi^+$ 
(this approximation is equivalent to Tamm-Dancoff and breaks chiral symmetry).
Then Fourier transforming to coordinate space by means of
\begin{equation}
V(r)= \int\frac{d{\bf q}}{(2\pi)^3} V(q) e^{i {\bf qr}}
\end{equation}
and
\begin{equation}
E_i(r) = \int\frac{d{\bf q}}{(2\pi)^3} e^{i {\bf qr}} (E_i(q)-E_i(0))
\end{equation}
one finally obtains a Schroedinger-type equation (the infrared cancellation
for color singlets is written explicitly)
\begin{equation} \label{schroedinger}
M \psi(r) = \int d^3{\bf r}' \left[
E_q({\bf r}-{\bf r}') + E_{\bar{q}}({\bf r}-{\bf r}') \right] \psi(r')
+[V(r)-E_q({\bf q}=0)-E_{\bar{q}}({\bf q}=0)]\psi(r) \ .
\end{equation}
The potential $V(r)$ in this equation will be considered further 
in the next subsection. 

The insight one gains from the Salpeter construction so devised is that the
static potential obtained from a Born-approximation is correct (modulo the 
infrared subtraction as needed for bound state equations with self-energies 
included)\footnote{Moreover one learns that in deriving eq. (\ref{schroedinger}) 
a collective mode, namely the back-propagating Salpeter wavefunction, has 
been neglected. Taking this mode into account gives rise to additional
energy-dependent hyperfine-terms proportional to $1/M^2$ in the 
non-relativistic Schroedinger equation.}.
Thus one can test the quark-quark kernel constructed in Landau gauge QCD by 
evaluating the static potential in the Salpeter approximation. 
However, it is difficult to estimate the systematic error of this procedure.
One way to estimate this is to cross-check with the Gross approximation, 
detailed in appendix \ref{Gross}. Relativistic corrections to the potential
of $O(v^2)$ could be investigated by appropriately expanding the spinors
contracting the kernel and employing a Fierz rearrangement. These could also
be evaluated in the Salpeter approach and compared to results from the lattice
\cite{Bali:1997am}. In general, the systematic error of the Salpeter equation 
is controllable in $1/M$ if one can independently establish that the BS wave 
amplitude has limited support in the quark off-shellness\footnote{We thank 
A. Szczepaniak for calling our attention to this point.}.

%%%%%%%%%%%%%%%%%%%%%%%%%%%%%%%%%%%%%%%%%%%%%%%%%%%
\subsection{Salpeter string tension \label{stringmass}}
%%%%%%%%%%%%%%%%%%%%%%%%%%%%%%%%%%%%%%%%%%%%%%%%%%%

We are now using the results of the last subsection to relate
phenomenology with the fundamental building blocks of our quark-quark
scattering kernel, i.e. the quark-gluon vertex and the gluon propagator.
To this end we will take the limit of our kernel with the quark-legs
going on-shell, yielding $q_0=0$ for the exchanged energy (ignoring
possible singularities of the amputated kernel). The resulting
non-relativistic potential is the appropriate one to be used in the context
of the Salpeter (RPA) equation and is known to be about $\sigma=\frac{3}{4}
0.18 \ GeV^2$ \cite{LlanesEstrada:1999uh,Godfrey:1985xj}.

We denote the momenta of the incoming quarks by $p_1$, $k_1$ and the 
outgoing $p_2$, $k_2$. The momentum transferred by the gluon rung is $q$. 
We choose the center of mass frame, where ${\bf p}_1=-{\bf k}_1$. We let $W$
be the total energy and $p_1^0=W/2+E/2$, $k_1^0=W/2-E/2$ the energy of the 
two incoming quarks ($E'$ for the outgoing). From the on-shell conditions
for the quarks one finds that $E=E'=0$ in the center of mass frame. The
gluon momentum is given by $q=(0,{\bf q})$.
\be
  \begin{picture}(300,90)(-60,-30)
    \SetWidth{1.5}
    \SetScale{0.5}
    \ArrowLine(0,80)(200,80)
    \ArrowLine(0,-6)(200,-6)
    \Vertex(85,80){7.07}
    \Vertex(85,-8){7.07}
    \Gluon(85,77)(85,-5){7.5}{4.43}
    \put(-60,-20){\large{${k_1=(\frac{W}{2},-{\bf p_1}})$}}
    \put(-60,50){\large{${p_1=(\frac{W}{2},{\bf p_1}})$}}
    \put(65,-20){\large{${k_2=(\frac{W}{2},-{\bf p_2})}$}}
    \put(65,50){\large{${p_2=(\frac{W}{2},{\bf p_2}})$}}
    \put(55,17){\large{${q=(0,{\bf q}})$}}
\end{picture}\label{one-rung}
\ee
The one-rung skeleton amputated kernel (\ref{one-rung}) then becomes 
\begin{equation}\label{stringamputatedkernel}
K_{IR} = g^2 \, C_F\, \sum_{i,j=1}^{12} \Gamma_i^\mu(p_1,p_2) \lambda_i(M^2,{\bf q}^2) 
 \, \, D_{\mu\nu}(q)\, \, 
\Gamma_j^\nu (k_1,k_2) \lambda_j(M^2,{\bf q}^2)
\end{equation}
where the Casimir $C_F=\frac{N^2-1}{2N}$ stems from the color trace and the quark-gluon
vertex is denoted by $\Gamma_\mu = \sum_{i=1}^{12} \Gamma_i^\mu \lambda_i$.

Next we contract each vertex of the kernel with external spinor
wavefunctions $(u^\dagger
\Gamma^\mu u)/Z_2$ (should the
kernel be part of a loop, the propagator attached to the right vertices
needs to be decomposed as $S= uu^\dagger/(D_+) + vv^\dagger/(D_-)$, and we
keep the positive energy pole). Note that the spinors
need to be multiplied with the inverse of the wavefunction
renormalization constant $Z_2$ in order to make the
construction renormalization-group invariant.
With the intent of obtaining the central potential we average over initial and
sum over final spins.
This means that at each vertex we have
$
\frac{1}{2} \sum_{\lambda_1 \lambda_2} Tr\left (u_{\lambda_2} u_{\lambda_1}^\dagger
\Gamma^\mu \right)/Z_2
$.
In the extreme non-relativistic limit, the $1/M$ lower components vanish.
In terms of Pauli bispinors we are then left with
\begin{eqnarray}
\frac{1}{2} \sum_{\lambda_1 \lambda_2} \left(
\begin{tabular}{c} $\chi_{\lambda_2}$\\ 0 \end{tabular} \right) 
\left( \chi_{\lambda_1}^\dagger \ 0 \right) = 
\frac{1}{2}\left(
\begin{tabular}{cccc} 1& 1& 0 & 0 \\ 1& 1& 0 & 0 \\ 0& 0& 0 & 0\\ 0& 0& 0 &
0
\end{tabular}
\right)
\frac{1}{2} \left( \frac{\mathbb{I}+\gamma_0}{2}\right) \left(
\mathbb{I}+\sigma^{yz} \right) \ .
\end{eqnarray}
Herein, we have made a choice of quantization axis for the spin so that
rotational invariance is no more manifest. If necessary this can be 
avoided by employing a generic vector $\hat{n}$.
Plugging this spinor combination into eq.~(\ref{stringamputatedkernel}) and
tracing each of the two vertices (with twelve tensors each) 
is conveniently carried out with the help of a symbol manipulation 
program such as FORM \cite{Vermaseren:2000nd}.
We furthermore define $p_1 \equiv p$ and perform the Wick-rotation to 
Euclidean space.

Employing momentum conservation, $k_2=k_1+q$, $p_2=p_1-q$, only terms with 
the four \lq \lq longitudinal'' $\lambda_i$ pieces remain, with all
transverse tensors $\tau_i$ contributing to subleading parts of the potential.
For the $\lambda_i$ we may write 
$$
\lambda_i(M^2,M^2,q^2) = A_i (q^2)^{-1/2-\kappa}\,,
$$
using our results from previous sections.
After some algebra we are left with
\begin{eqnarray}
K'_{IR} = \frac{g^2 \, C_F Z }{q^4(Z_2)^2}\, 
\left(A_1^2+2A_1 W( A_2-A_3)+ 4(WA_2-A_3)^2
\left(4({\bf p}\cdot \hat{\bf q})^2
-{\bf p}^2-\frac{W^2}{4}\right) 
-4A_4^2\left( (\hat{\bf q}\times{\bf p})_x^2-({\bf p}^2-p_x^2)
\right)\right)\,.\label{sigma}
\end{eqnarray}
where the factor $Z$ is the coefficient of the power law 
$Z(p^2)=Z \times (p^2)^{2\kappa}$
of the gluon propagator in the infrared.
Note that the total energy $W \sim M$ and $p\propto 1/M$ such that the 
last term is suppressed 
for large quark masses and the frame dependence disappears from the 
central potential.
To extract the central, static potential in the center of mass frame, 
we keep the first three terms and take $W=2M$. 
Comparing (\ref{sigma}) to the Fourier transform of a linear potential 
$\sigma r\to -8\pi\sigma/q^4$, we read off
\begin{equation}
\label{eq:st}
\sigma= \frac{\alpha_s Z}{2(Z_2)^2} \left(
A_1^2 + 4MA_1(2MA_2-A_3)-4M^2(2MA_2-A_3)^2
\right)
\end{equation}
and $\alpha_s=g^2/4\pi$.

This string tension has to match the one extracted from the
quarkonium static potential, $\sigma_{exp}=(0.424GeV)^2$.
$Z_2$ is very close to 1 and $Z\simeq 20 (GeV)^{-4\kappa}$ can be read-
off the solution for the vacuum polarization DSE, with a renormalization
point $\mu\simeq 13GeV$, such that $\alpha_s(\mu)\simeq 0.5$. 
Ignoring $A_2$ and $A_3$, we then have a phenomenological estimate $A_1
\simeq 0.16\ GeV^{1+2\kappa}$ where the energy units are $GeV$ but the
precise mass dimension depends on $\kappa$, and is here fixed from the
gluon propagator $Z$ to $\kappa=0.595$.

Of course, this number is just an order-of-magnitude guess, since the
string tension obtained from the quark-quark 2PI needs to be obtained by
summing a skeleton series as previously discussed in length.
The second term in the skeleton is the crossed-rung, one-loop diagram
that causes a correction to the one-gluon string tension 
\begin{equation}
\sigma\simeq \sigma^{1g}\left(1-
\frac{1}{6}\frac{\alpha_s A_1^2 Z}{M^2} c
\right)
\end{equation}
where $c$ is a pure number containing integration factorials, phase
space and angular factor. We explicitly show the color factor, and $M$
is a quark mass scale showing convergence of the skeleton expansion in
its presence. The corrections to the string tension can be organized as
\begin{equation}
\sigma \simeq \sigma^{1g}\left(
1-c_1 \frac{\sigma^{1g}}{M^2}+\dots
\right)
\end{equation}

To conclude this subsection, we note in passing that in the light of
our calculation the longstanding discussion on whether confinement is 
of a vector $\gamma \gamma$ or of a scalar nature seems oversimplified. 
Each of the quark-gluon vertices in our expansion of the quark-quark 
scattering kernel contains a full set of twelve Dirac tensors with vector,
scalar and tensor structures. After contraction with the gluon propagator 
one obtains a large number of terms and non-trivial cancellations can 
arise. The detailed structure of the resulting quark-anti-quark 
interaction needs to be investigated further and is certainly of great 
interest due to recent improved lattice data \cite{Koma:2007jq}.

%%%%%%%%%%%%%%%%%%%%%%%%%%%%%%%%%%%%%%%%%%%%%%%%%%%
\subsection{Mass dependence of the string tension}
%%%%%%%%%%%%%%%%%%%%%%%%%%%%%%%%%%%%%%%%%%%%%%%%%%%

In this subsection we consider the possible mass dependence on the
string tension from a phenomenological point of view. To this end
it seems sufficient to consider the non-relativistic $q\bar{q}$ 
Schroedinger equation, that contains some of the features exploited 
in more elaborate schemes as for example pNRQCD \cite{Brambilla:2004jw}.
The reduced radial wave function $u(r)$ satisfies 
\begin{equation} \label{schroedingerqq}
\frac{\hbar^2}{2\mu} u''(r) +\left[ E- \lambda r^\nu -
\frac{l(l+1)\hbar^2}{2\mu r^2}\right]u(r) =0
\end{equation}
with a power-law potential $V(r)=\lambda r^\nu$ and 
the reduced quark mass $\mu=M_Q/2$. 
In particular, $\nu=1$ and $\lambda=-\sigma$ corresponds to the linear
potential, $\nu=-1$ and $\lambda=\alpha_s$ to the (attractive) Coulomb
potential, and the limit $\nu\to 0$ can be used to represent a logarithmic
confining potential. The equation can be transformed to a scaleless form
\cite{Quigg:1977xe,Quigg:1997bx} by performing the change of variables
$(r,E) \rightarrow (\rho,\varepsilon)$:
\begin{equation} \label{scalingL}
r= \rho \left(\frac{\hbar^2}{2\mu \arrowvert \lambda \arrowvert}
\right)^{\frac{1}{2+\nu}}\,, \hspace*{1cm} 
E= \varepsilon \left(\frac{\hbar^2}{2\mu \arrowvert \lambda \arrowvert}
\right)^{\frac{-2}{2+\nu}} \frac{\hbar^2}{2\mu} \; .
\end{equation}

Therefore all the mass and parameter scaling of the spectrum and momenta
(inverse length) needs to reside in eq.~(\ref{scalingL}).  
For the Coulomb potential $\nu=-1$ one obtains
\begin{equation}
p\propto \mu \alpha_s\,,\ \ \ E\propto \mu \alpha_s^2 \ ,
\end{equation}
stating that all splittings grow with the quark mass at large $M_Q$ since the
(low-lying) states become smaller and see mostly the short-range Coulomb
potential of pQCD. A linear potential, however, causes splittings slowly decreasing  
with the quark mass and growing with the string tension
\begin{equation} \label{splittingslinear}
p\propto(\mu \sigma)^{1/3}\,,  \ \ E\propto \mu^{-1/3} \sigma^{2/3} \ .
\end{equation}

\begin{figure}[t]
\parbox[t]{9cm}{\vspace*{-3cm}
\includegraphics[width=7cm,angle=-90]{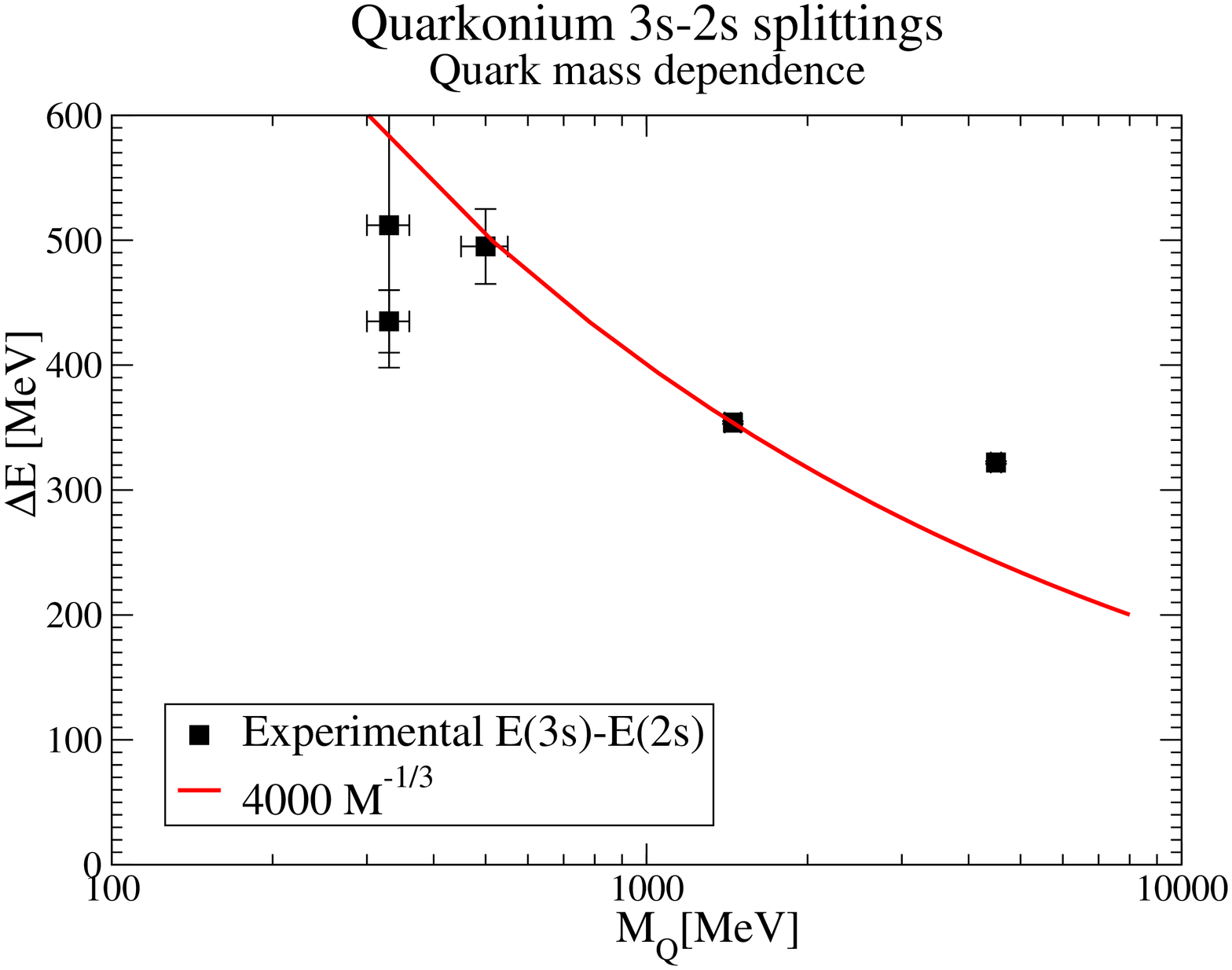}}
\parbox[t]{8cm}{
\begin{tabular}{|cc|cc|}
\hline
Charmonium & $E-E_{J/\psi}$ & Bottomonium & $E-E_{\Upsilon}$ \\
\hline
$\eta_c$       & -117    & $\eta_b$            & unknown \\
$J/\psi$       &    0    & $\Upsilon$          & 0       \\
$\chi_{c0}(1p)$&  318    & $\chi_{b0}(1p)$     & 399     \\
$\chi_{c1}(1p)$&  414    & $\chi_{b1}(1p)$     & 433     \\
$h_c$          &  429    & $h_b$               & unknown \\
$\chi_{c2}(1p)$&  459    & $\chi_{b2}(1p)$     & 452     \\
\hline
$\eta_c(2s)$   &  540(4) & $\eta_b(2s)$        & unknown \\
$\psi(2s)$     &  589    & $\Upsilon(2s)$      & 563     \\
$\psi(3770)$   &  675    & $\Upsilon(1d)$      & 701     \\
$X(3872)$      &  768    & No equivalent       & known   \\
$\chi_{c2}(2p)$&  832    & $\chi_{b2}(2p)$     & 809     \\
$X(3940)$      &  843(5) & $\chi_{b0}(2p)$     & 772     \\   
$X(3945)$      &  848(11)& $\chi_{b0}(2p)$     & 795     \\
\hline
$\psi(4040)$   &  942(1) & $\Upsilon(3s)$      & 895     \\
$\psi(4160)$   & 1056(3) & $\Upsilon(2d)$      & unknown \\
$\psi(4260)$   & 1166(12)& $\Upsilon(4s)$      & 1119    \\
$\psi(4320)$   & 1223    & $\Upsilon(3d)$      & unknown \\
$\psi(4415)$   & 1324(4) & $\Upsilon(5s)$      & 1405    \\
$\psi(4660)$   & 1563    & $\Upsilon(4d)$      & 1559(8) \\
\hline
\end{tabular}}
\caption{\label{fig:splittings} 
Left diagram: The dependence of the splitting $E(3s)-E(2s)$ on the quark mass. The
experimental assignments for $u\bar{u}$-, $s\bar{s}$-, $c\bar{c}$-, and
$b\bar{b}$-states are educated guesses based on the non-relativistic
quark model. The thick line represents a scaling based on eq.(\ref{scalingL})
where a mass independent string tension is assumed, i.e. $\nu=1$.\\
Right table:
Charmonium and bottomonium spectra (quoted are the splittings to
the $J/\psi(3097)$ and $\Upsilon(9460)$). Data from  \cite{Yao:2006px}, 
with new states reported from \cite{Aubert:2005rm,Aubert:2006ge,He:2006kg,:2007ea}.
}
\end{figure}

In fig.~(\ref{fig:splittings}) we show splittings between the $2s$ and $3s$ 
quarkonia for $\Upsilon$, $\psi$, $\phi$ and $\rho\pi$ systems. Ignoring 
$s-d$ wave mixing a typical assignment for these states are $\Upsilon(10355)$,
$\Upsilon(10023)$, $\psi(4040)$, $\psi(3686)$, $\phi(2170)$, $\phi(1680)$.
There is no candidate for $I=0$ $\omega(3s)$ so we give instead the $I=1$
$\pi$, $\rho$ splitting using $\rho(1900)$, $\rho(1465)$, $\pi(1800)$,
$\pi(1300)$. However, one has to bear in mind that large errors in the mass 
determination and the general failure of a non-relativistic quark model for 
light quarks make these splittings less useful. The thick line in our plot 
represents a scaling based on eq.(\ref{scalingL}) where a mass independent 
string tension is assumed, i.e. $\nu=1$. 

With respect to these data we now discuss whether a strong mass-dependence of
the string tension is possible. The $\phi$ and $\psi$ states fall nicely on
the $\mu^{-1/3}$ dependence suggested by a linear-potential with
a mass-independent string tension (thick line). 
However, as also noted e.g. in \cite{Rosner:2006jz}, charmonium and
bottomonium have a congruent spectrum, i.e. the splittings are almost the 
same in both systems. This suggests either 
(i) a logarithmic potential, 
(ii) a soft mass-dependence of the string tension of a linear potential of
the type $\sigma \propto \sqrt{M_Q}$,
(iii) the beginning of a transition to the Coulomb regime around the $\Upsilon$.

On the other hand, our analytical analysis of the mass dependence of the 
quark-gluon vertex in subsection \ref{IRmass} yielded three possibilities: 
either the vertex-dressing is proportional to $M_Q$ which results in $\sigma \propto M_Q^2$,
or it is proportional to $1/M_Q$, yielding $\sigma \propto M_Q^{-2}$, or
it is independent of the quark mass which gives $\sigma \propto M_Q^{0}$.
The first case together with eq.~(\ref{splittingslinear}) would give a 
Coulomb-like behavior $E\propto M_Q$ that is clearly ruled-out by the data.
The decreasing self-consistent solution $\sigma \propto 1/M_Q^2$ 
would give $E\propto M_Q^{-5/3}$ that is too strong a decrease for the data.

On the other hand, the third solution has a quark mass independent
string tension, scaling as the square of the QCD scale, just like in pure
Yang-Mills theories \cite{Szczepaniak:2001rg}. This behavior could explain 
the experimental splittings if the Coulomb potential has a large effect in 
the bottomonium spectrum. It would certainly be desirable to confirm this 
behavior in a lattice computation of the dependence of the string tension 
with the quark mass.

Moreover, many of the experimental splittings given in the table in fig.~\ref{fig:splittings} 
agree with a quark mass-independent scaling, including many of the
exciting new states. Of course, in the absence of coupled open decay
channels one would like to make the comparison for the highest possible
states, since the linear potential would then clearly dominate, but the open
channels make the loss of precision high in the spectrum comparable to the
loss of precision due to the Coulomb potential for low-lying states. 
Also, some of the states could be hidden exotics in nature, that is, composed
of wavefunctions $q\bar{q}g$, $q\bar{q}q\bar{q}$ and one needs careful
on-resonance data to be able to apply the Franck-Condon principle to
distinguish them \cite{General:2006ed,General:2007bk}.

%%%%%%%%%%%%%%%%%%%%%%%%%%%%%%%%%%%%%%%%%%%%%%%%%%%%%%%%%%%%%%%%%%%%%%
\section{Summary and conclusions \label{sec:concl}}
%%%%%%%%%%%%%%%%%%%%%%%%%%%%%%%%%%%%%%%%%%%%%%%%%%%%%%%%%%%%%%%%%%%%%%%

In this work we performed a thorough investigation of dynamical chiral 
symmetry breaking in the quark-gluon interaction of quenched Landau 
gauge QCD. We considered the coupled dynamical system of Dyson-Schwinger 
equations for the quark-propagator and the quark-gluon vertex and 
determined selfconsistent solutions both analytically and numerically. 
We identified the infrared leading diagrams in this equation and found 
these to agree with the leading ones in an $1/N_c$ counting scheme. This 
result is alternative to a number of treatments of the vertex-DSE (see e.g.  
\cite{Bender:2002as,Bhagwat:2004hn,Bhagwat:2004kj,Matevosyan:2006bk}, 
that picked out the subleading, Abelian parts of the vertex-DSE.
The most important results from our selfconsistent treatment are
\begin{itemize}
\item 
The qqg vertex uniform singularities have, in a milder form, been 
proposed in the 
past, in Ball-Chiu-type constructions based on the Slavnov-Taylor 
Identities.
We now find an overall (all external scales go to zero) as well as a 
collinear
(the external gluon momentum goes to zero) infrared divergence of the vertex
scaling with $(p^2)^{-1/2-\kappa}$, provided we take into account dynamical
chiral symmetry breaking. This includes corresponding chirally non-symmetric
tensor structures in the quark propagator and the quark-gluon vertex, as e.g.
Dirac scalars. In our numerical analysis we find a scalar part of the vertex which
is of the same strength in the infrared as the conventional $\gamma_\mu$-part.
\item This infrared divergence implies an infrared divergent running coupling
(defined from the quark-gluon vertex) in contrast to the infrared finite fixed
point behaviors of the couplings in the Yang-Mills sector of the theory. Our
findings agree with the old notion of infrared slavery \cite{Weinberg:1973un},
although the mechanism at work is different. In early works it was 
believed that
infrared slavery is caused by an infrared divergent gluon propagator. In 
contrast we find an infrared vanishing (or finite) gluon, that is accompanied by
an infrared divergent quark-gluon vertex. Both quantities together drive
the coupling to infinity in the zero momentum limit\footnote{Certainly, there is
no Landau pole at intermediate momenta in a nonperturbative treatment of the
running coupling. We do find a plateau at mid-to-low momentum, although in a 
model-dependent manner.}. 
\item The collinear infrared divergence also triggers an infrared divergent one-gluon
exchange kernel between two quarks, leading to a linear rising static 
quark potential which implies quark confinement.
\item If chiral symmetry is restored by hand, i.e. if the chiral symmetry breaking
tensor structures in the quark propagator and the quark-gluon vertex are 
arbitrarily set
to zero, we find a different selfconsistent solution of the quark propagator 
and vertex-DSEs: the vertex then has an infrared singularity of order
$(p^2)^{-\kappa}$ which is reduced in strength compared to the chirally broken
case. 
\item This reduced infrared strength in the chirally symmetric case leads an
infrared constant coupling from the quark-gluon vertex (i.e. similar to the
couplings in the Yang-Mills sector) and to a Coulomb-type static quark-antiquark potential,
which is not confining. We have therefore uncovered a mechanism which directly links
confinement to dynamical chiral symmetry breaking and vice versa.
\end{itemize}
These are the rigorous results reported in this paper. We have also, in a less
rigorous but nevertheless meaningful way, discussed some aspects and consequences
of our findings. In particular we gave a prescription to construct a 
confining and chiral symmetry preserving
Bethe-Salpeter kernel, which includes the infrared divergencies found in
our analysis. We discussed the mass dependence of the resulting string tension
in our approach and found a surprising behavior for the quark mass functions:
taking all tensor structures of the quark-gluon vertex into
account it seems that the infrared results for the quark mass $M(p^2=0)$ are much
less sensitive to the current quark mass than anticipated previously. Whereas
there is a change of $O(10^3)$ in the current quark mass from the up- to the
bottom quark, we only find changes of a factor 2-3 in $M(p^2=0)$ 
(somewhat larger changes could be expected in different truncations). 
As a result
we find a string tension which has no simple scaling law wrt. to the quark mass.
This result seems to be consistent with phenomenological expectations from
the splittings in quarkonia. 

Finally we discussed some aspects of the confining nature of our result. We presented
arguments in favor of the absence of long-ranged van-der-Waals forces even in the
presence of a strong singularity, such as the one suggested, in the quark-gluon interaction.
In a nutshell, the reason here is the absence of color-non-singlet exchanges between
color-singlet objects. The remaining color-singlet exchanges then always are gapped
such that the interaction vanishes at low momenta. This issue deserves further studies
in the future. In addition we shortly discussed the Casimir scaling property of 
one-gluon exchange and potential candidates of Green's functions which should
lead to the N-ality behavior of the string tension at large distances 
in unquenched QCD.

Finally, it is hard to see how the picture of confinement via a static 
linear potential may be also adequate for light quarks (see for example 
\cite{Roberts:2007ji}). In this respect 
it is reassuring that we find positivity violations for the light quark 
propagator, which in turn is a sufficient condition for the confinement 
of light quarks. At the current technical level of treating the DSE of 
the quark-gluon vertex this statement is truncation dependent. It 
remains to be seen, whether this finding can be corroborated in a 
truncation independent way.

In this work we gave a qualitative picture of what could be a confinement mechanism 
for quarks in Landau gauge QCD. Although we tried to make our arguments as rigorous
as possible, much work remains to be done. In particular it remains to be shown that
the quantitative aspects of the picture match expectations from phenomenology. Here
it seems desirable to develop a calculational scheme to determine the quark-antiquark
interaction in quantitative detail and confront the results with the experimental
findings as e.g. the spectra of charmonia and bottomonia.

%%%%%%%%%%%%%%%%%%%%%%%%%%%%%%%%%%%%%%%%%%%%%%%%%%%%%%%%%%%%%%%%%%%%%%%
\acknowledgments
%%%%%%%%%%%%%%%%%%%%%%%%%%%%%%%%%%%%%%%%%%%%%%%%%%%%%%%%%%%%%%%%%%%%%%%

The observation that the leading diagram in the vertex 
Dyson-Schwinger equation is
the non-Abelian piece of eq. (\ref{qqgnonabelian}) surfaced in 
conversations with Mandar
Bhagwat, Craig Roberts and Peter Tandy. We have enjoyed discussion with many
other colleagues such as Gunnar Bali, Stanley Brodsky, Jeff Greensite, 
Kurt Langfeld, Stefan Olejnik, Jan Pawlowski, Hugo Reinhardt, Lorenz von Smekal, 
Adam Szczepaniak, Daniel Zwanziger, and several graduate students
for the extended duration of this project.
We thank Richard Williams for a critical reading of the manuscript.
F. J. Llanes-Estrada thanks the 
members of the institutes of theoretical physics at T\"ubingen and Graz, and the 
Institute for Nuclear Physics at Darmstadt for their hospitality and 
acknowledges travel support from the DAAD, Univ. Complutense, Ministerio de
Educacion y Ciencia and project grants FPA 2004-02602, 2005-02327, 
PR27/05-13955-BSCH  (Spain). This work was also supported in part by grant 
M979-N16 of the Austrian Research Foundation FWF, by the DFG under contract 
AL 279/5-1\&2 and by the Helmholtz-University Young Investigator Grant VH-NG-332.

%%%%%%%%%%%%%%%%%%%%%%%%%%%%%%%%%%%%%%%%%%%%%%%%%%%%%%%%%%%%%%%%%%%%%%%
\appendix
%%%%%%%%%%%%%%%%%%%%%%%%%%%%%%%%%%%%%%%%%%%%%%%%%%%%%%%%%%%%%%%%%%%%%%

%%%%%%%%%%%%%%%%%%%%%%%%%%%%%%%%%%%%%%%%%%%%%%%%%%%%%%%%%%%%%%%%%%%%%
\section{Structure of the Quark-gluon vertex}
%%%%%%%%%%%%%%%%%%%%%%%%%%%%%%%%%%%%%%%%%%%%%%%%%%%%%%%%%%%%%%%%%%%%
The quark-gluon vertex is in general a function of 2 independent four 
momenta as well as its Lorentz and color indices. The tensor decomposition 
expresses it in a finite basis of the internal spaces given in Sec. \ref{app:tensor}. 
This also reduces the momentum dependence of the form factors to 3 
independent momentum squares. We will study several different kinematics 
defined in Sec. \ref{kin}.

\subsection{Tensor basis \label{app:tensor}}
In Landau gauge the quark gluon vertex has twelve different tensor structures
conveniently written as 
\be
\Gamma^\mu=\sum_{i=1}^4 \lambda_i L_i^\mu + \sum_{i=1}^8 \tau_i T_i^\mu \ , 
\ee
where the $T_i$ denote parts transverse to the gluon momentum.
The general basis for the quark-gluon vertex we employ here has been given 
by Ball and Chiu \cite{Ball:1980ay}. It reads
\ba \label{newdec}
L_1^\mu&=& \gamma^\mu \\ \nonumber
L_2^\mu&=&-(\not p_1 + \not p_2)(p_1+p_2)^\mu \\  \nonumber
L_3^\mu&=&-i(p_1+p_2)^\mu \\  \nonumber
L_4^\mu&=&-i \sigma^{\mu \nu}(p_1+p_2)_\nu \\  \nonumber
T_1^\mu&=& i(p_1^\mu p_2\cd p_3-p_2^\mu p_1 \cd p_3) \\  \nonumber
T_2^\mu&=&(p_1^\mu p_2\cd p_3-p_2^\mu p_1 \cd p_3)(\not p_1+\not
p_2)) \\  \nonumber
T_3^\mu&=& \not p_3 p_3^\mu -p_3^2 \gamma^\mu \\  \nonumber
T_4^\mu&=& -i(p_3^2 \sigma^{\mu \nu}(p_1+p_2)_\nu +2p_3^\mu
\sigma_{\lambda \nu} p_1^\lambda p_2^\nu ) \\  \nonumber
T_5^\mu&=& i \sigma^{\mu \nu} (p_3)_\nu \\  \nonumber
T_6^\mu&=&  (p_1^2-p_2^2)\gamma^\mu + (p_1+p_2)^\mu \not p_3
\\ \nonumber
T_7^\mu&=& \frac{i}{2}(p_1^2-p_2^2) [(\not p_1+\not p_2)\gamma^\mu
-(p_1+p_2)^\mu] -i (p_1+p_2)^\mu \sigma_{\lambda \nu} p_2^\lambda   
p_1^\nu\\  \nonumber
T_8^\mu&=& -\gamma^\mu \sigma_{\lambda \nu} p_2^\lambda p_1^\nu -\not
p_2 p_1^\mu +\not p_1 p_2^\mu
\ea
with $p_3=p_2-p_1$ and
$\sigma_{\mu \nu}=\frac{1}{2}(\gamma_\mu \gamma_\nu - \gamma_\nu
\gamma_\mu)$.
This basis is similar to that of 
\cite{skullerud:2002ge}, 
differing only in the 
momentum labels. The structures $L_1$, $L_2$, $T_2$, $T_3$ and $T_6$ 
each involve one Dirac matrix and therefore form the Dirac-vector 
part of the vertex. The structures $L_3$ and $T_1$ form the 
scalar part of the vertex, whereas the remaining structures
$L_4$, $T_4$, $T_5$ and $T_8$ have tensorial character. For the 
purpose of our power counting analysis as given in section \ref{irquarks}
the tensors $L_4$, $T_4$, $T_5$ with an even number of Dirac
matrices act similar to the scalar ones, whereas the tensor
$T_8$ with an odd number of Dirac matrices acts similar to the vector
part. 

%%%%%%%%%%%%%%%%%%%%%%%%%%%%%%%%%%%%%%%%%%%%%%%%%%%
\subsection{On the kinematics of the quark-gluon vertex \label{kin}}
%%%%%%%%%%%%%%%%%%%%%%%%%%%%%%%%%%%%%%%%%%%%%%%%%%%

Since the quark-gluon vertex Dirac amplitudes $\lambda_i$ and $\tau_i$, defined in eq. (\ref{eq:tensor-dec}), 
are functions of three Euclidean scalars $p_q^2$, $p_{\bar{q}}^2$, $p_g^2$,
it is customary to plot them in slices of this three-dimensional space, that
we denote \lq \lq kinematic sections''. Given the various choices in the 
literature,
we find it convenient to collect here definitions for the most used
sections. 

First, there are important kinematic sections that are \lq \lq degenerate'', 
in the
sense
that there is only one independent vector $p_\mu$ in four-dimensional Euclidean
space instead of the usual two vectors $p_q^\mu$, $p_{\bar{q}}^\mu$ (the
gluon
momentum being in the plane spanned by them). 

A popular degenerate section is the \lq \lq {\bf soft gluon}'' section, 
characterized by a gluon of exactly zero momentum

\begin{center}
  \begin{picture}(98,74) (-4,-16)
    \SetWidth{1.5}
    \ArrowLine(16,-16)(46,15)
    \ArrowLine(46,15)(75,-15)
    \Gluon(46,18)(46,58){5}{3}
    \Vertex(46,12){5}
    \put(55,33){\huge{$\mathbf 0$}}
    \put(-4,-10){\huge{$\mathbf p$}}
    \put(80,-10){\huge{$\mathbf p$}}
  \end{picture}
\end{center}

A similar section is the \lq \lq {\bf soft quark}'' section where one quark 
momentum vanishes. This corresponds in the chiral 
limit to an on-shell quark in Minkowski space: 
\begin{center}
  \begin{picture}(98,74) (-4,-16)
    \SetWidth{1.5}
    \ArrowLine(16,-16)(46,15)
    \ArrowLine(46,15)(75,-15)
    \Gluon(46,18)(46,58){5}{3}
    \Vertex(46,12){5}
    \put(55,33){\huge{$\mathbf p$}}
    \put(-4,-10){\huge{$\mathbf 0$}}
    \put(80,-10){\huge{$\mathbf p$}}
  \end{picture}
\end{center}

Next let us look at two interesting non-degenerate sections. We introduce
the so called
\lq \lq {\bf 1-2-3}'' or \lq \lq {\bf totally asymmetric}'' section by 

\begin{center}
  \begin{picture}(98,74) (-4,-16)
    \SetWidth{1.5}
    \ArrowLine(16,-16)(46,15)
    \ArrowLine(46,15)(75,-15)
    \Gluon(46,18)(46,58){5}{3}
    \Vertex(46,12){5}
    \put(55,33){\huge{$\mathbf{p_3}$}}
    \put(-4,-10){\huge{$\mathbf{p_1}$}}
    \put(80,-10){\huge{$\mathbf{p_2}$}}
  \end{picture}
\end{center}
imposing the restrictions (in terms of $p_1^2\equiv p^2$), $p_2^2 = 2 p^2$,
$p_3^2=3 p^2$ that fix the relative angle of $p_1$, $p_2$ and the quotient
of their Euclidean moduli.

The other section is the so called \lq \lq {\bf symmetric}'' section given by 
$p_1^2 = p_2^2=p_3^2\equiv  p^2$, in which all
scales flowing into the vertex have equal value (but $p_1$ and $p_2$ are not
parallel).

These four sections are in principle useful to study the infrared behavior
of the vertex, that is, the infrared exponents. To expose the angular
dependence of the vertex dressing functions, it is convenient to define a
\lq \lq {\bf fixed scale}'' section in which we vary the angle between $p_1$ and 
$p_2$ but fix $p_1^2 = p_2^2\equiv p^2\not = 0$. Thus we can plot 
$\lambda_i(x)$ in terms of the polar cosine
$$
x= \frac{p_1\cdot p_2}{\arrowvert p_1 \arrowvert \ \arrowvert p_2
\arrowvert}
=1-\frac{p_3^2}{2p^2} \ .
$$

%%%%%%%%%%%%%%%%%%%%%%%%%%%%%%%%%%%%%%%%%%%%%%%%%%%%%%%%%%%%%%%%%%%%%%%
\section{Second DSE formulation for the quark-gluon vertex\label{app:secondDSE}}
%%%%%%%%%%%%%%%%%%%%%%%%%%%%%%%%%%%%%%%%%%%%%%%%%%%%%%%%%%%%%%%%%%%%%%5

In section \ref{sec:self} in the main part of this work we gave a solution 
for the infrared behavior of the quark-gluon vertex in terms of power 
laws. There it turned out that the vertex has both on overall infrared
singularity of degree $(p^2)^{-1/2-\kappa}$ when all external momenta are
proportional to one scale $p$ which goes to zero 
and also a collinear singularity $(p_g^2)^{-1/2-\kappa}$ when only the gluon
momentum $p_g$ is small compared to $\Lambda_{\tt QCD}$.  

This solution has been derived from a version of the vertex-DSE, where the
bare vertex is always the one attached to the external gluon leg. This gives
the equation shown in fig. \ref{DS1} in the main body of this work, 
which includes all possible bare vertices from the Lagrangian of the theory. 
We will call this version of the DSE 's-channel version' in the following.
This solution also straightforwardly solves the corresponding equation
of motion for the vertex from an 3PI effective action, given in fig. \ref{DS3}. 
However, things are slightly more complicated for the second or
't-channel' version of the vertex-DSE. This
version has a bare vertex attached to an external quark line and is 
given in  fig.~\ref{DS2}.

The problem here is that the simple triangle diagrams involving only two-
and three-point functions do not reproduce the infrared singularity found 
from the other two equations. Let us carry through this exercise for example
for the diagram involving the fully dressed three-gluon vertex. Simple power
counting then results in an overall divergence of
\beq
(p^2)^{-3\kappa+1/2} \times (p^2)^{2\kappa-1} \times (p^2)^{2\kappa-1} 
\times (p^2)^{-1/2-\kappa} \times (p^2)^{2} \sim (p^2)^{0}\,,
\eeq
where the first term stems from the three-gluon vertex, the next two terms
from the gluon propagators followed by the scalar part of the remaining 
quark-gluon vertex and the loop integration, whereas the quark propagator is 
constant. To yield the previous IR divergence of the vector part of the vertex 
on the left hand side there is clearly infrared strength missing.

\begin{figure}[t]
\centerline{\epsfig{file=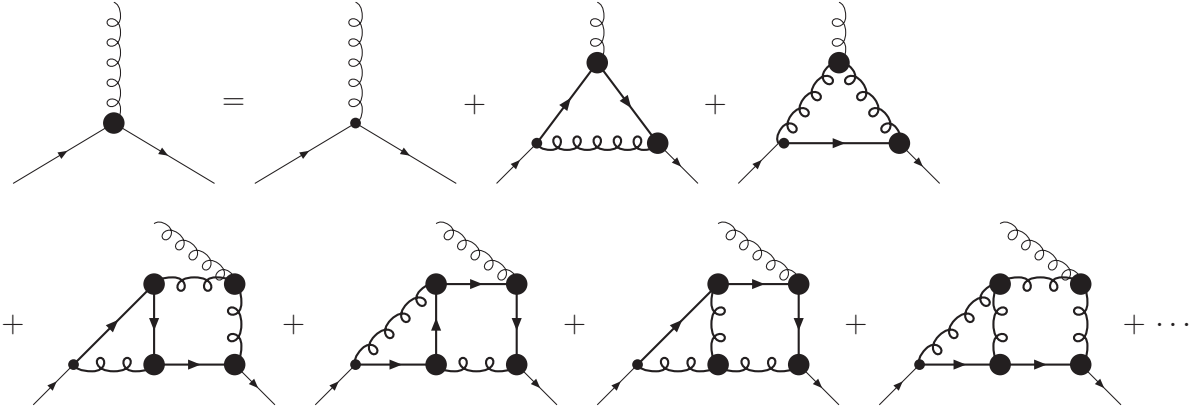,width=16cm}}
\caption{Leading contributions in the skeleton expansion of the 
t-channel-version of the DSE for the quark-gluon vertex. \label{DS2-skeleton}}
\end{figure}

The answer to this puzzle lies in the remaining diagram, that includes the
1PI gluon-quark four-point function. Since the ghost dynamics is the dominant 
dynamical contribution in the DSEs and there is no fundamental quark-ghost 
coupling, similarly to the case of the s-channel DSE discussed in detail in 
the main text, it is again necessary to perform a skeleton expansion to 
two-loop order as shown in fig. \ref{DS2-skeleton}.
 
The power counting analysis for this equation is complicated since not only 
soft modes of the order of the external momenta can dominate the different 
loop integrals but also hard modes of the order of the dynamically generated 
quark mass $M$ appearing in the quark propagators\footnote{We note that a very
similar integral appears in the DSE for the quark propagator, which is very 
well explored. These integrals are known to result in finite answers 
in the infrared for a wide range of possible dressings in the loops (see 
e.g. \cite{Fischer:2003rp}).}. 
In combination with the different possibilities to route any hard external momenta 
through the diagrams, the two-loop integrals can be decomposed into many different 
potentially IR sensitive regions \cite{Alkofer:2008jy}. The most singular 
contribution on the right hand side of the DSE dominates and will determine 
the scaling of the vertex on the left hand side. Naturally, GreenÔs functions 
involving only hard momenta which appear in a given graph do not scale with the 
soft external momentum.  Here, we will not give a complete power counting analysis 
analogous to the one for the first DSE presented in appendix \ref{sec:qpc}, but 
rather identify the contributions that provide the strong singularities found 
in the s-channel DSE discussed in the main text.
\begin{figure}[t]
\centerline{\epsfig{file=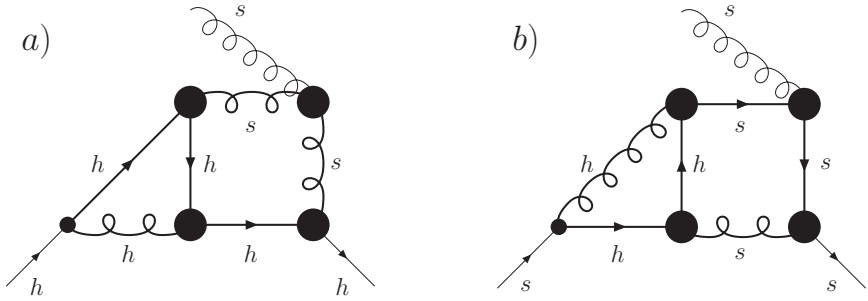,width=12cm}}
\caption{Kinematic configurations of the leading loop graphs that generate the 
strong soft-gluon divergence (a) and the uniform divergence (b) of the quark-gluon 
vertex. The labels $s$ and $h$ denote whether the momentum running through a line 
is soft or hard, respectively. In contrast to the s-channel-version of the DSE, 
here the soft-gluon singularity in graph (b) is necessary to induce the uniform 
divergence. \label{skel-kin}}
\end{figure}

Let us start with the quark-gluon vertex in the soft-gluon limit. One particular 
contribution arising from a specific kinematic region of the first two-loop graph 
in fig. \ref{DS2-skeleton} is given in fig. \ref{skel-kin} (a). Here, the loop 
integral of the initial graph (left part) arising in the unexpanded DSE 
fig.~\ref{DS2} extends over hard momenta whereas only the loop integral (right part) 
arising from the skeleton expansion involves soft scales of the order of the 
external gluon momentum\footnote{Note that we are only interested in the scaling 
with respect to the soft external momentum $p\ll \Lambda_{\tt QCD}$ and do not 
distinguish between any hard external momenta $q<\Lambda_{\tt QCD}$ and hard loop 
momenta $q\sim M$.}. In case that this contribution dominates it yields for the 
anomalous exponent of the quark-gluon vertex in the soft-gluon limit the self-consistent 
result (counting only IR exponents)
\begin{equation}
  \delta_{qg}^{gl}=2+\left(\frac{1}{2}+\delta_{3g}^{u}\right) 
  + 2\delta_{qg}^{gl}+2\left(\delta_{gl}-1\right)=2\delta_{qg}^{gl}
  +\frac{1}{2}+\kappa \quad \Rightarrow \quad \delta_{qg}^{gl}=-\frac{1}{2}-\kappa
\end{equation}
Accordingly the ghost-enhanced three-gluon vertex induces a strong soft-gluon 
singularity in full analogy to the s-channel version discussed in the main 
text. However, this mechanism does not generate a uniform divergence in the 
absence of soft divergences. To see how the latter is generated self-consistently 
consider graph (b) of fig. \ref{skel-kin} which presents one of the possible 
contributions. If it dominates it yields for the corresponding anomalous 
exponent
\begin{equation}
  \delta_{qg}^{u}=2+2\delta_{qg}^{u}+\delta_{qg}^{gl}+\delta_{qg}^q
  +\left(\delta_{gl}-1\right)+2\delta_q=2\delta_{qg}^{u}+\frac{1}{2}
  +\kappa \quad \Rightarrow \quad \delta_{qg}^{u} =-\frac{1}{2}-\kappa
\end{equation}
We therefore conclude that the structure of the loop integrals in the
last term of fig.~\ref{DS2} together with the collinear singularity
in the external gluon leg provides exactly such strength in the infrared
to reproduce our results for the quark-gluon vertex in the s-channel version
of the vertex-DSE also for the t-channel version. 

It is interesting to note a few properties of this result: First, in contrast 
to the s-channel DSE, here regions of the loop integral with hard momenta of 
the order of the quark mass dominate even in the uniform limit.
Further, the uniform divergence in the t-channel DSE is not directly induced 
by the three-gluon vertex but only indirectly once the strong soft-gluon divergence 
is present. Therefore, in the present case these two divergences are not totally 
independent but imply each other. Furthermore, in the t-channel version it is 
not necessary to distinguish between the different Dirac components of the 
vertex to recover the proper uniform scaling. The reason is that the corresponding 
graphs involve many fermionic Greens functions that involve hard momenta and do 
not scale with the soft external momentum, independent of their Dirac-structure. 
Correspondingly, in contrast to the case of the non-Abelian graph studied in 
the main text, there are no corresponding restrictions, here. 

Finally, from a practical point of view the s-channel version of the DSE lends 
itself much better to a numerical analysis, since, as explained in the main 
text, the IR dominant two-loop graph can be approximately replaced by a 1-loop 
graph with a dressed three-gluon vertex. In the quenched approximation this three-gluon 
vertex can be computed beforehand or, as done in the present analysis, 
simply be implemented by some ansatz. In the t-channel version of the DSE 
this is not possible since the initial loop dresses the desired quark-gluon 
vertex itself and therefore the expensive two-loop integrals indeed would have 
to be analyzed numerically.

%%%%%%%%%%%%%%%%%%%%%%%%%%%%%%%%%%%%%%%%%%%%%%%%%%%%%%%%%%%%%%%%%%%%%%%%%%
\section{Semi-perturbative analysis of the quark-gluon vertex
\label{section:semipert}}
%%%%%%%%%%%%%%%%%%%%%%%%%%%%%%%%%%%%%%%%%%%%%%%%%%%%%%%%%%%%%%%%%%%%%%%%%%

%%%%%%%%%%%%%%%%%%%%%%%%%%%%%%%%%%%%%%%%%%%%%%%%%%%%%%%%%%%%%%%%%%%%%%%%%%
\subsection{Non-Abelian diagram}
%%%%%%%%%%%%%%%%%%%%%%%%%%%%%%%%%%%%%%%%%%%%%%%%%%%%%%%%%%%%%%%%%%%%%%%%%

We start by evaluating the one-loop diagram involving the non-Abelian
three gluon vertex
\begin{equation} \label{qqgnonabelian}
\begin{minipage}[c][1\totalheight]{0.3\columnwidth}
$i g T^a_{ij} \Gamma^{\mu }_{\rm NA}(p1^2,p_2^2,(p_2-p_1)^2) \ = \ \ $
\end{minipage}
\begin{minipage}[c][1\totalheight]{0.3\columnwidth}
\begin{picture}(150,150)(0,0)
\ArrowLine(0,0)(50,50) \put(0,15){$p_1$}
\ArrowLine(100,50)(150,0) \put(140,15){$p_2$}
\Gluon(75,150)(75,100){3}{6}
\Gluon(50,50)(75,100){3}{6} \Gluon(75,100)(100,50){3}{6}
\ArrowLine(50,50)(75,50) \ArrowLine(75,50)(100,50)
\GCirc(75,50){3}{0} \GCirc(62,75){3}{0} \GCirc(87,75){3}{0}
\put(12,0){$i$} \put(135,0){$j$} \put(73,42){$q$} \put(63,130){$a$}
\put(82,130){$k_3^\mu=(p_2-p_1)^\mu$}
\put(10,75){$k_1=p_1 \! \! -\! \! q$} \put(98,75){$k_2=q\! \! -\! \! p_2$}
\end{picture} \ .
\end{minipage}
\end{equation}
The color structure for this diagram is 
$$
\frac{iN_c}{2} T^a_{ij}
$$
where $T^a_{ij}$ is the generator in the fundamental representation, and the
factor $N_c$ results from the contraction of the color matrices.
The full expression reads
\ba \label{spelledNAvertex}
\Gamma^{\mu}_{\rm NA}(p1^2,p_2^2,(p_2-p_1)^2) \ = 
\left( \frac{-i N_c}{2} g^2 Z_{1F}^2 Z_3\right) \int \frac{d^4q}{(2\pi)^4}
\left\{  \gamma^\nu(i \! \not q \sigma_v(q^2) +  \sigma_s(q^2))
\gamma^\rho \frac{Z(k_1^2)}{k_1^2} \frac{Z(k_2^2)}{k_2^2}
\right. \\ \nonumber \left.
 (\delta_{\sigma\theta}(k_1-k_2)_\mu +
  \delta_{\theta\mu}(k_2-k_3)_\sigma +
  \delta_{\sigma\mu}(k_3-k_1)_\theta )
\left( \delta_{\nu\sigma}-\frac{k_{1\nu}k_{1\sigma}}{k_1^2}\right)
\left( \delta_{\rho\theta}-\frac{k_{2\rho}k_{2\theta}}{k_2^2}\right)
\right\},
\ea
with the renormalized coupling $g^2$ and the renormalization constants $Z_{1F}$
and $Z_3$ for the bare quark-gluon vertices and the three-gluon vertex.
Dimensional analysis of eq. (\ref{spelledNAvertex}) immediately shows
a logarithmic divergence which appears only in $\lambda_1$.
All other pieces of the vertex, containing powers of either $p_i$, are 
free of ultraviolet divergencies. 
Our renormalization prescription for $\lambda_1$ is given 
in subsection \ref{numerics}.

%%%%%%%%%%%%%%%%%%%%%%%%%%%%%%%%%%%%%%%%%%%%%%%%%%%%%%%%%%%%%%%%%%%%%%%%%%
\subsection{Abelian vertex diagram}
%%%%%%%%%%%%%%%%%%%%%%%%%%%%%%%%%%%%%%%%%%%%%%%%%%%%%%%%%%%%%%%%%%%%%%%%%%
Next we turn our attention to the  \lq \lq QED-like'' diagram
\begin{equation} \label{qqgabelian}
\begin{minipage}[c][1\totalheight]{0.3\columnwidth}
$i g T^a_{ij} \Gamma^{\mu }_{\rm A}(p1^2,p_2^2,(p_2-p_1)^2) \ = \ \ $
\end{minipage}
\begin{minipage}[c][1\totalheight]{0.3\columnwidth}
\begin{picture}(150,150)(0,0)
\ArrowLine(0,0)(50,50) \put(0,15){$p_1$}
\ArrowLine(100,50)(150,0) \put(140,15){$p_2$}
\Gluon(75,150)(75,100){3}{6}
\ArrowLine(50,50)(62,75)  \ArrowLine(62,75)(75,100) \GCirc(62,75){3}{0} 
\ArrowLine(75,100)(87,75)  \ArrowLine(87,75)(100,50)\GCirc(87,75){3}{0}
\Gluon(50,50)(100,50){3}{6} \GCirc(75,50){3}{0}
\put(12,0){$i$} \put(135,0){$j$} \put(73,42){$q$} \put(63,130){$a$}
\put(82,130){$k_3^\mu=(p_2-p_1)^\mu$}
\put(10,75){$k_1=p_1 \! \! -\! \! q$} \put(98,75){$-k_2=p_2\! \! -\! \! q$}
\end{picture} \ .
\end{minipage}
\end{equation}
This has color structure 
$$
\frac{-1}{2N_c} T^a_{ij}
$$
and reads
\ba \label{spelledAvertex}
\Gamma^{\mu}_A(p_1^2,p_2^2,(p_2-p_1)^2) = \left( \frac{1}{2N_c}
g^2 \right)  \int \frac{d^4q}{(2\pi)^4} \left\{
\frac{Z(q^2)}{q^2} \gamma^\nu (i \! \not k_1 \sigma_v(k_1^2) +\sigma_s(k_1^2) 
) \gamma^\mu 
\right. \\ \nonumber \left.
(-i \! \not k_2 \sigma_v(k_2^2)  +\sigma_s(k_2^2)) \gamma^\rho \left(
\delta_{\nu \rho} - \frac{q_\nu q_\rho}{q^2}
\right) \right\}
\ea

Both the Abelian and the non-Abelian diagrams reproduce important
properties of the vertex; as discussed in \cite{davydychev:2000rt}, 
charge conjugation invariance requires
\ba \nonumber
C  \gamma^\mu C^{-1} = - \gamma^{\mu \ T} \ \ \ 
C i \Gamma^\mu C^{-1} = i \Gamma^{\mu \ T}
\ea
where $C$ is the charge conjugation operator and the superscript $T$ denotes 
transposition. Referring to the WT basis (\ref{newdec}) above, we see that the
$\lambda_4$, $\tau_4$, and $\tau_6$ Dirac amplitudes should be odd under
the
interchange of $p_1$ and $p_2$, whereas the other nine should be even. This
is indeed the case for both diagrams in our calculation.
Also, the anti-unitarity of the time reversal operator means that all the
Dirac amplitudes should be either real or imaginary. 
With the explicit factors of $i$ in eq.~(\ref{qqgdressed}) and the 
conventions of (\ref{newdec}) all the $\lambda$ and $\tau $ are strictly real.

%%%%%%%%%%%%%%%%%%%%%%%%%%%%%%%%%%%%%%%%%%%%%%%%%%%%%%%%%%%%%%%%%%%%%%
\subsection{Input for gluon and quark propagators \label{propinput}}
%%%%%%%%%%%%%%%%%%%%%%%%%%%%%%%%%%%%%%%%%%%%%%%%%%%%%%%%%%%%%%%%%%%%%%%

The dressing functions for the gluon $Z$ and quark $\sigma_v, \ \sigma_s$
propagators are taken as input from reference \cite{Fischer:2002hna}. 
We employ analytical fits to the numerical solution of the Schwinger-Dyson 
equations from this work corresponding to a choice $\alpha_s=\frac{g^2}{4\pi}=0.2$. 
The gluon propagator is well represented by:
\be \label{gluonprop}
Z(x)=\left( \frac{\alpha(x)}{\alpha(\mu)} \right)^{1+2\delta} R^2(x)
\ee
and the ghost propagator by
\be \label{ghostprop}
G(x)=\left( \frac{\alpha(x)}{\alpha(\mu)} \right)^{-\delta} R^{-1}(x)
\ee
where the low momentum behavior is captured by the irrational function
\be
R(x)=\frac{c\left(\frac{x}{\Lambda^2}\right)^\kappa + d 
\left(\frac{x}{\Lambda^2}\right)^{2\kappa}}{1+c\left(\frac{x}{\Lambda^2} 
\right)^\kappa +d \left(\frac{x}{\Lambda^2}\right)^{2\kappa}}. 
\ee
The parameters appearing in $R(x)$, solution of the DSE are
$c=1.269$ and $d=2.105$, and in the quenched approximation for the gluon 
and ghost equations $\delta(N_f=0)=-9/44$. 
The value for the exponent $\kappa \simeq 0.595$ in this fit is 
approximately known from the infrared analysis of the DSEs \cite{Lerche:2002ep}.

Finally the running coupling is well reproduced by
\be \label{alphas}
\alpha(x)=\frac{\alpha(0)}{\log \left( e^1 + 
a_1\left(\frac{x}{\Lambda^2}\right)^{a_2} +b_1 
\left(\frac{x}{\Lambda^2}\right)^{b_2}
\right)}
\ee
with $a_1=1.106$, $a_2=2.324$, $b_1=0.004$, $b_2=3.169$.
Using a $\widetilde{MOM}$
scheme and fitting $\Lambda$ to only the ultraviolet behavior, a value
$\Lambda_{\tt QCD}=0.714\,\mbox{GeV}$ has been given in
Ref.~\cite{Fischer:2003rp}.
Note, that these fits are for faster computing and to offer a closed
form: all results for the various two-point functions are solutions
of their respective Schwinger-Dyson equations.
These fits will be used for all numerical calculations presented in the 
course of this work.

In the same way we can parameterize the quark propagator functions,
$\sigma_{s,v}(p^2)$, and fit these to the numerical DSE solutions.
Again, we want to make explicit the leading logarithmic corrections to
$Z^f(p^2)$ and $M(p^2)$, achieved with
\begin{eqnarray}
\sigma_s(p^2) &=& \frac{B_{\hbox{\scriptsize chiral}}(p^2)}{p^2 + 
\msing^2} +
                  \frac{C_{cqm} \; \alpha(p^2 +
\msing^2)^{\gamma_m}}{p^2 + \msing^2} \,,
\label{eq:fitsigs}
\\
\sigma_v(p^2) &=& \frac{1}{p^2 + \msing^2} \left(1
                - \frac{\alpha(p^2 + \msing^2)}{2\pi}
                + \tilde{C} \; B_{\hbox{\scriptsize chiral}}(p^2)
                  \right) \, ,
\label{eq:fitsigv}
\end{eqnarray}
with 
\be
B_{\hbox{\scriptsize chiral}}(p^2) =
        C_{dcsb} \; \frac{\alpha(p^2 + \msing^2)^{1-\gamma_m}}
                         {p^2 + \msing^2 + \Lambda^2}
                  + \frac{C_4}{(p^2 + \msing^2 + \Lambda^2)^2} \; .
\label{eq:fitBchiral}
\ee
This parameterization ensures the correct asymptotic
behavior for $\sigma_{s,v}(p^2)$ and the quark functions
$M(p^2)$ and $Z^f(p^2)$ 
\cite{Fischer:2003rp,Alkofer:2003jj} 
in the
chiral limit. It is important to note that we use this fit for the quark 
propagator {\bf in this appendix only}! In the main text, where we 
back-feed the vertex into the quark-DSE, given in fig.~\ref{quarkdse0}, 
the corresponding quark propagator is solved self-consistently and 
dynamically from the set of coupled DSEs for the quark propagator 
and the quark-gluon vertex. With this input for the propagators the loop 
corrections eq. (\ref{qqgnonabelian}) and (\ref{qqgabelian}) are 
straightforwardly evaluated and yield the result shown in 
fig. \ref{fig:compareAnotA} in the main text.

%%%%%%%%%%%%%%%%%%%%%%%%%%%%%%%%%%%%%%%%%%%%%%%%%%%%%%%%%%%%%%%%%%%%%%%%%
\section{Power counting and Dirac structure: an explicit example} \label{IRanalysis}
%%%%%%%%%%%%%%%%%%%%%%%%%%%%%%%%%%%%%%%%%%%%%%%%%%%%%%%%%%%%%%%%%%%%%%%%%

In this appendix we demonstrate the fact that the power counting analysis 
depends only on the Dirac-scalar or vector nature of the individual tensors 
of the quark-gluon vertex explicitly in a special kinematic limit.
Due to momentum  conservation the vertex depends only on
the two external quark momenta $(p_1)_\mu$, $(p_2)_\mu$ or three Lorentz
invariants
$p_1^2, p_2^2, p_1\cdot p_2$. To analyze the infrared  limit of this
vertex in the
presence of only one external scale $p_1^2 \equiv p^2 \ll \Lambda_{\tt
QCD}^2$, we can set
$(p_2)_\mu = 2(p_1)_\mu =: 2p_\mu$ without loss of generality. (Note that
the case 
$(p_2)_\mu = (p_1)_\mu$ 
corresponds to the collinear limit where the gluon momentum goes to zero,
while 
the other momenta stay finite. We treat this limit separately in subsection
\ref{softgluondiv}.)

The choice $(p_2)_\mu = 2(p_1)_\mu$ leaves us with only four possible tensor
structures which can be denoted by
\beq
\Gamma_\mu (p) = i g \sum_{i=1}^4 \xi_i (p^2) G_{\mu}^i \label{vertex}
\eeq
with $\xi_i $ being Lorentz and Dirac scalar functions and
\beq
G_{\mu}^1 = \gamma_\mu \,, \ \
G_{\mu}^2 = \hat{p}_\mu \,, \ \
G_{\mu}^3 = \hat{\pslash} \hat{p}_\mu \,, \ \
G_{\mu}^4 = \hat{\pslash} \gamma_\mu\,, 
\eeq
where we have normalized the momentum,
$\hat{p}_\mu =p_\mu / \sqrt{p^2}$, to ease power counting. 
Note that $G_{\mu}^1$ and $G_{\mu}^3$ have an odd number of 
$\gamma$-matrices, whereas $G_{\mu}^2$ and $G_{\mu}^4$ have an even number. 
Therefore $\xi_{2,4}\not=0$ only if chiral symmetry is broken. 

To project out the respective dressing functions $\xi_{1..4}(p^2)$ from the
vertex-DSE 
we multiply the equation with $\gamma_\mu, \hat{p}_\mu, \hat{\pslash}
\hat{p}_\mu,
\hat{\pslash} \gamma_\mu$ respectively and take the Dirac-trace. Linear
combinations of these four basis elements then give the desired $\xi_{1..4}(p^2)$. 

For the internal quark-gluon vertices we can employ the expression
(\ref{vertex}) with any internal momentum as argument: it contains all
possible 
types of Dirac structures (vector, scalar, and tensor). Any more complicated
dependence
on external and internal momenta will generate the same powers of external
momenta in 
$\xi_{1..4}$ after integration (for dimensional reasons). To determine
the infrared exponents 
of the quark-gluon vertex we employ the ansatz
\beq
\xi_1(p^2) = \chi_1 \cdot (p^2)^{\beta_1} \,, \ \
\xi_2(p^2) = \chi_2 \cdot (p^2)^{\beta_2} \,, \ \
\xi_3(p^2) = \chi_3 \cdot (p^2)^{\beta_3} \,, \ \
\xi_4(p^2) = \chi_4 \cdot (p^2)^{\beta_4}		\label{vertex-exp}
\eeq
for the scaling of the dressing functions of the quark-gluon vertex with the 
momentum scale, where the pre-factors $\chi_i$ depend only on momentum ratios.

Since we contracted the right hand side of the DSE with appropriate tensor
structures,
see above, we eventually have to deal with scalar integrals only, which
depend on powers
of internal and external momenta. The evaluation of such diagrams has been
described in \cite{Anastasiou:1999ui}. Most important for our purpose is the fact that
all powers
of internal momenta transform to powers of external momenta after
integration. This
is also clear from simple dimensional consideration, provided there is only
one external scale present. In our case this is ensured by the condition 
$p^2 \ll \Lambda_{\tt QCD}^2 \le M^2$, which leaves only one scale $p^2$ in
the deep
infrared. The following analysis rests on the assumption that there is no
second small scale present. In principle this assumption may not be true 
for heavy quarks, since then one has a potential additional small scale 
$\Lambda_{\tt QCD}/M^2$. We have investigated this possibility in our
numerical calculations of the quark-gluon vertex and did NOT see any
influence of such a quantity. The corresponding results of our quark mass
study are presented in subsection \ref{massstudy}, where we discuss this 
issue further.

A similar analysis can be done using all four dressing functions
$\xi_{1..4}$. The explicit
integral equations are quite lengthy and add nothing to the power counting
argument, so we omit them here.
One then obtains:
\beqa
(p^2)^{\beta_1} \sim max &&\left\{ Z_f \, \frac{(p^2)^{1+\kappa+2\beta_1}}{M^2},
                             Z_f \, \frac{(p^2)^{1+\kappa+2\beta_2}}{M^2},
                             Z_f \, \frac{(p^2)^{1+\kappa+2\beta_3}}{M^2},
                             Z_f \, \frac{(p^2)^{1+\kappa+2\beta_4}}{M^2},\right.
\nonumber\\
                           &&Z_f \, \frac{(p^2)^{1+\kappa+\beta_1+\beta_3}}{M^2},
                             Z_f \, \frac{(p^2)^{1+\kappa+\beta_2+\beta_4}}{M^2},
			     Z_f \, \frac{(p^2)^{1/2+\kappa+\beta_1+\beta_2}}{M},\nonumber\\
	             &&\left.Z_f \, \frac{(p^2)^{1/2+\kappa+\beta_1+\beta_4}}{M},
			     Z_f \, \frac{(p^2)^{1/2+\kappa+\beta_2+\beta_3}}{M},
			     Z_f \, \frac{(p^2)^{1/2+\kappa+\beta_3+\beta_4}}{M}\right\}\\ 
(p^2)^{\beta_2} \sim max &&\left\{
\frac{(p^2)^{1+\kappa+\beta_1+\beta_2}}{M^2},
                             Z_f \, \frac{(p^2)^{1+\kappa+\beta_1+\beta_4}}{M^2},
                             Z_f \, \frac{(p^2)^{1+\kappa+\beta_2+\beta_3}}{M^2},
                             Z_f \, \frac{(p^2)^{1+\kappa+\beta_3+\beta_4}}{M^2},\right.
\nonumber\\
			   &&Z_f \, \frac{(p^2)^{1/2+\kappa+2\beta_1}}{M},
			     Z_f \, \frac{(p^2)^{1/2+\kappa+2\beta_2}}{M},
			     Z_f \, \frac{(p^2)^{1/2+\kappa+2\beta_3}}{M},
\nonumber\\
		     &&\left.Z_f \, \frac{(p^2)^{1/2+\kappa+2\beta_4}}{M},
			     Z_f \, \frac{(p^2)^{1/2+\kappa+\beta_1+\beta_3}}{M},
			     Z_f \, \frac{(p^2)^{1/2+\kappa+\beta_2+\beta_4}}{M}\right\}\\ 
(p^2)^{\beta_3} \sim max&&\left\{ Z_f \, \frac{(p^2)^{1+\kappa+2\beta_1}}{M^2},
                             Z_f \, \frac{(p^2)^{1+\kappa+2\beta_2}}{M^2},
                             Z_f \, \frac{(p^2)^{1+\kappa+2\beta_3}}{M^2},
                             Z_f \, \frac{(p^2)^{1+\kappa+2\beta_4}}{M^2},\right.
\nonumber\\
                           &&Z_f \, \frac{(p^2)^{1+\kappa+\beta_1+\beta_3}}{M^2},
                             Z_f \, \frac{(p^2)^{1+\kappa+\beta_2+\beta_4}}{M^2},
			     Z_f \, \frac{(p^2)^{1/2+\kappa+\beta_1+\beta_2}}{M},
\nonumber\\
		     &&\left.Z_f \, \frac{(p^2)^{1/2+\kappa+\beta_1+\beta_4}}{M},
			     Z_f \, \frac{(p^2)^{1/2+\kappa+\beta_2+\beta_3}}{M},
			     Z_f \, \frac{(p^2)^{1/2+\kappa+\beta_3+\beta_4}}{M}\right\}\\ 
(p^2)^{\beta_4} \sim max&&\left\{
Z_f \, \frac{(p^2)^{1+\kappa+\beta_1+\beta_2}}{M^2},
                             Z_f \, \frac{(p^2)^{1+\kappa+\beta_1+\beta_4}}{M^2},
                             Z_f \, \frac{(p^2)^{1+\kappa+\beta_2+\beta_3}}{M^2},
                             Z_f \, \frac{(p^2)^{1+\kappa+\beta_3+\beta_4}}{M^2},\right.
\nonumber\\
			   &&Z_f \, \frac{(p^2)^{1/2+\kappa+2\beta_1}}{M},
			     Z_f \, \frac{(p^2)^{1/2+\kappa+2\beta_3}}{M},
\nonumber\\
	             &&\left.Z_f \, \frac{(p^2)^{1/2+\kappa+2\beta_4}}{M},
			     Z_f \, \frac{(p^2)^{1/2+\kappa+\beta_1+\beta_3}}{M},
			     Z_f \, \frac{(p^2)^{1/2+\kappa+\beta_2+\beta_4}}{M}\right\} \label{fullsol}
\eeqa
which is solved by
\beqa \label{mainlawqqg}
\beta_2 = -1/2-\kappa \,,  && \beta_1,\beta_3 \in [-1/2-\kappa ,
-\kappa] \nonumber \\
&& \beta_4 \in [-1/2-\kappa , 1/2-\kappa]\,. \label{power}
\eeqa
This leaves a range of possibilities with the common trait that the most
infrared singular piece diverges as $(p^2)^{-1/2-\kappa}$ in the infrared. We have analyzed
this analytic solution in numerical calculations, presented  in section 
\ref{sec:numsec},
and found that only the case  
$\beta_2=\beta_1=\beta_3=\beta_4=-1/2-\kappa$ is realized.

%%%%%%%%%%%%%%%%%%%%%%%%%%%%%%%%%%%%%%%%%%%%%%%%%%%
\section{A simple mechanical model \label{model}}
%%%%%%%%%%%%%%%%%%%%%%%%%%%%%%%%%%%%%%%%%%%%%%%%%%%

\begin{center}
\begin{figure}[h]
\includegraphics[width=5cm]{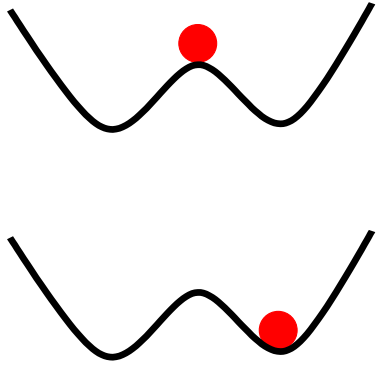}
\caption{\label{fig:spontaneous1}
A bead on a rigid ramp with two degenerate minima rolls down to either. 
The parity symmetry of the system is broken down spontaneously at the 
ground state.
}
\end{figure}
\end{center}

The simplest and well-known example of spontaneous symmetry breaking is given in fig.~
\ref{fig:spontaneous1}. A parity-symmetric potential  $V(x)=x^2-x^4$
has two minima at $x_m=\pm 1/\sqrt{2}$ and a classical bead at a minimum 
of the
potential has to choose between left or right. Therefore the 
solution to the system in its
ground state spontaneously breaks the parity symmetry, but the 
potential perceived by the bead remains parity symmetric.
This example is analogous to the spontaneous chiral symmetry breaking in the
Dyson-Schwinger equation for the propagator with a fixed quark-gluon vertex
and gluon propagator, where a quark mass gap is generated as a function of
the potential constants.

\begin{center}
\begin{figure}[h]
\includegraphics[width=5cm]{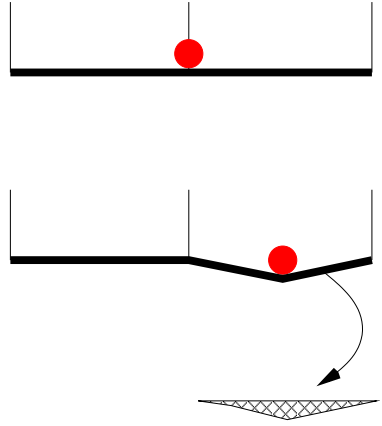}
\caption{\label{fig:sp}
A bead on an elastic band held in position at its middle point.
The displacement of the bead (spontaneous parity symmetry
violation by the \lq \lq reduced'' bead system in its ground state) is
proportional to the stretch of the band (spontaneous parity symmetry
violation by the potential that the bead sees), as measured for example 
by
the area between the band and the horizontal. Both are coupled and it is 
not
possible to understand one without the other.
}
\end{figure}
\end{center}

In this paper we have considered a more general situation, and it is 
worth
illustrating it with the following simple example. Consider now a bead of
mass $m$ deposited on top of an elastic band of elastic constant $k$ 
and
half-length $l_0$. The half-band is held fixed at the two ends and the 
mid point. This
situation is represented in fig.~\ref{fig:sp}
The bead can of course stay at the middle of the band, with its weight
compensated by the clamp holding the band by the middle, but this is a point
of unstable equilibrium. A random perturbation will make the bead  
roll left or right.
If the radius of the bead is small compared to $l_0$, the band will adopt a
triangle shape, with length $l_1$ left of the bead and $l_2$ right of the
bead. The bead will be displaced from the horizontal a height $y$ and from
the center a distance $x$.
The potential energy is now
$$
V(x,y) = -mgy + \frac{k}{2} \left( (l_1+l_2)-l_0\right)^2
$$
with the geometric constraints 
$$
x^2+y^2 = l_1^2;\ \  \ \ \ (l_0-x)^2 + y^2 = l_2^2 \ .
$$
Minimizing the potential yields
$$
\frac{\partial V(x,y)}{\partial x}=0 \to x=\frac{l_0}{2}
$$
and, for small stretchings,
$$
\frac{\partial V(x,y)}{\partial y}=0 \to y = \frac{1}{2}\left(
l_0^2mg/k \right)^{1/3} 
$$
so that the bead in this minimum sees an effective potential that 
is approximately
\ba \nonumber
V(x)=-mg \left( \frac{mg}{kl_0} \right)^{1/3} \arrowvert l_0/2 -x \arrowvert
\ \ \ \ \ & x>0\ ; \\ \nonumber
V(x)=0\ \ \ \ \ & x<0
\ea
reflecting the back-reaction of the mass of the bead on the stretchable
band (if $m=0$, $V(x)=0$ remains parity symmetric under $x\to -x$).

The position of the bead is again analogous to the quark mass function in
the propagator after spontaneous chiral symmetry breaking, and the slope of
the potential (or equivalently the area between the band and the horizontal)
analogous to a scalar piece in the vertex function, that also appears after
spontaneous symmetry breaking. One cannot understand one without the other,
and both are proportional.

In a field theory of course, the number of degrees of freedom is infinite
and not only two, and so are the coupled Schwinger-Dyson equations, all of
which have to simultaneously display spontaneous symmetry breaking.

%%%%%%%%%%%%%%%%%%%%%%%%%%%%%%%%%%%%%%%%%%%%%%%%%%%%%%%%%%%%%%%%
\section{Power counting with kinematic singularities\label{sec:qpc}}
%%%%%%%%%%%%%%%%%%%%%%%%%%%%%%%%%%%%%%%%%%%%%%%%%%%%%%%%%%%%%%%%

In this appendix we present the full power counting analysis for the quark sector 
of quenched QCD, given by the coupled system of the quark propagator and the quark-gluon vertex DSE. In particular we take into account the possibility of kinematic singularities of the vertex and do not make assumptions on the IR limit of the quark propagator. The analysis of kinematic singularities follows the corresponding analysis for the gauge sector of the theory presented in \cite{Alkofer:2008jy}. 

Whereas the propagator is generally given by a single IR exponent $\delta_q$, the quark gluon vertex has in addition to the uniform limit that all momenta vanish $\delta_{qg}^{u}$ also the possibility that only the gluon $\delta_{qg}^{gl}$ or only the quark leg vanishes $\delta_{qg}^{q}$. As noted in subsection \ref{irquarks} in the main text it is crucial to distinguish between the scalar and vector components in the analysis. We distinguish them by additional lower case letters $v$ for the vector and $s$ for the scalar part, so that the full set of possibly distinct IR exponents is given by $\delta_{qv}$, $\delta_{qs}$, $\delta_{qgv}^{u}$, $\delta_{qgs}^{u}$, $\delta_{qgv}^{gl}$, $\delta_{qgs}^{gl}$, $\delta_{qgv}^{q}$ and $\delta_{qgs}^{q}$. Where the exponents without a tensor specification are used, they stand for both Dirac structures so that terms with all possible combinations arise after decomposition. Finally, we define the canonical exponent of the quark propagator $\chi_q$ which is given by $\chi_{qs}=0$ and 
$\chi_{qv}=1/2$ for the two cases, respectively. 

\begin{figure}[h]
\centerline{\epsfig{file=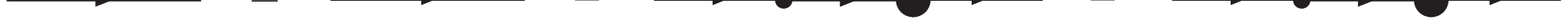,width=12cm}}
\caption{\label{fig:qv}Decomposed DSE for the vector part of the quark 
propagator. Vector and scalar components are denoted by $V$ and $S$, respectively.}
\end{figure}

\begin{figure}[h]
\centerline{\epsfig{file=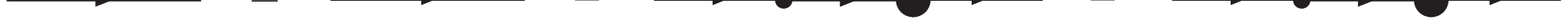,width=12cm}}
\caption{\label{fig:qs}Decomposed DSE for the scalar part of the quark 
propagator}
\end{figure}

The distinction of the tensor components is already important for the 
quark 
propagator, where the different components feature a different scaling already at the tree level. In order to perform the power counting analysis for the DSE of the individual components of the quark propagator we first have to consider the scaling behavior of the appearing inverse of the dressed propagator which is given for the parameterization eq. (\ref{qq}) by
\begin{equation}
S^{-1}\left(p\right)=\left(i \sigma_{v}\not\! p+\sigma_{s}\right)^{-1}=i\frac{\sigma_{v}}{\sigma_{v}^{2}p^{2}+\sigma_{s}^{2}}\not\! p-\frac{\sigma_{s}}{\sigma_{v}^{2}p^{2}+\sigma_{s}^{2}} \to i\left(p^{2}\right)^{\delta_{qv}-\min\left( 2\delta_{qv}+1,2\delta_{qs}\right)}\not\! p-\left(p^{2}\right)^{\delta_{qs}-\min\left( 2\delta_{qv}+1,2\delta_{qs}\right)}\, .
 \end{equation} 
The DSE for the vector  part of the
quark propagator is given in fig. \ref{fig:qv}, where the labels $V$ and $S$ next to the propagator and vertices denote vector and scalar components, respectively.
This DSE leads to an equation for the corresponding IR exponent
\begin{equation}
\label{eq:v-prop}
\delta_{qv}+\frac{1}{2}-\min\left( 2\delta_{qv}+1,2\delta_{qs}\right)=\min\left(\frac{1}{2},\delta_{qgv}^{u}+\delta_{qv}+\frac{3}{2}+
2\kappa,\delta_{qgs}^{u}+\delta_{qs}+1+2\kappa,\delta_{qgv}^{q}+\frac{1}{2},
\delta_{qgs}^{q}+\frac{1}{2} \right) \, .
\end{equation}
Here it has been taken into account that due to a possible quark mass the loop integral can in addition to the IR regime also be dominated by hard modes reflected by the last two terms, where the $\delta_{qg}^{q}$ arises due to the possibility of a soft quark divergence of the vertex and the $1/2$ due to a further suppression arising from cancellations owing to gluon transversality.
The corresponding DSE for the
scalar part of the propagator is shown in fig. \ref{fig:qs}
which leads to
\begin{equation}
\label{eq:s-prop}
\delta_{qs}-\min\left( 2\delta_{qv}+1,2\delta_{qs}\right)=\min\left(0,\delta_{qgs}^{u}+\delta_{qv}+ 
\frac{3}{2}+2\kappa,\delta_{qgv}^{u}+\delta_{qs}+1+2\kappa, 
\delta_{qgs}^{q},\delta_{qgv}^{q}\right) \, .
\end{equation}
The DSE for the quark-gluon vertex in a skeleton expansion was given in fig. \ref{vertex-DSE}.
When hard scales are present that do not scale to zero there are different regions of the loop integration that can be IR sensitive. In the uniform case there are two possibilities corresponding to a soft or hard loop momentum respectively. In the soft-particle limits different propagators in the graphs can involve both soft and hard momenta and there is one additional IR sensitive region of the loop integral given by a narrow window of hard modes. This corresponds to a second, not equivalent way to route the large momentum through the loop. When assessing the counting
of the vertex, each of these regions could dominate and determine the scaling of the corresponding Green«s function on the left hand side. 
Employing the IR exponents from the Yang-Mills sector in table \ref{tab:IR-scaling-YM} yields the equations for the IR-exponents of the quark-gluon vertex in the different limits
\begin{align}
\delta_{qg}^{u}=&\min \left( 0; 2\delta_{qg}^{u}\!+\!\delta_{q}\!+\!\chi_{q}\!+\!\frac{1}{2}\!+\!\kappa, 2\delta_{qg}^{q}\!+\!1\!-\!2\kappa; 2\delta_{qg}^{u}\!+\!\delta_{q}\!+\!\chi_{q}\!+\!\frac{1}{2}\!+\!4\kappa, 2\delta_{qg}^{q} ; 2\delta_{qg}^{u}\!+\!2\delta_{q}\!+\!2\chi_{q}\!+\!1\!+\!2\kappa ,2\delta_{qg}^{q} \right) \, , \\
\delta_{qg}^{gl}=&\min \left( 0; 2\delta_{qg}^{gl}\!+\!\frac{1}{2}\!+\!\kappa , 2\delta_{qg}^{q}\!+\!\delta_{q}\!+\!\chi_{q}\!+\!3\!-\!2\kappa , 1\!-\!2\kappa; 2\delta_{qg}^{gl}\!+\!\frac{1}{2}\!+\!4\kappa , 2\delta_{qg}^{q}\!+\!\delta_{q}\!+\!\chi_{q}\!+\!2,0; \right. \nonumber \\
&\left. \qquad\qquad 2\delta_{qg}^{gl}\!+\!1\!+\!2\kappa, 2\delta_{qg}^{q}\!+\!2\delta_{q}\!+\!2\chi_{q}\!+\!2, 0 \right) \, , \\
\delta_{qg}^{q}=&  \min\left(0;\delta_{qg}^{u}\!+\!\delta_{qg}^{q}\!+\!\delta_{q}\!+\!\chi_q\!+\!2,\delta_{qg}^{gl}\!+\!\delta_{qg}^{q}\!+\!2,\delta_{qg}^{q};\delta_{qg}^{u}\!+\!\delta_{qg}^{q}\!+\!\delta_{q}\!+\!\chi_q\!+\!1\!+\!2\kappa,\delta_{qg}^{gl}\!+\!\delta_{qg}^{q}\!+\!1\!+\!2\kappa,\delta_{qg}^{q}; \right. \nonumber \\ 
&\left. \qquad \qquad 2\delta_{qg}^{q}\!+\!\delta_{q}\!+\!\chi_{q}\!+\!2,\delta_{qg}^{u}\!+\!\delta_{qg}^{gl}\!+\!\delta_{q}\!+\!\chi_{q}\!+\!1\!+\!2\kappa,\delta_{qg}^{q}\right) \, .
\end{align}
where semicolons separate the contributions from the first four graphs in fig. \ref{vertex-DSE} and commas those from different regions of the same loop integration. Since the ghost loop contribution in the first diagram in fig. \ref{vertex-DSE} is IR divergent it is immediately clear, and reflected in the power counting above, that the second (non-Abelian) graph is subleading. 

Although the remaining system of equations for the quark sector looks extremely 
involved, it is largely simplified by powerful constraints arising from it. In particular, the non-linear terms on the right hand sides of the vertex equations yield the constraints
\begin{equation}
\delta_{qg}^{u} +\delta_{q}+\chi_{q}+\frac{1}{2}+\kappa \geq 0 \quad , \quad \delta_{qg}^{gl} +\frac{1}{2}+\kappa \geq 0 \quad , \quad \delta_{qg}^{q}+\delta_q+\chi_q+2 \geq 0 \; , \label{constraints}
\end{equation} 
Using these constraints in the above equations obtained from the DSEs as well as the known possible range for $\kappa$ \cite{Alkofer:2008jy} they precisely cancel the non-linear contributions from the Abelian graph in the uniform and the soft-gluon limit as well as all non-trivial contributions in the soft-quark limit and the propagator equations. Dropping the distinction between contributions from the different graphs, the remaining system reads
\begin{align}
\delta_{qv}-\min\left( 2\delta_{qv}+1,2\delta_{qs}\right)=&\min\left(0,\delta_{qgv}^{q}, \delta_{qgs}^{q} \right) \, , \\
\delta_{qs}-\min\left( 2\delta_{qv}+1,2\delta_{qs}\right)=&\min\left(0, \delta_{qgs}^{q},\delta_{qgv}^{q}\right) \, , \\
\delta_{qg}^{u}=&\min \left( 0, 2\delta_{qg}^{u}\!+\!\delta_{q}\!+\!\chi_{q}\!+\!\frac{1}{2}\!+\!\kappa, 2\delta_{qg}^{q}\!+\!1\!-\!2\kappa \right) \, , \label{eq:qguni} \\
\delta_{qg}^{gl}=&\min \left( 2\delta_{qg}^{gl}\!+\!\frac{1}{2}\!+\!\kappa , 2\delta_{qg}^{q}\!+\!\delta_{q}\!+\!\chi_{q}\!+\!3\!-\!2\kappa, 2\delta_{qg}^{q}\!+\!2\delta_{q}\!+\!2\chi_{q}\!+\!2,1\!-\!2\kappa \right) \, , \label{eq:qgsg} \\
\delta_{qg}^{q}=&  \min\left(0,\delta_{qg}^{q}\right) \, .
\end{align}

To solve the system let us start with the equation 
for the soft-quark limit of the quark-gluon vertex. 
\begin{comment}
The last entry in 
the equation shows that the corresponding IR exponents for the scalar and 
vector part have to be identical. Moreover, there are constraints from the other equations
\begin{equation}
\delta_{qg}^{q} \geq -\delta_{qs}\quad, \quad \delta_{qg}^{u} \geq -\delta_{q}-\chi_{q}-\frac{1}{2}-\kappa \quad , \quad \delta_{qg}^{gl} \geq -\frac{1}{2}-\kappa \; ,
\end{equation}
that cancel all non-trivial contributions and the equation becomes trivial
\begin{equation}
\delta_{qg}^{q}=  \min\left(0,\delta_{qg}^{q}\right) \, .
\end{equation}
\end{comment}
From a purely mathematical point of view, the second element in the minimum function yields a condition which is trivially fulfilled for any $\delta_{qg}^{q}<0$ that is compatible with all constraints from the other equations. However, since the perturbative vertex does not contain a soft-quark divergence and there is no dynamical contribution that could induce a singularity we conclude that $\delta_{qg}^{q}=0$ is the only possible physical solution. 

Next consider the equations for the propagators. Due to the trivial vertex in the soft-quark limit they reduce to
\begin{equation}
\delta_{qv}=\min\left( 2\delta_{qv}+1,2\delta_{qs}\right) \quad , \quad \delta_{qs}=\min\left( 2\delta_{qv}+1,2\delta_{qs}\right)
\end{equation}
It is easy to see that this system only has the trivial solution $\delta_{q}=\delta_{qv}=\delta_{qs}=0$.
Correspondingly the self-consistent solution has the same structure as the bare propagator with a mass term that suppresses IR propagation. This mass is naturally expected to be enhanced by spontaneous symmetry breaking.

As can be seen form eqs. (\ref{eq:qguni}) and (\ref{eq:qgsg}) the vertex equation in the soft-gluon limit is totally decoupled from the equation in the uniform limit. Therefore, it 
is possible to solve the soft-gluon equation first which is considerably simplified by the trivial solution for the propagators and the vertex in the soft-quark limit. Due to the absence of propagators in the equation there are no restrictions from the Dirac structure in this case. Thereby the exponents for the scalar and vector part are equal and it is not necessary to decompose the equation into different tensor components. The remaining equation reads 
\begin{align}
\delta_{qg}^{gl}&=  \min\left(2\delta_{qg}^{gl}\!+\!\frac{1}{2}\!+\!\kappa,1\!-\!2\kappa\right)\end{align}
and has the two solutions
\begin{align}
\delta_{qg}^{gl}&=-\frac{1}{2}-\kappa \quad \vee \quad \delta_{qg}^{gl}=1-2\kappa
\end{align}
The first one is a strong IR divergence that is self-consistently 
enhanced by the quark dynamics and which precisely saturates the 
above bound, whereas the second one is the mild singularity fed in 
from the corresponding soft singularities in the gauge sector.

\begin{figure}[h]
\centerline{\epsfig{file=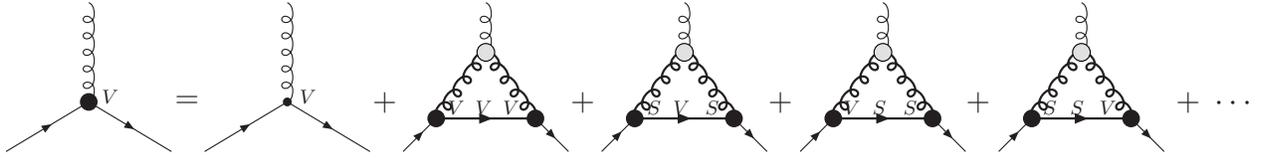,width=17cm}}
\caption{\label{fig:qgv} The vector part in the decomposition of the IR-leading order of the skeleton expansion of the quark-gluon vertex. The open blob represents the ghost triangle in fig. \ref{vertex-DSE} that represents the IR-leading contribution to the 3-gluon vertex.}
\end{figure}

\begin{figure}[h]
\centerline{\epsfig{file=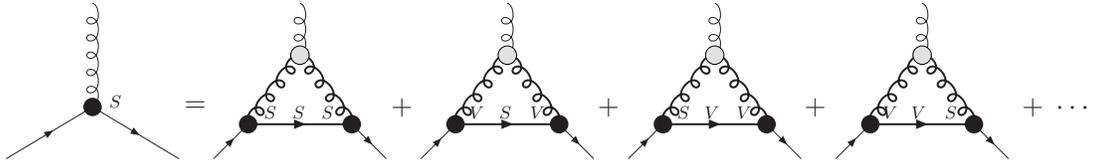,width=14.8cm}}
\caption{\label{fig:qgs}The scalar part in the decomposition of the IR-leading order of the skeleton expansion of the quark-gluon vertex.}
\end{figure}

As has been noted before in the main text, due to the different scaling of the two components of the propagators, the remaining uniform vertex equation has to be decomposed into two equations for scalar and vector components. This decomposition is performed for the remaining IR leading diagram in figs. \ref{fig:qgv} and \ref{fig:qgs}.

Correspondingly, the remaining system for the components of the vertex in the uniform limit is given by
\begin{align}
\delta_{qgv}^{u}&=  \min\left( 2\delta_{qgv}^{u}\!+\!\delta_{qv}\!+\!1\!+\!\kappa,2\delta_{qgs}^{u}\!+\!\delta_{qv}\!+\!1\!+\!\kappa,\delta_{qgv}^{u}\!+\!\delta_{qgs}^{u}\!+\!\delta_{qs}\!+\!\frac{1}{2}\!+\!\kappa,1\!-\!2\kappa\right)\\
\delta_{qgs}^{u}&=  \min\left(2\delta_{qgs}^{u}\!+\!\delta_{qs}\!+\!\frac{1}{2}\!+\!\kappa,2\delta_{qgv}^{u}\!+\!\delta_{qs}\!+\!\frac{1}{2}\!+\!\kappa,\delta_{qgv}^{u}\!+\!\delta_{qgs}^{u}\!+\!\delta_{qv}\!+\!1\!+\!\kappa,1\!-\!2\kappa\right)
\end{align}
Aside from the last terms in these equations arising from the mild soft gluon divergence of the 3-gluon vertex, this is the system that has been discussed in detail in section \ref{irquarks} and it is not affected by the soft singularities. As in the case of the soft-gluon singularity discussed above, these latter terms in principle also allow for the possibility of a weaker singularity $\delta_{qg}^{u}\!=\!-\!1/2\!-\!\kappa$ that is not self-consistently enhanced but only induced from the gauge sector. However, our numerical results presented in sec. \ref{sec:numsec} clearly favor the strong divergences as a solution of the full DSE system.

So far we have distinguished between the scalar and vector components
of quark correlation functions since this was crucial to obtain the
correct IR exponents that properly take into account the Dirac structure.
This distinction is surely also important to assess the chiral properties
of the above solution and we recall once more that the scalar part is crucial. 
Aside from this, however, now that the solutions have 
been found these are just different parts of the same Green's function.
Thus analogously to the power counting in the gauge sector where only
the IR exponents of the leading tensor structure(s) are considered, the solutions for
the IR-leading behavior of the quark sector of quenched QCD is given in 
Table \ref{tab:IR-scaling-qQCD}.

\begin{table}[h]
\begin{tabular}{|c|c|c|c|}
\hline 
$\delta_{q}$ & $\delta_{qg}^{u}$ & $\delta_{qg}^{q}$ & $\delta_{qg}^{gl}$\tabularnewline
\hline 
$0$ & $-\frac{1}{2}-\kappa\;\vee\;1\!-\!2\kappa$ & $0$ & $-\frac{1}{2}-\kappa\;\vee\;1\!-\!2\kappa$ \tabularnewline
\hline
\end{tabular}
\caption{\label{tab:IR-scaling-qQCD}The IR-exponents for the leading 
Green's
functions of the possible IR fixed points of the quark sector of quenched QCD. The scalar Dirac parts are of leading order in the IR.}
\end{table}

%%%%%%%%%%%%%%%%%%%%%%%%%%%%%%%%%%%%%%%%%%%%%%%%%%%%%%%%%%%%%%%%%%%%%
\section{Dirac Algebra \label{app:dirac}}
%%%%%%%%%%%%%%%%%%%%%%%%%%%%%%%%%%%%%%%%%%%%%%%%%%%%%%%%%%%%%%%%%%%%
Skipping for now the
color index to  concentrate on the Dirac space indices, the
vertex in eq.  (\ref{spelledNAvertex}) can be
decomposed with the help of Lorentz invariance
and the sign convention in eq. (\ref{newdec}) as
\ba \label{vertexdec} \nonumber
\Gamma^\mu_{\rm NA} =  \Gamma_1 p_1^\mu + \Gamma_2 p_2^\mu
+\Gamma_3 \gamma^\mu + \Gamma_4 \not p_1 p_1^\mu +\Gamma_5 \not p_1 p_2^\mu
+\Gamma_6 \not p_2 p_1^\mu + \Gamma_7 \not p_2 p_2^\mu +
  \\
\Gamma_8 \not p_1 \gamma^\mu
+\Gamma_9 \not p_2 \gamma^\mu + \Gamma_{10} \not p_1 \not p_2 p_1^\mu +
\Gamma_{11} \not p_1 \not p_2p_2^\mu + \Gamma_{12} \not p_1 \not p_2 
\gamma_\mu
\ea

Our computer codes will calculate the vertex as a Dirac matrix each of whose
elements is in turn an integral over momentum space. Once this matrix is
obtained we need to project out the various structures in eq. 
(\ref{vertexdec}).
Now we briefly describe the algorithm that achieves this. First choose
(without loss of generality) a coordinate frame in which
$p_1=(p_1,0,0,0)$ and $p_2=(p_2 \cos\chi,0,0,p_2 \sin \chi)$.
Then pick a vector $R$ of unit norm $R\cd R=1$ along the axis 1,2, 4
alternatively and numerically compute $R\cd \Gamma$.
The possible structures appearing in this projected vertex are now:
\be
R\cd \ov{\Gamma}_{\rm NA}= \ov{\Gamma}_1 {\bf 1} + \ov{\Gamma}_2 \not R +
\ov{\Gamma}_3 \not p_1 + \ov{\Gamma}_4 \not p_2 + \ov{\Gamma}_5 \not p_1 \not 
R
+\ov{\Gamma}_6 \not p_2 \not R + \ov{\Gamma}_7 \not p_1 \not p_2 +
\ov{\Gamma}_8  \not p_1 \not p_2 \not R \ ,
\ee
where $R\cd \ov{\Gamma}_{\rm NA}$ takes different values after projection  
and subtraction in the following three step algorithm, like in eq.
(\ref{step2},\ref{step3}) below,
and the various $\ov{\Gamma}_i$ are functions of the
(Euclidean) scalars $(p1^2,p2^2,p1\cd p2)$.

They can easily be obtained by tracing the vertex $R\cd \ov{\Gamma}$ with
various projectors:
\ba \! \!
\left[
\begin{tabular}{c}
$\! \ov{\Gamma}_1\! $ \\
$\! \ov{\Gamma}_5\! $ \\
$\! \ov{\Gamma}_6\! $ \\
$\! \ov{\Gamma}_7\! $ \\
\end{tabular}
\right]\!  =\!  \frac{1}{4\Delta} \! \! \left[
\begin{tabular}{cccc}
$p_1^2 p_2^2$ & $( 2 p_1\cd p_2 R\cd p_2\! -\! p_2^2 R\cd p_1)$ & $-p_1^2
R\cd p_2$ & $-p_1\cd p_2$   \\
$( 2 p_1\cd p_2 R\cd p_2 \! - \! p_2^2 R\cd p_1)$ &
$p_2^2$ & $-p_1\cd p_2$ & $-R\cd p_2$ \\
$-p_1^2 R\cd p_2$ & $-p_1\cd p_2$ & $p_1^2$ & $R\cd p_1$ \\
$-p_1\cd p_2$ & $-R\cd p_2$ & $R\cd p_1$  & $1$
\end{tabular} \right] \! \!
\left[
\begin{tabular}{c}
$\! {\rm Tr}(R\cd \ov{\Gamma}_{NA})\! $ \\
$\! {\rm Tr}(\not R \! \not p_1 R\cd \ov{\Gamma}_{NA}) \! $ \\
$\! {\rm Tr}(\not R \! \not p_2 R\cd \ov{\Gamma}_{NA})\! $ \\
$\! {\rm Tr}(\not p_2 \! \not p_1 R\cd \ov{\Gamma}_{NA})\! $
\end{tabular}
\right] \nonumber \\ \nonumber \\ \nonumber
\! \! 
\left[
\begin{tabular}{c}
$\ov{\Gamma}_2$ \\
$\ov{\Gamma}_3$ \\
$\ov{\Gamma}_4$ \\
$\ov{\Gamma}_8$ \\
\end{tabular}
\right] = \frac{1}{4\Delta}\left[
\begin{tabular}{cccc}
$p_1^2 p_2^2$ & $( 2 p_1\cd p_2 R\cd p_2 \! -\! p_2^2 R\cd p_1)$ &
$-p_1^2 R\cd p_2$ & $-p_1\cd p_2$  \\
$( 2 p_1\cd p_2 R\cd p_2\! -\! p_2^2 R\cd p_1)$ &
$p_2^2$ & $-p_1\cd p_2$ & $-R\cd p_2$ \\
$-p_1^2 R\cd p_2$ & $-p_1\cd p_2$ & $p_1^2$ & $R\cd p_1$ \\
$-p_1\cd p_2$ & $-R\cd p_2$ & $R\cd p_1$  & $1$
\end{tabular} \right] \! \!
\left[
\begin{tabular}{c}
${\rm Tr}(\not R R\cd \ov{\Gamma}_{NA})$ \\
${\rm Tr}(\not p_1 R\cd \ov{\Gamma}_{NA})$ \\
${\rm Tr}(\not p_2 R \cd \ov{\Gamma}_{NA})$ \\
$\! {\rm Tr}(\! \not R \! \not p_2 \! \not p_1 \! R \cd
\ov{\Gamma}_{NA})\! $
\end{tabular}
\right]
\ea
with
\be \label{detsystem}
\Delta =  p_1^2p_2^2-(p_1\cd p_2)^2+ 2p_1\cd p_2 R\cd p_1 R\cd p_2
-p_1^2 (R\cd p_2)^2 - p_2^2 (R\cd p_1)^2
\ee
In case the vector $R$ is in the $p_1$, $p_2$ plane, the system
degenerates. Then the expansion
\be
R\cd \ov{\Gamma} = \ov{\Gamma}_1 {\bf 1} + \ov{\Gamma}_3 \not p_1 +
\ov{\Gamma}_4 \not p_2 + \ov{\Gamma}_7 \not p_1 \not p_2
\ee
suffices, and the components $\ov{\Gamma}_i$ can be read off
\ba
\left[
\begin{tabular}{c}
$\ov{\Gamma}_1$ \\
$\ov{\Gamma}_7$ \\
\end{tabular}
\right] = \frac{1}{4(p_1^2 p_2^2 -(p_1\cd p_2)^2)}\left[
\begin{tabular}{cc}
$p_1^2 p_2^2$  & $-p_1\cd p_2$ \\
$-p_1 \cd p_2$ & $1$
\end{tabular} \right] \! \!
\left[
\begin{tabular}{c}
${\rm Tr}(R\cd \ov{\Gamma}_{NA})$ \\
${\rm Tr}(\not p_2 \not p_1 R\cd \ov{\Gamma}_{NA})$
\end{tabular}
\right]
\\ \nonumber
\left[
\begin{tabular}{c}
$\ov{\Gamma}_3$ \\
$\ov{\Gamma}_4$ \\
\end{tabular}
\right] = \frac{1}{4(p_1^2 p_2^2 -(p_1\cd p_2)^2)}\left[
\begin{tabular}{cc}
$p_2^2$  & $-p_1\cd p_2$ \\
$-p_1 \cd p_2$ & $p_1^2$
\end{tabular} \right] \! \!
\left[
\begin{tabular}{c}
${\rm Tr}(\not p_1 R\cd \ov{\Gamma}_{NA})$ \\
${\rm Tr}(\not p_2 R\cd \ov{\Gamma}_{NA})$
\end{tabular}
\right]
\ea

If one chooses first a vector along the 2 (or 3) axis, and employs
the full projection above, the combinations $\Gamma_3$,  $\Gamma_8$,
 $\Gamma_9$,  $\Gamma_{12}$ can be separated.
Then choosing $R$ along the fourth axis picks up in addition the
$p_2^\mu$ contributions, and projecting now
\begin{equation} \label{step2}
R\cd \ov{\Gamma}=\Gamma^4 - \Gamma_3 \gamma^4 - \Gamma_8 \not p_1 \gamma^4
-
\Gamma_9 \not p_2 \gamma^4- \Gamma_{12} \not p_1 \not p_2 \gamma^4
\end{equation}
we can reconstruct $\Gamma_2$,  $\Gamma_5$, $\Gamma_7$, $\Gamma_{11}$.
Once these are known we can calculate along the first axis
to pick up the $p_1^\mu$ contributions extracted from
\ba \label{step3}
R\cd \ov{\Gamma}=\Gamma^1 -\Gamma_2 p_2^1 -\Gamma_3 \gamma^1 - \Gamma_5
\not p_1 p_2^1
-\Gamma_7 \not p_2 p_2^1 -\Gamma_8 \not p_1 \gamma^1 \\ \nonumber -
\Gamma_9 \not p_2 \gamma^1- \Gamma_{11} \not p_1 \not p_2 p_2^1
-\Gamma_{12} \not p_1 \not p_2 \gamma^1   
\ea
which finally allows to calculate the remaining
$\Gamma_1$, $\Gamma_4$, $\Gamma_6$, $\Gamma_{10}$.

%%%%%%%%%%%%%%%%%%%%%%%%%%%%%%%%%%%%%%%%%%%%%%%%%%%%%%%%%%%%%%%%%%%%%55
\subsection{Degenerate points $p_1=p_2$}
%%%%%%%%%%%%%%%%%%%%%%%%%%%%%%%%%%%%%%%%%%%%%%%%%%%%%%%%%%%%%%%%%%%%%%
The kinematical limit $p_1=p_2:=p$, which corresponds to the infrared limit of the
soft gluon section, cf. \ref{kin}, requires special attention.
Here $\Delta$ defined in eq. (\ref{detsystem})
vanishes and the linear system becomes ill-defined: this can be tracked to
a reduction of the linear space spanned by the vertex, now
\ba \label{symmetricnaive}
\Gamma=\Gamma_1^{sym} p^\mu + \Gamma_3^{sym} \gamma_\mu
+ \Gamma_4^{sym} \not p p^\mu + \Gamma_8^{sym} \not p \gamma^\mu \\
R\cd \ov{\Gamma} = \ov{\Gamma}_1^{sym} {\mathbf I}+
 \ov{\Gamma}_2^{sym} \not R + \ov{\Gamma}_3^{sym} \not p +
 \ov{\Gamma}_5^{sym} \not p \not R \ .
\ea
The general linear system now can be easily solved:
\be \label{degsystem}
\left[
\begin{tabular}{c}
$\ov{\Gamma}_1^{sym}$ \\
$\ov{\Gamma}_5^{sym}$ \\
$\ov{\Gamma}_2^{sym}$ \\
$\ov{\Gamma}_3^{sym}$ \\
\end{tabular}
\right]
=\frac{1}{4(p^2-(R\cd p)^2)} \left[
\begin{tabular}{cccc}
$p^2$      & $-R\cd p$ & $0$ & $0$ \\
$-R \cd p$ & $1$       & $0$ & $0$ \\
$0$ & $0$  & $p^2$      & $-R\cd p$\\
$0$ & $0$  & $-R \cd p$ & $1$
\end{tabular}
\right] \left[
\begin{tabular}{c}
${\rm Tr}(R\cd \ov{\Gamma}) $\\
${\rm Tr}(\not R \not p R \cd \ov{\Gamma}) $\\
${\rm Tr}(\not R R \cd \ov{\Gamma}) $\\
${\rm Tr}(\not p R \cd \ov{\Gamma}) $ 
\end{tabular}
\right]
\ee
To obtain the various projected functions the algorithm is in all
similar to the general case. With $p$ along the 1 axis, choose R along
the 2 axis ($R\cd \ov{\Gamma}=\Gamma^2$) and solve for the barred
$\ov{\Gamma}$ by means of eq.
(\ref{degsystem}), then equating
$$
\Gamma_3^{sym} = \ov{\Gamma}_2^{sym} \ \ \
\Gamma_8^{sym} = \ov{\Gamma}_5^{sym} \ .
$$
Next choose $R$ along the $p$ (first in our kinematics) axis and employ
\be
R\cd \ov{\Gamma}=\Gamma^1-\Gamma_3^{sym}\gamma^1-\Gamma_8^{sym}
\not p \gamma^1
\ee
after solving
\be
\ov{\Gamma}_1^{sym} = \frac{1}{4} {\rm Tr} (R\cd \ov{\Gamma})\ \ \
\ov{\Gamma}_3^{sym} = \frac{1}{4p^2} {\rm Tr}
(\not p R\cd \ov{\Gamma})
\ee
the last needed terms can be extracted
\be
\Gamma_1^{sym}=\ov{\Gamma}_1^{sym}/p \ \ \
\Gamma_4^{sym}=\ov{\Gamma}_3^{sym}/p
\ee
%%%%%%%%%%%%%%%%%%%%%%%%%%%%%%%%%%%%%%%%%%%%%%%%%%%%%%%%%%%%%%%%
\subsection{Transformation to the Ball-Chiu basis.}
%%%%%%%%%%%%%%%%%%%%%%%%%%%%%%%%%%%%%%%%%%%%%%%%%%%%%%%%%%%%%%%%

If we denote collectively the various $L_i$ and $T_i$ by $H_i$, and the
tensors in Dirac space with one index $\mu$ that appear in eq.
(\ref{vertexdec}) above by $\ov{H}_i$ then, since the vertex accepts a
decomposition in both basis, we have
\be \label{twodec}
\Gamma = h_i H_i = \Gamma_i \ov{H}_i \ .
\ee
We wish to obtain the new coefficient functions $h_i$ (split into the four
WT functions $\lambda_1\ ... \ \lambda_4$ and the eight transverse
functions $\tau_1\ ... \ \tau_8$) in terms of our calculated $\Gamma_i$
$i\subset (1,12)$.
This is given in terms of a linear transformation (not unitary since
neither basis is orthonormal)
$$
h_i=A_{ij} \Gamma_j \ .
$$
Observing eq. (\ref{twodec}) it is obvious that
$$
H_i=A^{-1\ {\rm T}}_{\ \ i j} \ov{H}_j
$$
but the matrix $A^{-1\ {\rm T}}_{\ \ i j}$ is simple to obtain by
directly expanding  eq. (\ref{newdec}) and passing $\gamma_\mu$ to
the right to compare with eq. (\ref{vertexdec}). The resulting matrix
has two separate blocks, one connecting the terms with an odd number
of $\gamma$ matrices between themselves, another with the even terms.  
Inverting it we immediately find
\ba \! \!
\left[
\begin{tabular}{c}
$\! \lambda_1 \! $ \\
$\! \lambda_2 \! $ \\
$\! \tau_2    \! $ \\
$\! \tau_3    \! $ \\
$\! \tau_6    \! $ \\
$\! \tau_8    \! $
\end{tabular}
\right]\!  &=&\!  \frac{1}{4} \! \! \left[
\begin{tabular}{cccccc}
$4$ & $-2 p_1\cd p_3 $ &  $-2 p_2\cd p_3$&  $2 p_1\cd p_3$&
 $2 p_2\cd p_3$& $2(2p_1^2-p_3^2) $ \\
$0$ &  $\frac{-2 p_1\cd p_3}{\Delta} $ & $\frac{-2 p_2\cd p_3}{\Delta} $ &
$\frac{-2 p_1\cd p_3}{\Delta} $ &  $\frac{-2 p_2\cd p_3}{\Delta} $ &
$\frac{-2p_3^2}{\Delta}$ \\
$0$ & $\frac{2}{\Delta}$ & $\frac{-2}{\Delta}$ &  $\frac{2}{\Delta}$ &
 $\frac{-2}{\Delta}$ &  $\frac{-4}{\Delta}$ \\
$0$ & $1$ & $-1$ &  $-1$ & $1$ & $0$ \\
$0$ & $-1$ & $-1$ &  $1$ & $1$ & $-2$ \\
$0$ & $0$ & $0$ &  $0$ & $0$ & $4$
\end{tabular} \right] \! \!
\left[
\begin{tabular}{c}
$\! \Gamma_3 \! $ \\
$\! \Gamma_4 \! $ \\
$\! \Gamma_5 \! $ \\ 
$\! \Gamma_6 \! $ \\ 
$\! \Gamma_7 \! $ \\ 
$\! \Gamma_{12} \! $ 
\end{tabular} \right] \\
\left[
\begin{tabular}{c}
$\! \lambda_3 \! $ \\
$\! \lambda_4 \! $ \\  
$\! \tau_1    \! $ \\
$\! \tau_4    \! $ \\
$\! \tau_5    \! $ \\
$\! \tau_7    \! $
\end{tabular}
\right]\!  &=& \!  \frac{i}{4} \! \! \left[
\begin{tabular}{cccccc}
$\frac{4p_1\cd p_3}{\Delta}$ & $\frac{4p_2\cd p_3}{\Delta}$ &
$\frac{4p_1\cd p_3}{\Delta}$ & $\frac{4p_2\cd p_3}{\Delta}$ &
$\frac{4p_1\cd p_2 p_1\cd p_3}{\Delta}$ &
$\frac{4p_1\cd p_2 p_2\cd p_3}{\Delta}$
\\
$0$ & $0$ & $-2$ & $-2$ & $-2p_1\cd p_3$ & $-2p_2\cd p_3$
\\
$\frac{-4}{\Delta}$ & $\frac{4}{\Delta}$ & $\frac{-4}{\Delta}$
& $\frac{4}{\Delta}$ & $\frac{-4p_1\cd p_2}{\Delta}$ &
$\frac{4p_1\cd p_2}{\Delta}$ \\
$0$ & $0$ & $0$ & $0$ & $-1$ & $1$ \\
$0$ & $0$ & $-2$ & $2$ & $0$ & $0$ \\
$0$ & $0$ & $0$ & $0$ & $-2$ & $-2$ \\
\end{tabular} \right] \! \!
\left[
\begin{tabular}{c}   
$\! \Gamma_1 \! $ \\   
$\! \Gamma_2 \! $ \\ 
$\! \Gamma_8 \! $ \\ 
$\! \Gamma_9 \! $ \\ 
$\! \Gamma_{10} \! $ \\
$\! \Gamma_{11} \! $
\end{tabular} \right]\\
\ea
with $\Delta=(p_2^2-p_1^2)$.
We have to keep in mind that there is a singular kinematical limit. 
When $p_1=p_2:=p$ the vertex expansion in the naive basis
eq. (\ref{vertexdec}) degenerates to
\be
(\Gamma_1+\Gamma_2+p^2(\Gamma_{10}+\Gamma_{11})) p^\mu +
(\Gamma_3+p^2\Gamma_{12})\gamma^\mu +
(\Gamma_4+\Gamma_5+\Gamma_6+\Gamma_7)\not p p^\mu +   
(\Gamma_8 + \Gamma_9)\not p \gamma^\mu
\ee
and since all the transverse tensors vanish $T_i(p1=p2)=0$
the decomposition in the basis of eq. (\ref{newdec}) referred to
the naive basis in eq. (\ref{symmetricnaive}) reduces to
\be   
\left[ \begin{tabular}{c}
$\lambda_1$  \\ $\lambda_2$ \\ $\lambda_3$  \\ $\lambda_4$
\end{tabular} \right] =
\left[ \begin{tabular}{cccc}
$1$ & $0$ & $0$ & $0$ \\
$0$ & $-\frac{1}{4}$ & $0$ & $0$ \\
$0$ & $0$ & $\frac{i}{2}$ &  $\frac{i}{2}$ \\
 $0$ & $0$ & $0$ & $\frac{-i}{2}$
\end{tabular} \right] \left[ \begin{tabular}{c}
$\Gamma_3^{\rm sym}$ \\
$\Gamma_4^{\rm sym}$ \\
$\Gamma_1^{\rm sym}$ \\
$\Gamma_8^{\rm sym}$
\end{tabular}\right]
\ee

%%%%%%%%%%%%%%%%%%%%%%%%%%%%%%%%%%%%%%%%%%%%%%%%%%%%%%%%%%%%%%%%%%%%%%
\section{Diagrammatic construction of the quark-quark scattering kernel \label{app:scattering}}
%%%%%%%%%%%%%%%%%%%%%%%%%%%%%%%%%%%%%%%%%%%%%%%%%%%%%%%%%%%%%%%%%%%%%%%%%%%%%%

While one does not need to be particularly careful when constructing a
truncation of the heavy-quark-heavy-quark scattering kernel, the chiral Ward
identities need to be respected if one is to extend such model to light quarks. 
This implies an intimate relation between the scattering kernel and the gap
equation and forbids the naive skeleton expansion used in the previous subsection.

A systematic recipe to construct a scattering kernel that is consistent with
a given quark propagator, respecting the chiral Ward identities, has been
given by Munczek \cite{Munczek:1994zz}. The basic idea is to take functional 
derivatives of the quark self-energy with respect to the quark propagator.
This procedure can be iterated for any type of semi-perturbative or
diagrammatic construction of the quark-gluon vertex by cutting all
additional quark lines in the contributing diagrams. 

In the most general case, i.e. with an arbitrary vertex, Munczek's construction
would entail cutting the selfenergy diagrams
\begin{center}
  \begin{picture}(253,45) (34,-20)
    \SetScale{0.6}
    \SetWidth{1.5}
    \Line(76,7)(56,-15)
    \Line(82,7)(62,-15)
    \put(90,0){\large{$\mathbf{+}$}}
    \ArrowLine(26,-6)(136,-6)
    \GlueArc(80.4,2.14)(35.14,-10,191.72){7.5}{7.41}
    \Vertex(120,-4){9.85}
    \ArrowLine(183,-7)(293,-7)
    \GlueArc(237.4,1.14)(35.14,-10,191.72){7.5}{7.41}
    \Vertex(277,-5){9.85}
    \Line(284,2)(264,-20)
    \Line(288,1)(268,-21)
  \end{picture} \ .
\end{center}
This would naturally lead to a Dyson-Schwinger equation for the connected 
quark-quark scattering kernel
\begin{equation} \label{DSEqqkernel}
  \begin{picture}(345,55) (3,3)
    \SetScale{0.6}
    \SetWidth{1.5}
    \GBox(25,11)(54,67){0.882}
    \ArrowLine(4,57)(24,58)
    \ArrowLine(3,20)(23,21)
    \ArrowLine(55,57)(75,58)
    \ArrowLine(54,20)(74,21)
    \put(56,22){\large{$\mathbf{=}$}}
    \ArrowLine(131,61)(201,61)
    \ArrowLine(129,21)(199,21)
    \Gluon(151,62)(191,23){7.5}{2.56}
    \Vertex(152,64){7.07}
    \put(130,22){\large{$\mathbf{+}$}}
    \ArrowLine(313,29)(334,3)
    \GBox(289,28)(324,55){0.882}
    \ArrowLine(279,4)(300,28)
    \ArrowLine(326,49)(348,57)
    \ArrowLine(257,74)(287,50)
    \GlueArc(286,58.84)(27.01,-2,148.39){7.5}{3.63}
  \end{picture} \ ,
\end{equation}
that couples the four-point quark scattering kernel to a higher five-point 
function. This has been recently and independently stressed by 
Matevosyan, Tandy and Thomas \cite{Matevosyan:2006bk}. However, the confining 
properties of this kernel are not apparent. Indeed if we analyze the scaling
of the first term on the right hand side as a function of the gluon momentum
we obtain 
$kernel(k^2) \sim (k^2)^{-1/2-\kappa} (k^2)^{2\kappa-1} \sim (k^2)^{\kappa-3/2}$
which is not singular enough to represent a confining potential. If, however, we
would have two dressed quark-gluon vertices in this diagram we obtain
$kernel(k^2) \sim (k^2)^{-1-2\kappa} (k^2)^{2\kappa-1} \sim (k^2)^{-2}$, 
leading to a linear rising potential as discussed in subsection \ref{kernel-skel}.

In order to make this behavior of the kernel explicit we propose the following
scheme: (1) we employ a skeleton expansion for the (partially) 2PI quark-gluon 
scattering kernel to obtain a selfconsistent equation for the quark-gluon vertex
which reproduces our results from the main part of this paper, 
(2) we employ a symmetric formulation of the quark self-energy to obtain
Munczek's quark-quark scattering kernel with appropriate propagator cuts.
This symmetric formulation will lead to disagreement with perturbation
theory, but this can be systematically controlled. Then in step (3)  
the higher n-point-functions appearing in Munczek's construction can be
obtained from a closed system of equations by cutting the system in step
(1).

Let us give all these equations in diagrammatic form. Step (1) is materialized 
by giving the exact equation for the quark-gluon vertex in t-channel formulation, 
fig.\ref{DS2}, the BS equation for the connected kernel appearing there, and 
a model for the 2PI scattering kernel in terms of the quark-gluon vertex:
\begin{equation} \label{mysystem1}
  \begin{picture}(400,177) (11,-9)
    \SetScale{0.6}
    \SetWidth{1.5}
    \GBox(48,91)(77,145){0.882}
    \put(45,130){\large{$=$}}
    \ArrowLine(17,141)(48,141)
    \ArrowLine(77,142)(108,142)
    \Gluon(20,99)(47,99){5}{2}
    \Gluon(78,98)(105,98){5}{2}
    \ArrowLine(262,141)(293,141)
    \Gluon(265,99)(292,99){5}{2}
    \ArrowLine(322,142)(353,142)
    \Gluon(323,98)(350,98){5}{2}
    \GBox(352,93)(381,147){0.882}
    \ArrowLine(380,141)(411,141)
    \Gluon(381,97)(408,97){5}{2}
    \GOval(308,119)(39,16)(0){0.882}
    \put(90,130){\large{$+$}}
    \Vertex(48,219){5.66}
    \Gluon(124,261)(124,224){5}{2}
    \ArrowLine(87,189)(123,222)
    \ArrowLine(125,222)(151,190)
    \put(70,70){\large{$=$}}
    \put(145,70){\large{$+$}}
    \Gluon(48,260)(48,223){5}{2}
    \ArrowLine(11,188)(47,221)
    \ArrowLine(49,221)(75,189)
    \ArrowLine(176,189)(212,222)
    \GBox(212,199)(241,253){0.882}
    \ArrowLine(241,220)(267,188)
    \GlueArc(204.62,222.9)(21.09,72,233.24){5}{3}
    \GlueArc(251,254.41)(16.15,-124,36.41){5}{3}
    \ArrowLine(153,140)(184,140)
    \ArrowLine(213,141)(244,141)
    \Gluon(156,98)(183,98){5}{2}
    \Gluon(214,97)(241,97){5}{2}
    \GOval(198,116)(39,16)(0){0.882}
    \ArrowLine(17,53)(48,53)
    \ArrowLine(77,54)(108,54)
    \Gluon(20,11)(47,11){5}{2}
    \Gluon(78,10)(105,10){5}{2}
    \GOval(62,29)(39,16)(0){0.882}
    \put(75,15){\large{$=$}}
    \ArrowLine(172,53)(240,54)
    \Gluon(171,9)(239,9){5}{4}
    \Gluon(199,54)(204,15){5}{2}
    \Vertex(205,13){6}
    \Vertex(202,56){6}
  \end{picture}
\end{equation}

These three equations form a closed system whose input are the dressed
quark propagator, address below, and the dressed gluon propagator and
three-gluon vertex. In the quenched approximation considered in this work
these functions do not depend on the quark propagator. 

For step (2) let us express the self energy in a more symmetric form. By substituting 
the quark-gluon vertex DSE in t-channel form, fig.\ref{DS2}, into the self
energy one obtains
\begin{equation}\label{symmetricsE}
\begin{picture}(380,40) (17,-21)
    \SetScale{0.6}
    \SetWidth{1.5}
    \put(15,-5){\large{$\mathbf{\Sigma = }$}}
    \ArrowLine(88,-18)(166,-18)
    \GlueArc(126,-33)(41,21,157){5}{5}
    \put(110,-5){\large{$+$}}
    \GBox(254,-22)(287,19){0.882}
    \ArrowLine(215,-18)(254,-19)
    \ArrowLine(287,-18)(328,-19)
    \GlueArc(243,-14)(26,67,189){5}{3}
    \GlueArc(302,-14)(25,-15,125){5}{3}
\end{picture}.
\end{equation}
We now cast this self-energy in a form that is amenable to systematic improvement. 
\begin{equation}\label{symmetricsF}
  \begin{picture}(408,54) (17,10)
    \SetScale{0.6}
    \SetWidth{1.5}
    \put(-13,22){\large{$\mathbf{\Sigma =}$}}
    \ArrowLine(49,28)(128,28)
    \GlueArc(87,16)(42,26,152.06){5}{5}
    \Vertex(47,34){5.66}
    \Vertex(128,34){5.66}
    \put(80,22){\large{$\mathbf{-\sum_{n=0}^\infty n}$}}
    \GOval(320,51)(41,16)(0){0.882}
    \Gluon(269,76)(307,75){5}{2}
    \Gluon(334,77)(372,76){5}{2}
    \Vertex(354,84){7.21}
    \ArrowLine(270,28)(305,29)
    \ArrowLine(334,30)(369,31)
    \Vertex(354,31){7.21}
    \put(225,55){\large{$\mathbf{n}$}}
    \ArrowLine(224,27)(259,28)
    \Gluon(224,27)(245,65){5}{2}
    \Vertex(246,29){6}
    \Vertex(246,29){6}
    \Vertex(231,54){6}
    \Gluon(399,65)(418,30){5}{2}
    \ArrowLine(381,31)(418,30)
    \Line(281,90)(291,70)
    \Line(281,40)(291,20)
    \Line(395,40)(405,20)
    \Line(405,40)(415,60)
  \end{picture}
\end{equation}
This equation is easily obtained by substituting the DSE for the
quark-gluon four-point function,  second of eqs.(\ref{mysystem1}), 
into eq. (\ref{symmetricsE}) above. 

Resumming the series in (\ref{symmetricsF}) using 
$
\sum_{n=0}^\infty n x^n = x \left(\frac{1}{1-x}\right)^2.
$
would bring us back to the asymmetric formulation of the self-energy. 
In the present form eq.(\ref{symmetricsF}) is useful since it explicitly
contains the confining part of the interaction given by the diagram
with two dressed quark-gluon vertices. A truncation of eq.(\ref{symmetricsF})
at a given $n$ serves as a practical model for a confining kernel which
can be worked upon using Munczek's construction. The obvious approximation 
is to keep the first term with two dressed vertices and discard all terms 
in the sum. The resulting self-energy disagrees with perturbation theory 
at the two-loop level. But this is no worse than the often employed 
approximation of keeping the self-energy formally exact but approximating 
the quark-gluon vertex at one-loop in the skeleton expansion as in the 
third part of eq. (\ref{mysystem1}). 
Moreover, the inclusion of terms with a higher $n$ iterations of the 2PI
kernel systematically improves the perturbative properties of 
eq.(\ref{symmetricsF}). Note that also the 2PI kernel was approximated 
at first order in the skeleton expansion, so it also disagrees with 
perturbation theory at one loop (two-loops in the self-energy). This 
also needs to be systematically improved with the help of the skeleton 
expansion to finally match perturbation theory at any given order.

Cutting the self-energy equation then yields an equation for the
quark-quark scattering kernel:
\begin{equation} \label{mysystem2}
\begin{picture}(364,70) (50,0)
    \SetScale{0.6}
    \SetWidth{1.5}
    \GBox(32,52)(62,99){0.882}
    \ArrowLine(4,95)(32,95)    
    \ArrowLine(4,57)(32,57)
    \ArrowLine(61,95)(89,95)    
    \ArrowLine(61,57)(89,57)
    \Vertex(167,50){5.39}    
    \Vertex(188,95){5.39}
    \ArrowLine(286,97)(314,96)    
    \ArrowLine(340,96)(368,95)
    \ArrowLine(339,76)(367,75)
    \Vertex(295,46){5.39}
    \ArrowLine(294,48)(314,81)
    \GlueArc(308,62)(17,-120,37){5}{3}
    \GBox(314,73)(339,98){0.882}
    \ArrowLine(266,46)(294,46)    
    \ArrowLine(418,96)(446,96)
    \Vertex(448,96){5.40}
    \ArrowLine(447,96)(458,65)
    \GBox(458,44)(483,69){0.882}
    \ArrowLine(430,46)(458,46)    
    \ArrowLine(482,46)(510,46)
    \GlueArc(454,75)(24,-15,105){5}{3}
    \ArrowLine(483,67)(511,67)    
    \ArrowLine(137,50)(216,50)
    \ArrowLine(139,94)(218,94)
    \Gluon(166,52)(186,94){5}{3}
    \ArrowLine(564,50)(643,50)    
    \ArrowLine(566,94)(645,94)
    \Gluon(593,52)(613,94){5}{3}
    \put(320,40){\large{$-$}}
    \put(230,40){\large{$+$}}
    \put(150,40){\large{$+$}}
    \put(65,40){\large{$=$}}
  \end{picture} 
\end{equation}               
where we have used the derivative of the vertex with respect to the
propagator yielding a five-point function. To obtain an equation for 
this five point function we take a derivative of eq.~(\ref{mysystem1})
with respect to the quark propagator and yielding

\begin{equation} \label{mysystem3}
  \begin{picture}(381,170) (-2,40)
    \SetScale{0.6}
    \SetWidth{1.5}
    \GBox(30,309)(55,334){0.882}
    \ArrowLine(3,332)(30,331)    
    \ArrowLine(2,312)(29,311)
    \ArrowLine(56,332)(83,331)    
    \ArrowLine(57,310)(84,309)
    \Gluon(42,334)(41,365){5}{2}    
    \GBox(166,280)(191,305){0.882}
    \ArrowLine(139,285)(166,284)    
    \ArrowLine(191,282)(218,281)
    \Gluon(192,298)(219,296){5}{2}    
    \Gluon(177,308)(175,337){5}{2}
    \ArrowLine(135,340)(209,339)
    \GBox(304,291)(338,341){0.882}
    \ArrowLine(277,297)(304,296)    
    \ArrowLine(276,340)(303,339)
    \ArrowLine(339,336)(366,335)    
    \ArrowLine(340,294)(367,293)
    \Gluon(282,298)(303,320){5}{2}
    \Gluon(337,313)(366,310){5}{2}    
    \Gluon(135,205)(214,203){5}{5}    
    \Vertex(173,209){5.66}
    \GBox(160,233)(185,258){0.882}    
    \Gluon(168,234)(171,206){5}{2}
    \ArrowLine(186,254)(213,253)    
    \ArrowLine(185,238)(212,237)
    \ArrowLine(134,255)(161,254)    
    \ArrowLine(131,239)(158,238)
    \GBox(27,108)(61,158){0.882}
    \ArrowLine(0,114)(27,113)    
    \ArrowLine(-1,157)(26,156)
    \ArrowLine(62,153)(89,152)    
    \ArrowLine(63,111)(90,110)
    \Gluon(60,130)(89,127){5}{2}    
    \Gluon(-2,133)(27,130){5}{2}
    \GOval(43,223)(27,10)(0){0.882}
    \ArrowLine(50,242)(77,241)    
    \ArrowLine(9,244)(36,243)
    \Gluon(55,222)(82,220){5}{2}
    \ArrowLine(53,206)(80,205)    
    \ArrowLine(5,205)(32,204)
    \Gluon(6,224)(33,222){5}{2}    
    \GOval(171,130)(27,10)(0){0.882}
    \ArrowLine(178,149)(205,148)    
    \ArrowLine(137,151)(164,150)
    \Gluon(183,129)(210,127){5}{2}
    \ArrowLine(181,113)(208,112)    
    \ArrowLine(133,112)(160,111)
    \Gluon(134,131)(161,129){5}{2}
    \GBox(324,83)(349,108){0.882}
    \ArrowLine(295,87)(322,87)    
    \ArrowLine(349,87)(376,87)
    \Gluon(350,102)(379,99){5}{2}    
    \Gluon(313,143)(332,109){5}{2}
    \GOval(304,147)(22,8)(0){0.882}
    \ArrowLine(311,165)(338,165)    
    \ArrowLine(271,165)(298,165)
    \Gluon(268,141)(297,138){5}{2}    
    \GOval(483,124)(22,8)(0){0.882}
    \ArrowLine(490,141)(517,141)    
    \ArrowLine(450,141)(477,141)
    \Gluon(447,115)(476,115){5}{2}    
    \Gluon(492,115)(519,115){5}{2}
    \GBox(517,91)(551,141){0.882}
    \ArrowLine(492,93)(519,93)    
    \ArrowLine(552,93)(579,93)
    \Gluon(552,113)(579,113){5}{2}    
    \ArrowLine(552,136)(579,136)
    \put(64,190){\large{$=$}}
    \put(138,190){\large{$+$}}    
    \put(64,130){\large{$=$}}
    \put(64,75){\large{$=$}}    
    \put(140,75){\large{$+$}}    
    \put(235,75){\large{$+$}}
  \end{picture}
\end{equation}

It is thus apparent that we have achieved a truncation that is \emph{both}
chirally symmetric and confining.
The quark propagator can be obtained from the first term in equation (\ref{symmetricsF}) with
quark and gluon vertex given by simultaneously solving the two-equation system
(\ref{mysystem1}).
The matching quark scattering kernel is then obtained from equation
(\ref{mysystem2}) with the attending five-point function given by the two
equations in (\ref{mysystem3}).  One has two systems of two equations 
that need to be solved sequentially.

It is also apparent how the truncation can be matched to perturbation
theory to any given order in a systematic way by expanding a geometric
series of 2PI quark-gluon scattering kernels, and expanding the 2PI kernel
itself. The number of terms as usual grows very rapidly (but also in
perturbation theory). 
This is a hard task, and it is far from our intention to 
develop this program in the near future, but at least we have found an 
algorithm for the wanted construction that may allow statements of 
principle.

Alternatively, and especially for heavy quark applications, one is not so
interested in preserving chiral symmetry explicitly, since it is already badly 
broken by the large current quark mass. Then a simple skeleton truncation 
of the kernel may represent a suitable approximation.

%%%%%%%%%%%%%%%%%%%%%%%%%%%%%%%%%%%%%%%%%%%%%%%%%%%
\section{Coupling to photons}
%%%%%%%%%%%%%%%%%%%%%%%%%%%%%%%%%%%%%%%%%%%%%%%%%%%

In this appendix let us comment on the quark-photon vertex.
Hard electromagnetic probes (Deep Inelastic Scattering off the nucleon, 
jets in
electron-positron collisions) and final states (Drell-Yan processes) are the
tell-tale experiments about the existence of quarks. Moreover,
electromagnetic form factors also provide a glimpse of hadron structure at
intermediate energies. It is therefore worth paying a moment's attention to
the quark-photon coupling to see what effect if any does the infrared
counting have on it. 

The  Bethe-Salpeter equation for the quark-photon
coupling to flavor $i$, that for $N_f$ active flavors, and in term of the
$j\to i$ (possibly flavor changing) scattering kernels reads

\begin{center}
  \begin{picture}(361,133) (6,-2)
    \SetWidth{1.5}
    \SetScale{1.0}
    \Vertex(42,52){7.21}
    \put(74,55){\huge{\Black{$=$}}}
    \Photon(45,119)(45,56){5}{4}
    \ArrowLine(8,15)(43,55)
    \ArrowLine(43,55)(71,15)
    \Photon(142,118)(142,55){5}{4}
    \ArrowLine(105,14)(140,54)
    \ArrowLine(140,54)(168,14)
    \put(175,55){\huge{$\mathbf{+ \sum_j^{N_f}}$}}
    \GBox(268,43)(322,75){0.882}
    \Photon(296,131)(299,100){5}{3}
    \ArrowLine(274,75)(297,99)
    \ArrowLine(297,99)(314,75)
    \ArrowLine(242,5)(278,42)
    \ArrowLine(308,42)(336,7)
    \put(270,90){\huge{$\mathbf{j}$}}
    \put(320,90){\huge{$\mathbf{\bar{j}}$}}
    \put(6,-2){\huge{$\mathbf{i}$}}
    \put(63,-2){\huge{$\mathbf{\bar{i}}$}}
    \put(105,-2){\huge{$\mathbf{i}$}}
    \put(170,-2){\huge{$\mathbf{\bar{i}}$}}
    \put(254,-2){\huge{$\mathbf{i}$}}
    \put(337,-2){\huge{$\mathbf{\bar{i}}$}}
  \end{picture}
\end{center}              
The first term is irrelevant for the
purpose of infrared counting. The second displays the quark-quark scattering 
kernels, that have an infrared dimension of $(p^2)^{-2}$
This is compensated by the integration measure 
$d^4p$, so that the infrared counting becomes trivial. Thus the equation
is consistent with an infrared finite solution. This is also what one would
expect from the Abelian Ward-identity and vector meson dominance. In
addition, past and recent phenomenological studies of various form factors 
\cite{Maris:2000sk,Oettel:2000jj,Holl:2005vu,Bhagwat:2006pu} seem to do 
well without the need for an infrared enhanced vertex.

A further question is whether any experiment so far is actually sensitive to
the infrared divergences. We cannot think of an observable that really
constrains this. For example, if one thinks of time-like pair production or
jets at $e^- e^+$ colliders, the diagram that would show our
enhanced quark-gluon vertex is
\begin{center}
  \begin{picture}(110,53) (14,-23)
    \SetWidth{1.5}
    \Photon(14,6)(64,9){5}{3}
    \ArrowLine(65,10)(117,30)
    \ArrowLine(64,9)(109,-23)
    \Vertex(65,9){5.66}
    \Vertex(93,21){5.66}
    \Gluon(91,21)(122,-3){5}{3}
  \end{picture}
\end{center}
One sees immediately that, by the very definition of a jet, the soft-gluon kinematic
point is cut out of the data.

It does appear that one cannot constrain the infrared counting easily from
experiment. A first attempt in this direction is given in 
ref.~\cite{Aguilar:2004td}, 
but much more work is necessary.
Similar considerations apply to the coupling of quarks to other external
currents, such as axial currents necessary for the weak interactions of
hadrons.

%%%%%%%%%%%%%%%%%%%%%%%%%%%%%%%%%%%%%%%%%%%%%%%%%%%%%%%%%%%%%%%%%%%%%%%%%
\section{The Gross-Yennie argument and its failure in QCD \label{GY}}
%%%%%%%%%%%%%%%%%%%%%%%%%%%%%%%%%%%%%%%%%%%%%%%%%%%%%%%%%%%%%%%%%%%%%%%%%
\subsection{Crossed ladders simplify $k_0$ pole analysis in QED}
%%%%%%%%%%%%%%%%%%%%%%%%%%%%%%%%%%%%%%%%%%%%%%%%%%%%%%%%%%%%%%%%%%%%%%

In the following we explain why the familiar Gross-Yennie construction of a 
potential for photon exchange in QED cannot be applied to gluon exchange within QCD.

In an equal-time description, all physical particles are put on-shell and
energy is not conserved in the interaction vertices. Therefore we start by considering
the first skeleton diagram of our (heavy) quark-quark scattering kernel in
the center of momentum (CoM) frame, with total energy $W=P^0$ as customary,
\be
  \begin{picture}(300,100)(-60,-30)
    \SetWidth{1.5}
    \SetScale{0.5}
    \ArrowLine(0,80)(200,80)
    \ArrowLine(0,-6)(200,-6)
    \Vertex(85,80){7.07}
    \Vertex(85,-8){7.07}
    \Gluon(85,77)(85,-5){7.5}{4.43}
    \put(-40,-20){\large{$\mathbf{(\frac{W}{2}-q^0,-{\bf q}})$}}
    \put(-40,50){\large{$\mathbf{(\frac{W}{2}+q^0,{\bf q}})$}}
    \put(65,-20){\large{$\mathbf{(\frac{W}{2}-q^{0'},-{\bf q}')}$}}
    \put(65,50){\large{$\mathbf{(\frac{W}{2}+q^{0'},{\bf q}'})$}}
\end{picture}
\ee
and apply our IR counting. In Euclidean space, if all scales were
simultaneously sent to zero one obtains 
$(P_E^2)^{2(-1/2-\kappa)-1+2\kappa} = \frac{1}{P_E^4}$
i.e. a strong IR kernel in terms of the Euclidean momentum $P_E$.
With $(\frac{W}{2}+q_0)$ (and all others) fixed at its on-shell value by 
a Dirac delta function (see also
below), all other singularities in the $(\frac{W}{2}+q_0)$ plane cancel and one can
perform a Wick rotation. 
The statement is then that all particles going on-shell simultaneously, the
skeleton diagram would be proportional to the inverse fourth-power of the
off-shell momentum $P$. 
However, to write down a Hamiltonian description we only put the four quarks
on-shell, the gluon rung giving rise to a potential interaction. Therefore
the condition that the gluon also goes on-shell for our vertex becoming
strongly enhanced, 
\begin{equation}
(q^0-q^{0'})^2 -({\bf q-q}')^2 \to 0
\end{equation}
becomes, taking into account the mass-shell condition for the four quark
legs that has been imposed beforehand,
\begin{equation}
(\frac{W}{2}-q_0)^2-{\bf q}^2 =M^2
\end{equation}
(and three others), 
\begin{equation}
(E_q-E_{q'})^2-({\bf q}-{\bf q}')^2 \to 0
\end{equation}
which therefore leads to the statement
\begin{equation}
\hat{V}({\bf q}-{\bf q}') = \frac{8\pi\sigma_1}{({\bf q}-{\bf q}')^4}
\end{equation}
that is, a strong confining potential with string tension named $\sigma_1$
(for the first term in the skeleton expansion) and Fourier transform to
coordinate space linear in the relative coordinate
\begin{equation}
V(r)= \sigma_1\ \  r \ .
\end{equation}
Here $\sigma_1$ includes the Casimir color factor $4/3$.

The string tension can be extracted from our kernel by contracting all its
indices with appropriate spinors. 
One can of course not trust the Born approximation since the kernel is
strong, and therefore an iteration mandatory. 

Upon iteration of the kernel (following diagram on the left), 
the quarks in the intermediate state are now integrated over the off-mass 
shell region
\begin{center} 
  \begin{picture}(328,90) (12,-13)
    \SetScale{0.7}
    \SetWidth{3}
    \ArrowLine(20,14)(169,14)
    \ArrowLine(240,14)(386,14)
    \SetWidth{1}
    \Vertex(50,84){5}
    \Vertex(50,12){5}
    \Vertex(132,12){5}
    \Vertex(132,82){5}
    \ArrowLine(240,82)(386,82)
    \ArrowLine(19,82)(166,82)
    \Vertex(270,12){5}
    \Vertex(352,12){5}
    \Vertex(352,82){5}
    \Vertex(270,82){5}
    \Gluon(50,82)(50,14){5}{5}
    \Gluon(132,82)(132,12){5}{5}
    \Gluon(270,82)(352,12){5}{8}
    \Gluon(270,12)(352,82){5}{8}
    \put(12,64){\large{$\mathbf{q}$}}
    \put(60,64){\large{$\mathbf{k}$}}
    \put(105,64){\large{$\mathbf{q'}$}}
    \put(4,-8){\large{$\mathbf{P-q}$}}
    \put(50,-8){\large{$\mathbf{P-k}$}}
    \put(95,-8){\large{$\mathbf{P-q'}$}}
    \put(170,64){\large{$\mathbf{q}$}}
    \put(222,64){\large{$\mathbf{k}$}}
    \put(265,64){\large{$\mathbf{q'}$}}
    \put(151,-12){\large{$\mathbf{P-q}$}}
    \put(264,-12){\large{$\mathbf{P-q'}$}}
    \put(184,-8){\large{$\mathbf{(P\! +\! k\!-\!q\!-\! q')}$}}
    \ArrowArc(95,54)(24,-116,106)
    \put(55,33){\large{$ \mathbf{-k_0}$}}
    \put(140,30){\large{${\mathbf +}$}}
  \end{picture}
\end{center}
This equation is strictly valid only in QED, with fermions interacting via
photon exchange. There, it has been shown by Gross and Yennie how the addition of the
crossed-rung diagram leads to the correct one-body limit (the second particle
moves in a static potential created by the first particle whose mass is
large). To see this, assume that the thicker bottom line in the diagrams 
above corresponds to a heavy particle of mass $M\to \infty$. Then the two
propagators of the heavy particle in both diagrams can be combined. 
In the static limit, the physical intuition is that the total energy becomes 
the energy of the lighter  particle plus the mass of the heavy particle that
can absorb recoil momentum without absorbing energy, $P_0=e+M$, with
$e=q_0=q_0'$. For the heavy particle we will also ignore the negative energy
part of the propagator, suppressed by a further $1/M$. With this notation,
the difference of the two diagrams is given by a common factor times the
difference of the internal fermion propagators in the heavy mass limit, i.e. 
\begin{eqnarray}
\frac{1}{M^2-(P-k)^2-i\epsilon} + \frac{1}{M^2-(P+k-q-q' )^2-i\epsilon}
\to \\ \nonumber
\frac{1}{2M} \left(
\frac{1}{M-P_0+k_0-i\epsilon} + \frac{1}{M-P_0-k_0+q_0+q_0'-i\epsilon}
\right) &=& \\ \nonumber
\frac{1}{2M} \left(
\frac{1}{k_0-e-i\epsilon} - \frac{1}{k_0-e+i\epsilon}
\right) &=& 
\frac{2\pi i \delta (k_0-e)}{2M} \ .
\end{eqnarray}
Thus, if one of the particle's masses is sent to infinity, the intermediate
$k_0$ variables are fixed to their on-shell value $e$ as intuitively
expected. The Wick rotation is then possible, and one can take the large 
$M$ limit for the second particle if so wished. Gross and Yennie have given 
the induction step leading to the cancellation to all orders of perturbation 
theory \cite{Gross:1982nz}, by summing all ladders and crossed ladders. 
The argument is valid in the presence
of vertex and gluon-propagator dressing functions, if the naive 
vertex mass-counting of perturbation theory holds, and thus their proof 
is
essentially complete for QED in the static (automatically quenched) limit. 

In QCD, however, there appear extra color factors in each of the diagrams,
such that the simple cancellation mechanism of Gross and Yennie cannot work.
We are aware of no extension of the argument to include color factors as
well as QCD gluon radiation. The need for including all crossed ladders makes 
difficult to show the cancellation explicitly when one rewrites the perturbation 
series as an integral equation, 
\begin{equation} \label{BetheSalpeterinho}
  \begin{picture}(528,60) (-70,-41)
    \SetWidth{1.5}
    \ArrowLine(7,10)(40,10)
    \ArrowLine(7,-23)(39,-23)
    \GBox(40,-30)(74,15){0.882}
    \ArrowLine(74,-23)(107,-23)
    \ArrowLine(74,10)(107,10)
    \put(92,-12){\huge{$\mathbf =$}}
    \GOval(130,-8)(27,6)(0){0.882}
    \put(142,-13){\huge{$\mathbf +$}}
    \GBox(204,-30)(238,16){0.882}
    \ArrowLine(238,-23)(271,-23)
    \ArrowLine(238,10)(271,10)
    \ArrowLine(171,-23)(204,-23)
    \ArrowLine(171,10)(204,10)
    \GOval(170,-8)(27,6)(0){0.882}
\end{picture}
  \end{equation}
as the canceling poles are entangled between the explicit
propagators and the 2PI kernel. Therefore, no truncation of the
full Bethe-Salpeter equation is known that consistently takes the one-body
limit correctly.

%%%%%%%%%%%%%%%%%%%%%%%%%%%%%%%%%%%%%%%%%%%%%%%%%%%%%%%%%%%%%%%%%%%%%%%%
\subsection{The Gross equation \label{Gross}}
%%%%%%%%%%%%%%%%%%%%%%%%%%%%%%%%%%%%%%%%%%%%%%%%%%%%%%%%%%%%%%%%%%%%%%%%%

Given that the Salpeter equation (see section \ref{salpeter} in the main part 
of this paper) needs an inconsistent approximation of the
energy $k_0$ thus introducing an uncontrollable systematic error, Gross devised
an equation inspired by the Gross-Yennie cancellation mechanism. The idea is
to replace the bona-fide propagators $S_q S_{\bar{q}}$ from the field theory 
by an ad-hoc Gross function, containing a Dirac delta that puts the quarks 
on their mass-shell, and subsequently evaluating the kernel at the pole 
thus forced. 
The Gross function in the center of mass frame for undressed fermions is
simple and natural,
\begin{equation}
G(k,P)=  \frac{2\pi \delta((S_q^{-1}-S_{\bar{q}}^{-1})/2)}{S_q^{-1}
+S_{\bar{q}}^{-1}}
=\frac{2\pi \delta(k_0)}{2P^0(M^2+{\bf k}^2-P_0^2/4)} \ .
\end{equation}
However, in the presence of momentum-dependent dressing functions, one
obtains a quite tedious two-particle propagator, where we use the shorthands
$A_\pm=A((P^0/2\pm k^0)^2-{\bf k}^2)$ (sim. for B), 
\begin{eqnarray}
((\not k + \frac{P^0}{2}\gamma^0)A_+ - B_+)\times 
((-\not k + \frac{P^0}{2}\gamma^0)A_- - B_-) \delta(k^0 - \bar{k}^0)
\\ \nonumber
\left[ \left(
2(P_0+2k_0)(B_+ B'_+ +((P^0/2+ k^0)^2-{\bf k}^2) A_+ A'_+
)-2(P_0-2k_0)(B_- B'_- +((P^0/2- k^0)^2-{\bf k}^2) A_- A'_- ) 
\right. \right. \\ \nonumber \left. \left.
+(P_0+2k_0)A_+^2-(P_0-2k_0)A_-^2\right) 
\left(((P^0/2+ k^0)^2-{\bf k}^2)A_+^2+B_+^2
\right) \right]^{-1}
\end{eqnarray}
where the on-shell pole $\bar{k}^0$ is now located by solving the implicit
equation
\begin{equation}
B^2_+-B^2_- +((P^0/2+\bar{k}^0)-{\bf k}^2)A^2_+-((P^0/2-\bar{k}^0)-{\bf
k}^2)A^2_-=0
\end{equation}
(a numeric fixed-point method or Newton iteration is necessary since the
$B_+$, etc.
functions depend on $\bar{k}^0$).
The resulting local potential that would go into 
 eq.(\ref{schroedinger}) is again formally 
the same as with the Born or Salpeter constructions, except for the
different pole position, possibly causing a shift in the gluon momentum and
obscuring the confining soft-gluon divergence,
and in addition the resulting non-local kinetic energy for the equation 
is markedly more difficult.

It is difficult to see how a confining potential arises here except in the
limit $A\to 1,\ B\to M$ where $\bar{k}_0 \to 0$ reproducing essentially the
same result than the one of section \ref{salpeter}.

%%%%%%%%%%%%%%%%%%%%%%%%%%%%%%%%%%%%%%%%%%%%%%%%%%%%%%%%%%%%%%%

\end{document}